\documentclass[12pt]{iopart}

\usepackage{graphicx}
\usepackage{subfig}
\usepackage{amsmath}
\usepackage{xcolor}
\usepackage{hyperref}
\usepackage{svg}
\svgsetup{inkscape=/Applications/Inkscape.app/Contents/MacOS/inkscape}
\usepackage{booktabs}
\usepackage{multirow}
\usepackage{adjustbox}

\begin{document}

\title[Population of spiking neurons to balance an inverted pendulum]{Benchmarking Spiking Neurons for Linear Quadratic Regulator Control of Multi-linked Pole on a Cart: from Single Neuron to Ensemble}

\author{Shreyan Banerjee$^1$, Luna Gava$^2$, Aasifa Rounak$^1$, Vikram Pakrashi$^1$}
\address{$^1$ UCD Centre for Mechanics, Dynamical Systems and Risk Laboratory,
School of Mechanical and Materials Engineering, University College Dublin,
Belfield, Dublin 4, Dublin, Ireland.\\
$^2$ Event-Driven Perception for Robotics, Istituto Italiano di Tecnologia, Via San Quirico 19d, 16163, Genova, Italy.}
\ead{vikram.pakrashi@ucd.ie}
\vspace{10pt}
\begin{indented}
\item[] March 2024
\end{indented}

\begin{abstract}
The emerging field of neuromorphic computing for edge control applications poses the need to quantitatively estimate and limit the number of spiking neurons, to reduce network complexity and optimize the number of neurons per core and hence, the chip size, in an application-specific neuromorphic hardware. While rate-encoding for spiking neurons provides a robust way to encode signals with the same number of neurons as an ANN, it often lacks precision. To achieve the desired accuracy, a population of neurons is often needed to encode the complete range of input signals. However, using population encoding immensely increases the total number of neurons required for a particular application, thus increasing the power consumption and on-board resource utilization. A transition from two neurons to a population of neurons for the LQR control of a cartpole is shown in this work. The near-linear behavior of a Leaky-Integrate-and-Fire neuron can be exploited to achieve the Linear Quadratic Regulator (LQR) control of a cartpole system. This has been shown in simulation, followed by a demonstration on a single-neuron hardware, known as Lu.i. The improvement in control performance is then demonstrated by using a population of varying numbers of neurons for similar control in the Nengo Neural Engineering Framework, on CPU and on Intel's Loihi neuromorphic chip. Finally, linear control is demonstrated for four multi-linked pendula on cart systems, using a population of neurons in Nengo, followed by an implementation of the same on Loihi. This study compares LQR control in the NEF using $7$ control and $7$ neuromorphic performance metrics, followed by a comparison with other conventional spiking and non-spiking controllers. 
% This work lays the foundation of linear neuromorphic control with two to many neurons, both on simulation and hardware.
\end{abstract}


%\maketitle

\section{Introduction}
Neuromorphic technology, based on the hardware implementation of spiking neurons and the co-localization of memory and computation, is a promising venue for low-power, low-latency computation on the edge~\cite{dewolf2020nengo}. These properties make neuromorphic a key enabling technology for applications that require fast reactions to sensory stimuli with limited space and power budget.
While a lot has been done in the domain of sensing and perception~\cite{gallego2020event}, the work on spike-based control is still limited. Among these few works, some were applied to control different robotic platforms using neuromorphic processors~\cite{perez2013neuro, davies2018loihi, zhao2020closed, stagsted2020towards, paredesvalles2023fully, dewolf2016, michael2022, alex2022, marrero2024}.
A complete spike-based architecture to control the robot head movements from the dynamic vision sensor in open-loop was presented~\cite{perez2013neuro}.  
Conventional control techniques, such as the Proportional Integral and Derivative (PID) control have been applied to drones using neuromorphic computing~\cite{stagsted2020towards}, where a rate-encoded error signal is passed to populations of spiking neurons corresponding to the P, I, and D coefficients, and the control torque is obtained by summing up the output. This is done on the Loihi neuromorphic chip~\cite{davies2018loihi}. Similarly, a closed-loop PID controller for the humanoid robot iCub was presented ~\cite{zhao2020closed} (using an open-source iCub
humanoid platform simulator called iCubSim, controlled with Yet Another Robot Platform (YARP)), which encodes the space-of errors using a space-to-rate encoding scheme and implements the P, I, and D terms using three separate populations of spiking neurons. 

\textcolor{black}{DeWolf \textit{et al.} \cite{dewolf2016} demonstrates the adaptive control of a robotic arm using anatomically structured spiking neurons using the Neural Engineering Framework (NEF). The dynamics of the robotic arm is computed and the target to be reached is provided to the anatomically organized REACH model, that computes the desired joint forces. Volinski \textit{et al.} \cite{michael2022} also uses the REACH model in the NEF framework for adaptive control of a wheelchair-mounted robotic arm for people with upper body disabilities. Ehrlich \textit{et al.} \cite{alex2022} shows that SNNs require much less parameters and inference times compared to ANNs for the estimation of inverse kinematics in robotic systems. The authors used Intel's Loihi neuromorphic hardware and non-spiking NVIDIA’s Xavier to demonstrate their results on a 6-DOF robotic arm system. The use of  NEF to implement PID control on a 3-DOF robotic arm by mapping the PID coefficients on the synaptic weights has also been demonstrated in  \cite{marrero2024}. 
DeWolf \textit{et. Al.} \cite{Dewolf2023} implements an SNN running on Intel's Loihi neuromorphic chip to control a 7-DOF robotic arm in simulation.This work shows that the energy consumption and the computational latency in Loihi is very less compared to standard digital hardware like CPU and GPU. Zaidel \textit{et. Al.} \cite{Zaidel2021} demonstrates inverse kinematics and the PID control of 6-DOF robotic arm using the NEF. Arana \textit{et. Al.} \cite{Arana2023} demonstrates the regularization and trajectory tracking of an underactuated Ball and Plate platform using SNNs with NEF.}
%A fully neuromorphic vision-to-control pipeline~\cite{paredesvalles2023fully} was proposed for controlling a small flying robot, but the control part consists of a single decoding layer linearly mapping the visual spiking observations to control commands. 
Other studies focused on spiking-based control of canonical and extensively studied mechanical systems, such as the inverted pendulum. This system choice is preferred as a starting point for the control study because it can scale up in complexity by adding components, such as more links, and can modify behaviour by changing the physical characteristics of the system. Moreover, the cartpole represents an under-actuated system whose control is challenging and its dynamics form the basis of other kinds of systems such as self-balancing robots, human transporters, bipedal locomotion, and missile launching.
For example, a moving wheel inverted pendulum (MWIP) system was balanced using a
SNN-based linear quadratic regulator (LQR)~\cite{juarez2022spiking} implemented in the Neural Engineering Framework known as Nengo~\cite{bekolay2014nengo}. This work shows that the control signal is noisy, but it can still reduce steady-state errors for the system response variables over time. At the same time, another study related to the control of canonical mechanical systems~\cite{wiklendt2009small} addressed both swing-up and stabilization of a double pendulum using an SNN learnt through evolutionary computing, combined with an LQR controller. 
\textcolor{black}{A benchmark example of using cartpole as a system with multiple levels of difficulty, multiple parameter settings of the RISP neuroprocessor, and multiple spike encoding schemes for applying neuromorphic reinforcement learning (RL) agents have been demonstrated by Plank \textit {et. Al.} \cite{plank2025}. An optimal spike encoding algorithm called SECLOC is devised and used to control a cartpole both in simulation and hardware, using PID and LQR controllers by Diaz \textit{et. Al.} \cite{diaz2024}. A closed-loop optimal control using Linear Quadratic Gaussian (LQG) controller on a spring-mass-damper and a cartpole system using Spike Coding Network (SCN) is demonstrated by Slijkhuis \textit{et. Al.} \cite{slijkhuis2023}. A cloud edge-based online supervised learning framework is proposed by Ahmadvand \textit{et. Al.} \cite{ahmadvand2024cloud}, that can work under strict energy constraints. The effect of different decoding schemes and a learned encoding scheme on the evolutionary training of a cartpole and a lunar lander system is explored by Rafe \textit{et. Al.} \cite{rafe2021}. This work shows an inverse relationship between the state exposure times and the generations needed to reach the goal for the reinforcement learning agent. An evolutionary framework for training the connectivities in a spiking neural network using a synapse-level Bellman equation is proposed by Banerjee \textit{et. Al.} \cite{banerjee2024control}. An adaptive LQR-based control of a cartpole using Prescribed Error Sensitivity learning rule has been proposed by Banerjee \textit{et. Al.} \cite{banerjee2024adaptive}. Arana \textit{et. Al.} \cite{Arana2024} uses a controller designed using the Neural Engineering Framework (NEF) for the stabilization and tracking of an Unmanned Aircraft System (UAS), using PD-like control.}

Despite these contributions, existing literature lacks the demonstration of the effect of the number of neurons on vital control performance metrics. Topology evolution methods mostly rely on heuristics~\cite{qiu2018evolving} and limit their work to a double-pendulum. A significant effort in limiting the number of neurons needed for a control application has been taken by Cook~\cite{cook2004takes}, which suggests that only two spiking neurons may suffice for a controller to learn bicycle riding in simulation. 

However, there is a need to quantitatively analyze the number of neurons required for controlling more universal canonical systems to ensure scalability and reduce the footprint on neuromorphic hardware to be applied for edge control applications.

%This paper aims to study the effect of the number of spiking neurons on control performance and to limit the number needed for the linear control of the inverted pendulum on a cart system, with varying number of pendulum links, by moving from a rate-encoding scheme with $2$ neurons, to a population-encoded scheme for an improved performance. 
\textcolor{black}{The main objective of the paper is to show the variation of control performance metrics and on-board resource utilization, as one moves from rate encoding of
the input with very few neurons, to population encoding, by varying the number of neurons. This is demonstrated using a cartpole as a canonical system. This knowledge is also used to control a
multi-linked pendulum on a cart, both in simulation and on hardware. This is important because neuromorphic systems operate within strict constraints on power, and resource utilization. Limiting
the number of neurons is of prime importance when deploying neuromorphic chips on the edge, where there is a limited space and energy budget.}
Various spiking models of neurons have been proposed in the literature, each based on a particular type of biological neuron~\cite{pakdaman2001periodically,izhikevich2003simple,nelson1995hodgkin,rowe2017robotic}. In this work, the Leaky-Integrate-and-Fire (LIF) model is employed due to its near-linear behaviour and diverse applications. 

Then key contributions of the paper are:
\begin{itemize}
    \item Demonstrating LQR-based balancing of a cartpole using two LIF neurons, both in simulation and hardware, with a rate-based encoding scheme.
    \item Moving from rate-based to population-based encoding using an ensemble of spiking neurons in Nengo NEF to increase the precision of control for a cartpole. This is supported by an exploratory analysis of the effect of increasing the number of neurons on the Key Performance Indicators (KPIs) of neuromorphic computing (on CPU and Loihi), and control.
    \item Using population-encoded neurons to control 1 to 4-linked poles on a moving cart and studying the KPIs thereof.
    \item Implementing such an ensemble on Intel's Loihi neuromorphic chip to control 1 to 2-linked poles on a moving cart.
    \item Comparing this implementation with $3$ other spiking controllers and $4$ non-spiking controllers, and studying the KPIs for control and neuromorphic simulation thereof.
\end{itemize}

This study will benefit edge applications of neuromorphic boards by constraining the power requirements, computational overhead, onboard space allocation, and weight. Section \ref{sec:methods} explains the control method and the system equations used in designing the simulator. Section \ref{sec:experiments} outlines the experiments performed, both in simulation and hardware. Section \ref{sec:results} illustrate the outcomes of the experiments and section \ref{sec:conclusion} concludes the study with some future directions.

\section{Methods}\label{sec:methods}
The inverted pendulum on a cart~\cite{florian2007correct} is a canonical system often used to study linear and non-linear under-actuated control. A cart with a single pole hinged to it, rotating in a plane forms the cartpole. It has two degrees of freedom (DOFs) corresponding to the linear motion of the cart and the rotatory motion of the pole, respectively. Additionally, the study of chaotic dynamics and their control can be performed with multi-linked pendulums on a cart for large angles~\cite{kaheman2023experimental, he2022chaotic}. Xin et. al.~\cite{xin2018linear} demonstrated that an n-linked pendulum on a cart has linear controllability around its Unstable Equilibrium Position (UEP) for all physical parameters.
The subsequent subsection presents the equations of motion for the chosen system followed by the equations applicable to a general system with an n-linked pendulum on a cart.

\subsection{Cartpole Dynamics}\label{dynamics}
\begin{figure}
    \centering
    \includegraphics[scale = 0.65]{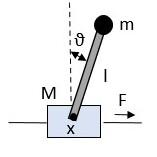}
    \caption{The cartpole system, where $x$ is the position of the cart on the rail, $\theta$ is the angle of the pole from the vertical, $\vec{F}$ is the force on the cart, $M$ is the mass of the of the cart, $m$ is the mass of the pole, and $l$ is the pole length.The other symbols hold their usual meanings.}
    \label{fig:cart-pole_scheme}
\end{figure}

Figure~\ref{fig:cart-pole_scheme} shows a schematic of the cartpole system with two state variables, namely, \(x\), which is the displacement of the cart on the rail centered at \(0\) and \(\theta\), which is the angular displacement of the pole about the vertical Unstable Equilibrium Position (UEP). The control objective is to balance the pole for small angle perturbations about the UEP, by applying a positive or negative force $F$ on the cart in one dimension along the rail to move it to the right or the left respectively. The general equations of motion for any system are: 
\begin{equation}\label{GEN}
\ddot{\mathbf{X}} = f(\mathbf{X},\mathbf{U})
\end{equation}
and its corresponding linearized model is:
\begin{equation}\label{GENlin}
\begin{cases}
    \dot{\mathbf{X}} = A \mathbf{X} + B \mathbf{U}\\
    \mathbf{Y} = C \mathbf{X} + D \mathbf{U}
\end{cases}
\end{equation}
where $X$ is the system state, $U$ is the control input, and $Y$ is the observable output, $A$,$B$,$C$ and $D$ are the system parameter matrices.

For the cartpole, we define the state and the control input as:

%\begin{equation}
\begin{equation}\label{cartpole_state}
    \mathbf{X} = 
\begin{bmatrix}
    x &
    \dot{x} &
    \theta &
    \dot{\theta} 
\end{bmatrix}\\
\end{equation}
\begin{equation}\label{cartpole_input}
U = F
\end{equation}
%\end{equation}

To get the values of the state parameter matrices $A$ and $B$, the Lagrangian equations are solved to derive the system-specific equations of motion for the cartpole:

\begin{equation}\label{cartpoleEOM}
\begin{bmatrix}
m+M & mlcos(\theta) \\
mcos(\theta) & ml
\end{bmatrix}
\begin{bmatrix}
\ddot{x} \\
\ddot{\theta} \\
\end{bmatrix}
=
\begin{bmatrix}
mlsin(\theta){\dot{\theta}}^2 + F \\
mgsin(\theta) \\
\end{bmatrix}
\end{equation}

where $g$ is the acceleration due to gravity, $l$ is the length of the pole, \(M\) is the mass of the cart, \(m\) is the mass of the pole, and \(F\) is the force on the cart.

Using small angle perturbations around the UEP ($\theta = 0$), allows us to substitute $sin(\theta)$ with $\theta$, and  $cos(\theta)$ with 1, while the squared terms approximate to zero. These approximations give a linearized state-space model of the cartpole:

\begin{equation}\label{StateSpace1} 
\begin{cases}
\begin{bmatrix}
\dot{x}\\
\ddot{x}\\
\dot{\theta}\\
\ddot{\theta}\\
\end{bmatrix}
=
\begin{bmatrix}
0 & 1 & 0 & 0 \\
0 & 0 & -\frac{mg}{M} & 0 \\
0 & 0 & 0 & 1 \\
0 & 0 & -\frac{(m+M)g}{ml} & 0 \\
\end{bmatrix}
\begin{bmatrix}
x\\
\dot{x}\\
\theta\\
\dot{\theta}\\
\end{bmatrix}
+
\begin{bmatrix}
0\\
\frac{1}{M}\\
0\\
-\frac{1}{Ml}\\
\end{bmatrix}
u
\vspace{0.5cm}
\\
\begin{bmatrix}
y_{1}\\
y_{2}\\
y_{3}\\
y_{4}\\
\end{bmatrix}
=
\begin{bmatrix}
1 & 0 & 0 & 0 \\
0 & 1 & 0 & 0 \\
0 & 0 & 1 & 0 \\
0 & 0 & 0 & 1 \\
\end{bmatrix}
\begin{bmatrix}
x\\
\dot{x}\\
\theta\\
\dot{\theta}\\
\end{bmatrix}
\end{cases}
\end{equation}

\begin{figure}
    \centering
    \includegraphics[scale = 0.65]{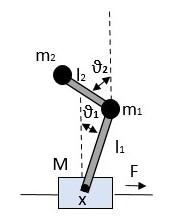}
    \caption{Double inverted pendulum on a cart. $\alpha$ and $\beta$ are respectively the angles of the first and second links with respect to the vertical(anticlockwise positive).}
    \label{fig:DPC}
\end{figure}

Extending this analysis to an n-linked pendulum on a cart (n+1 DOF) the general equations of motion are obtained:

\begin{equation}
    \dot{\mathbf{X}} = A_{n\times n} \textbf{X}_{n \times1} + B_{n\times m}\textbf{U}_{m \times 1}
\end{equation}
\begin{equation}
    \textbf{Y}_{p \times 1} = C_{p \times n} \textbf{U}_{n \times 1}
\end{equation}

The equations of motion for a n-linked pole on a cart has the following form,
\begin{equation}\label{GenEOM}
A\ddot{\mathbf{X}} = B + C 
\end{equation}
% (\sum_{i=1}^{n} m_i)l_1 & (\sum_{i=2}^{n} m_i)l_2 cos(\theta_1 - \theta_2) &  \dots 

where \\
\begin{align}
&A=\\
&\begin{bmatrix}
\sum_{i=1}^{n} m_i + M  & (\sum_{i=1}^{n} m_i) l_1 cos(\theta_1) & \dots &  m_n l_n cos(\theta_n)  \\
\\
(\sum_{i=1}^{n} m_i) cos(\theta_1) & \ddots & &  \\
%\vdots & \ddots & & \\
\vdots & & [a_{ij}] &\\
m_ncos(\theta_n) & & & \ddots \\
\end{bmatrix} 
\end{align}
\vspace{2mm}

where
\begin{equation}
a_{ij} = [\sum_{k=max(i,j)}^{n} m_k] l_j  cos(\theta_i - \theta_j)
\end{equation}
\begin{equation}
B = 
\begin{bmatrix}
\sum_{j=1}^{n} [(\sum_{i=j}^{n} m_i) l_j sin(\theta_j)(\dot{\theta_j})^2]  \\
\\
- \sum_{k=1}^{n} [(\sum_{j=max(1,k)}^{n} m_j) l_k (\dot{\theta_k})^2 sin(\theta_i - \theta_k)]\\
\vdots \\
\\
-m_n\sum_{k=1}^{n} [ l_k (\dot{\theta_k})^2 sin(\theta_i - \theta_k)]\\
\end{bmatrix}
\end{equation}

\begin{equation}
C =
\begin{bmatrix}
u \\
(\sum_{k=1}^{n} m_k) g sin(\theta_i)\\
\vdots\\
m_n g sin(\theta_n)\\
\end{bmatrix}
\end{equation}

Linearizing this model for small angles, the general equations of motion are derived:

\begin{equation}\label{GenLinEOM}
A_l \ddot{\mathbf{X}} = C_l
\end{equation}

where
\begin{equation}
A_l=
\begin{bmatrix}
\sum_{i=1}^{n} m_i + M  & (\sum_{i=1}^{n} m_i) l_1  & \dots &  m_n l_n  \\
\\
(\sum_{i=1}^{n} m_i)  & \ddots & &  \\
%\vdots & \ddots & & \\
\vdots & & [a_{l_{ij}}] &\\
m_n & & & \ddots \\
\end{bmatrix} 
\end{equation}
\vspace{2mm}

where
\begin{equation}
a_{l_{ij}} = [\sum_{k=max(i,j)}^{n} m_k] l_j 
\end{equation}
\begin{equation}
C =
\begin{bmatrix}
u \\
(\sum_{k=1}^{n} m_k) g \theta_i\\
\vdots\\
m_n g \theta_n\\
\end{bmatrix}
\end{equation}

Figure \ref{fig:DPC} shows a schematic of the
such a system with 3 DOFs, called the double inverted pendulum on a cart system.

From the controllability matrix~\cite{nise2020control}, it can be formally proved that these systems are fully controllable ~\cite{xin2018linear}.

%%%%%%%%%%%%%%%%%%%%%%%%%%%%%%%%%%%%%%%%%%%%%%%%%%%%%%%%%%%%%%%%%%%%%%%%%%%%%%
\subsection{LQR}
We apply linear control using state variable feedback for the cartpole described in Section~\ref{dynamics}. The control objective is to balance the pole for small-angle perturbations about the unstable equilibrium position (UEP), where the system behaves linearly and can be controlled (balanced) with a linear controller. The aim is to bring all the pole angles to 0, considering UEP as the reference for the poles. The chosen controller, known as the Linear Quadratic Regulator (LQR)~\cite{islam2019performance, 5629961} is preferred to classical Proportional Integral Derivative (PID) controller~\cite{saraf2020comparative} due to its robustness, superior control performances, and ability to deliver precise control metrics. LQR chooses a set of $m\times n$ state feedback coefficients $K$ for a state vector $\mathbf{X}$ of length $n$ by minimizing a quadratic cost function $J$:

\begin{equation}\label{cost}
\hspace{0.1cm}J = \int (\mathbf{X}^TQ\mathbf{X} + \mathbf{U}^TR\mathbf{U})\,dt
\end{equation}

In order to attain the desired transient and steady-state response characteristics for the cartpole, LQR allows for adjusting the values of the state and control cost matrices $Q$ and $R$, respectively. Elevating the values for the elements in $Q$ increases the speed of response allowing quicker and more aggressive control, while a higher value of $R$ reduces the energy to be spent on the control. 

Minimization of the cost function is obtained by solving the Algebraic Riccati Equation (ARE)~\cite{saraf2020comparative} to obtain the feedback vector $K_{m \times n}$. The control law is given by:

\begin{equation}\label{controllaw}
\mathbf{U}_{m\times 1} = -K_{m\times n}. \mathbf{X}_{n\times 1}
\end{equation}
where $n$ is the state vector dimension and $m$ is the dimension of the control input. In the cartpole case, there are four state variables and one control input,i.e., $n=4$, and $m=1$. 

This control law shows a matrix multiplication to be emulated using spiking neurons.

\subsection{LIF Neuron}\label{lif_def}
A spiking neuron uses a mixed-signal form of computation, with its data encoded as a sequence of spikes. These spikes are spatio-temporally distributed and the information within the signal can be encoded through methods like frequency (rate encoding), population encoding (where a population of neurons encode the entire signal), or latency encoding (precise timing of the spike). Among the many available spiking neuron models, the LIF neuron model was selected, showing that two spiking neurons or a spiking neuron with some pre and post-processing is sufficient for the application at hand. Incoming pre-synaptic spikes $\Theta$ are filtered at the synapse to produce an analog synaptic current $I_{syn}$ that induces a change in the membrane potential $V_{mem}$ of the neuron:

\begin{equation}\label{fig:LIF1}
C_{mem} \frac{dV_{mem}}{dt}= -g_{leak}(V_{mem}-V_{leak}) + {I_{syn}}
\end{equation}
where \(V_{leak}\) is the leakage voltage, \(g_{leak} = \frac{1}{R_{leak}}\) is the synaptic conductance, and \(C_{mem}\) is the membrane capacitance.

When $V_{mem}$ crosses a pre-defined threshold $V_{th}$, the neuron outputs spikes $\delta$:

\begin{equation}\label{LIF2}
I_{out} = 
\begin{Bmatrix}
\delta  \quad\quad\cdots V_{mem}\ge V_{th}\\
0  \quad \quad  \cdots otherwise
\end{Bmatrix}
\end{equation}

% \begin{table}
% \caption{\label{tab:LIFEqn}Equations for an LIF neuron with high $\tau_{mem}$ and low $\tau_{syn}$. Symbols carry the same meaning as described in Section ~\ref{sec:methods}.}
% %\begin{indented}
% %\item[]
% \begin{tabular}{@{}llll}
% \br
% \textbf{Input frequency} &  \textbf{Synaptic current}   & \textbf{Potential change}      &  \textbf{Output frequency} \\
% \mr

% \begin{align*}
% f_i &= \frac{1}{T}\Sigma_{t=1}^{T} \delta(t-t_i)\\
% t_i &= input \, spike \\
% & timing \, for \, the \\
% & i^{th} \, neuron \\
% \end{align*}

% %\vspace*{0.2cm}
% &
% {\color[HTML]{000000} 
% \begin{align*}
% I_{syn} &= \frac{1}{T}\Sigma_{t}^{T} \Sigma_{i}^{n} w_i \delta(t-t_i)\\
%  &= \Sigma_{i}^{n} w_i \times f_{i} I_{spk}
% \end{align*}
% }
% %\vspace*{0.2cm}          
% & {\color[HTML]{000000} 
% \begin{align*}
% \Delta V_{mem} = \frac{I_{syn}}{C_{mem}}
% \end{align*}
% }  
% %\vspace*{0.2cm}
% &{\color[HTML]{000000} 
% \begin{align*}
% &f_{out} = \frac{\Delta V_{mem}}{V_{th}}\\
% &= {\Sigma_{i}^{n} w_i \times f_{i} \times \frac{I_{mag}}
% {V_{th}C_{mem}}}
% \end{align*}
% } 
% \br
% \end{tabular}
% %\end{indented}
% \end{table}

\begin{figure}
\centering
\includegraphics[width=9cm]{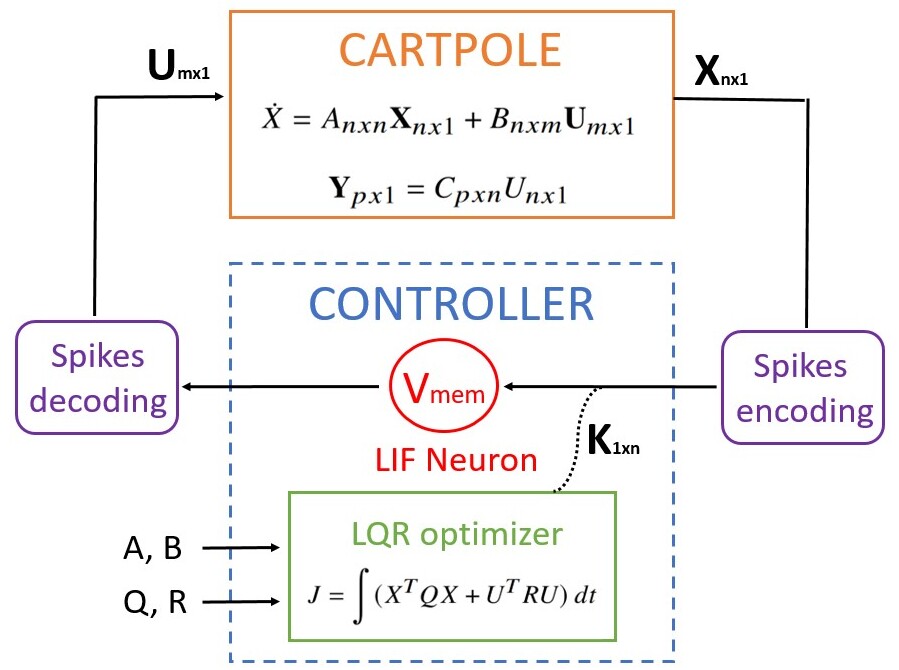}

\caption{The control pipeline has three two blocks: the CARTPOLE as the system to be balanced, and the CONTROLLER made of the LIF neurons as the computational units and the LQR optimizer as the control method.}
\label{fig:controlpipeline}
\end{figure} 

where \(I_{out}\) is the output current comprising spikes $\delta$. The rate of change of membrane potential with the synaptic current depends on the membrane time constant \(\tau_{mem} = \frac{1}{R_{leak}C_{mem}}\). 
If a neuron receives spike trains through $n$ synaptic inputs, then the synaptic current~\cite{Luiweb, gerstner2014neuronal} at the right-hand-side of Equation~\ref{fig:LIF1} at a time instant $t$ is obtained by convolving the input spikes with a synaptic kernel of length \(\tau_{syn}\):

\begin{equation}\label{syncurrent}
I_{syn} =  \Sigma_{x=-\infty}^{-\infty} \Sigma_{i}^{n}\Theta(x-t_{i})e^{-(\frac{t-x}{\tau_{syn}})}
\end{equation}

where $\Theta$ is the pre-synaptic spike computed as the Heaviside step function:
\begin{equation}
\Theta(t-t_i) = 
\begin{Bmatrix}
1  \quad\quad\  \cdots t\ge t_{i}\\
0  \quad\quad\  \cdots t < t_i 
\end{Bmatrix}
\end{equation}

%This model is present in the Lu.i hardware that will be discussed later. 
In the simulation environment, Equation~\ref{syncurrent} can be simplified assuming that $\tau_{syn}$ is very high and the synaptic kernel collapses to a summing junction. Considering the duty cycle of $\Theta$ to be very small, it converts into a Dirac-Delta function:

\begin{equation}\label{del}
\delta(t-T) = 
\begin{Bmatrix}
I_{spk}  \quad\quad\quad  \cdots t = T\\
0  \quad\quad\  \cdots otherwise 
\end{Bmatrix}
\end{equation}
where $I_{spk}$ is the amplitude of each spike which is equal to one.

The final simplified version of Equation ~\ref{syncurrent} is shown in Equation ~\ref{syncurr}.

\begin{equation}\label{syncurr}
I_{syn} = \Sigma_{t=-\infty}^{\infty}\Sigma_{j}^{n} \delta(t-t_{j})
\end{equation}

where $I_{syn}$ is the synaptic current of the $i^{th}$ neuron receiving spikes through $j$ synaptic inputs, $t_{j}$ is the time of spike from the $j^{th}$ synaptic input. 

%Table \ref{}

%%%%%%%%%%%%%%%%%%%%%%%%%%%%%%%%%%%%%%%%%%%%%%%%%%%%%%%%%%%%%%%%%%%

\subsection{LIF neuron in the cartpole feedback control loop}
Figure~\ref{fig:controlpipeline} shows the control system block diagram where the LIF neuron is placed in the feedback control after its weights are tuned to match the $K$ values obtained from the LQR optimizer block (see Equation~\ref{controllaw}). \textcolor{black}{The states of the system are frequency encoded into spike trains which are then fed to the dendrites of the LIF neuron. The values of the feedback LQR coefficients $K$ are matched with the weights on the dendrites. The spike output received is decoded using a linear decoding scheme by scaling the frequency of the output spike train obtained.}
As shown in Equations~\ref{fig:LIF1} and ~\ref{syncurrent}, a noteworthy observation emerges when setting $\tau_{mem}$ and $\tau_{syn}$ to a very high value. Under these conditions, the membrane potential $V_{mem}$ exhibits a linear relationship with the synaptic current $I_{syn}$ and the synapse behaves like a summing junction. This linearization forms the basis for implementing linear state feedback control with LIF neurons. By adopting this parameter configuration alongside a frequency encoding approach for the spikes, the resulting neuron output frequency becomes approximately proportional to the weighted sum of the input spike frequencies.
Table~\ref{tab:LIFEqn} shows the equations for the input-output frequency relationship of a LIF neuron with $n$ synaptic inputs.
The output frequency is proportional to the weighted sum of input frequencies. If $K=\frac{I_{mag}}{V_{th}C_{mem}}$ is chosen to be close to unity, then output frequency comes as equal to weighted input frequencies. Hence, a scaled version of the output can be used as the control signal when the input is the state and the weights are set equal to the feedback coefficients.

\begin{table}
\caption{Equations for an LIF neuron with high $\tau_{mem}$ and low $\tau_{syn}$. Symbols carry the same meaning as described in Section~\ref{sec:methods}.}
% \begin{indented}
\fontsize{11pt}{11pt}\selectfont
\begin{tabular}{@{}*{4}{l}}
\br
\textbf{Input \newline frequency} & \textbf{Synaptic current}   & \textbf{Potential change}      & \textbf{Output frequency}  
\\ \mr

$f_i = \frac{1}{T}\Sigma_{t=1}^{T} \delta(t-t_i)$ & $I_{syn} = \frac{1}{T}\Sigma_{t}^{T} \Sigma_{i}^{n} w_i \delta(t-t_i)$ & $\Delta V_{mem} = \frac{I_{syn}}{C_{mem}}$ & $f_{out} = \frac{\Delta V_{mem}}{V_{th}}$ \\
$t_i = input \, spike$  & $= \Sigma_{i}^{n} w_i \times f_{i} I_{mag}$ & &$=\Sigma_{i}^{n} w_i \times f_{i} \times \frac{I_{mag}}{V_{th}C_{m}}$\\
$timing \, for \, the$ &&\\
$i^{th} \, neuron$ && \\

%\vspace*{0.2cm}
\br
\end{tabular}
\label{tab:LIFEqn}
% \end{indented}
\end{table}

\subsection{Population of LIF neurons for the linear control of a higher DOF mechanical system}\label{sec:esemble_method}

To study the effect of using a population of neurons for the feedback control of a higher-order mechanical system, the Neural Engineering Framework (NEF) called Nengo is used, which allows the creation of an ensemble that integrates more than one LIF neuron, for signal representation and processing. The NEF is composed of 3 principles~\cite{10.3389/fninf.2013.00048}, namely, representation (for encoding and decoding time-varying signals), transformation (for applying functions to such signals), and dynamics (for representing dynamical equations using recurrent connections).
In this work, the representation principle has only been used. 
Each neuron in NEF is sensitive to a certain range of positive or negative input currents, known as the receptive field of the neuron. This sensitivity is determined by the tuning curves of the neurons.  

% To estimate the optimal number of neurons in an ensemble needed to achieve the desired feedback control of a system, we used a Fourier transform-based approach.
From the principle of the NEF, a greater number of neurons per ensemble divides the input space into multiple sub-spaces, increasing precision. However, an excessive number of neurons increases the network complexity, size, and power computation and slows down the computation. Hence, it is important to find a sweet spot for the number of neurons where the signal representation is adequately precise, yet the computational complexity is acceptable. The following subsection discusses the block diagram of the feedback control system designed to balance a multi-linked pendulum on a cart with spiking neurons in the loop, using Nengo.

\begin{figure}
\centering
\includegraphics[width=0.6\textwidth, trim = {0cm 0cm 0 0.2cm}, clip]{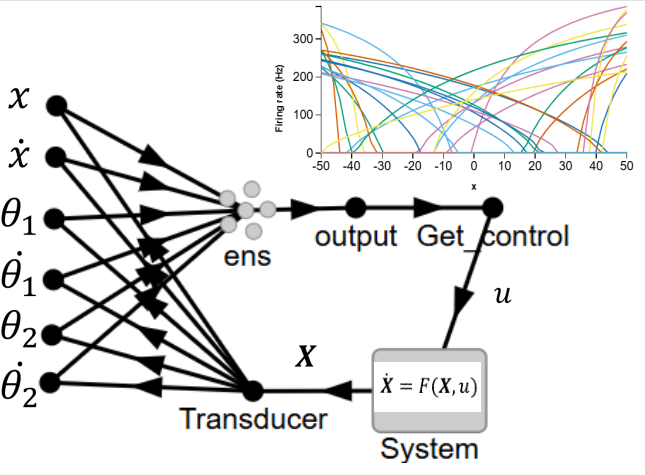}
\caption{\textcolor{black}{Nengo GUI block diagram for LQR control of a double pendulum on a cart (DPC). The System block represents the DPC simulator that takes in the control signal from the Get\_control node and passes the next state to the Transducer. The Transducer sends the six state variables to the nodes which when multiplied by the weights produce the input to the ensemble ens. Based on the tuning curves of the neurons in the ensemble The Get\_control node scales the output by $-1$ to produce the control signal.}}
\label{fig:GUIBD}
\end{figure}

\subsubsection{Control flow design in Nengo}
In order to design the control system in Nengo, with the neuronal ensemble in the loop, the dynamical equations of the system were implemented in Python and all the states were probed. Figure~\ref{fig:GUIBD} shows six states (for a double-pendulum on a cart system) being sent to the input of the neuron ensemble after being multiplied by the synaptic weights. The ensemble outputs the control signal, which is fed back to the system. \textcolor{black}{Here, the state values are fed through specific channels to the ensemble of neurons, and the decoder weight on each channel is matched to the corresponding value of the state feedback vector $K$. Based on the tuning curves of the neurons, each neuron actively spikes for a specific range of inputs. Thus, a collection of spike trains is obtained at the output of the ensemble. The control value is linearly decoded by calculating a weighted sum of the output firing rates of the neurons.}

\section{Experiments on neuromorphic hardwares}\label{sec:experiments}

Two approaches were chosen for the control of a cartpole: a) Use of fewer neurons for a single pendulum: two neurons were used in simulation and one neuron was deployed in hardware (Section~\ref{Single neuron}), and b) Use of an ensemble of spiking neurons for multilinked pendulums (Section~\ref{Multiple neuron}). The neuromorphic control implementation was done using the Loihi chip and the Lu.i board. Both the configurations and descriptions of the neuromorphic boards have been described in detail in the following sections. Section~\ref{Single neuron} first shows a preliminary experiment to test the efficiency of the Lu.i neuron board for a weighted sum of the input operation. Then, Lu.i is used to achieve real-time \textcolor{black}{weighted sum of inputs} for the feedback control of a cartpole. 
Table~\ref{tab:exp} summarizes the list of experiments conducted: two runs in the CPU using conventional LQR on MATLAB and one using the Lu.i neuromorphic hardware explained in Section~\ref{Single neuron}.
Section~\ref{Multiple neuron} first shows an implementation of LQR using an ensemble with a range of neurons from $2$ to $2048$ neurons in Nengo to balance a pole on a cart, along with various control and neuromorphic performance metrics on the CPU and the Loihi neuromorphic chip. Then a parametric study reflects the effect of varying the ensemble parameters on the control performance. This is followed by the deployment of an ensemble of $100$ neurons to achieve feedback control of a multi-linked pendulum on a cart system, backed by an implementation of the same on Loihi. The study concludes with a comparison between various existing models of control using neuromorphic and non-neuromorphic architectures.

\subsection{\textbf{Balancing a pole on a cart using two LIF neurons}}\label{Single neuron}
%This subsection shows the practical application of the weighted summation property across the neuron's dendrites in the context of linear state variable feedback (LSVF) control, specifically employing LQR.

% \begin{figure}
% \centering

% \caption{The Lu.i educational neuron board. It has 3 synaptic inputs with variable weight potentiometers and adjustable values for $\tau_{syn}$,$\tau_{mem}$ and $V_{leak}$.}
% \label{fig:Lui}
% \end{figure}

Initially, a validation with a simple numerical operation is performed. This is followed by an implementation on a cartpole with an inverted pendulum. The LIF neurons were used to control a cartpole simulator using MATLAB, followed by an experiment with the Lu.i physical board. %The Lu.i board was connected to a laptop - sending control values to the simulator running on MATLAB.
%and a laptop running the MATLAB cartpole simulator was set up to verify the simulation results on hardware. 

\subsubsection{\textbf{Simulations}}

%Before integrating Lu.i directly as a feedback controller, simulations for spiking neurons were used. 

\paragraph{\textbf{Cartpole simulator}}
The Cartpole simulator utilizes open-source code by Brunton {\it et al.} ~\cite{MATLAB_Cart}, which is useful to define the parameter values and state variables for the system. In contrast to the reference frames specified in the equations of motion in Section ~\ref{sec:methods}, the reference angle $\theta = 0$ is defined as the position where the pendulum is hanging vertically downward. Therefore, the control objective is to guide the system to the unstable equilibrium point (UEP), where $\theta = \pi$, starting from small perturbations around this position. During the initial phase of each experiment, the pole is displaced by an angle of $0.2$ radians from the UEP. The perturbations applied could be counter-clockwise (positive) and clockwise (negative). Both the cases have been taken into account for the following studies.

\paragraph{\textbf{Spiking neuron}}
A MATLAB model for a spiking neuron was developed following the Equations~\ref{fig:LIF1} and ~\ref{LIF2}. 
The value of \(\tau_{mem}\) was set to be very high ($100$ seconds) following the analysis in the Section \ref{lif_def}.

\paragraph{\textbf{LQR setup}}
The LQR feedback coefficients were computed in MATLAB as per the desired response characteristics.
The $Q$ and $R$ matrices for this experiment were set to
\begin{equation}
Q = 
\begin{bmatrix}
    1 & 0 & 0 & 0 \\
     0 & 1 & 0 & 0 \\
      0 & 0 & 10 & 0 \\
       0 & 0 & 0 & 10 \\
\end{bmatrix}
,
\vspace{2mm}
 R = 0.0001
\end{equation}

\begin{table}
\caption{List of the experiments for cartpole balancing using one or two neurons.}
\label{tab:exp}
\begin{center} 
%\item[] 
\begin{tabular}{@{}llllll}
\br
 &Hardware/&Controller&Number&Inputs&Controlled\\
 &Simulation& &of neurons &per neuron & States\\
\mr
1&Simulation&MATLAB LIF&2&4&$x$,$\theta$\\
%2&Simulation&MATLAB LIF&NA&NA&$\theta$\\
2&Simulation&MATLAB LIF&2&3&$\theta$\\
%4&Simulation&MATLAB LIF&1&3&$\theta$\\
3&Hardware&Lu.i&1&3&$\theta$\\
\br
\end{tabular}
\end{center}
\end{table}

\paragraph{\textbf{Cartpole feedback control in simulation}}
The cartpole has four state variables, necessitating the utilization of four synaptic inputs for full control of all the states. Due to the need for both positive and negative control forces, at least two neurons, one firing for positive synaptic inputs and the other one for negative inputs are needed for the simulations.  Thus, the first simulation involves using two neurons with $4$ dendrites each. One neuron spikes for a positive synaptic current and the other, for a negative synaptic current. 

As described above, corresponding to each state variable, four synaptic inputs to a neuron are necessary. However, there is a physical constraint of three inputs on the Lu.i board. To employ Lu.i for control, the information about the state $x$ may be omitted, as indicated in Equation~\ref{StateSpace1}, where the other states are independent of $x$. Consequently, only one neuron with three synaptic inputs is used to control the angle of the pole. Thus, the second simulation uses two neurons with $3$ dendrites each, for controlling the angle $\theta$ of the pole on the cart, without considering the cart position $x$. The data input to the dendrites were encoded using rate-based encoding, where the frequency of the spikes is proportional to the value of the state variable.

\subsubsection{\textbf{Hardware}}
\begin{figure}[ht]
\centering
\subfloat[]{\includegraphics[width=5cm]{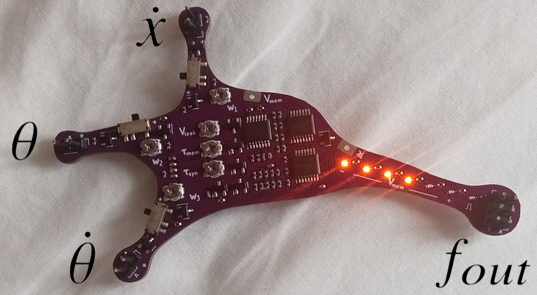}}
\hspace{1cm}
\subfloat[]{
\includegraphics[width=5cm, trim = {2cm 0 0 9cm}, clip]{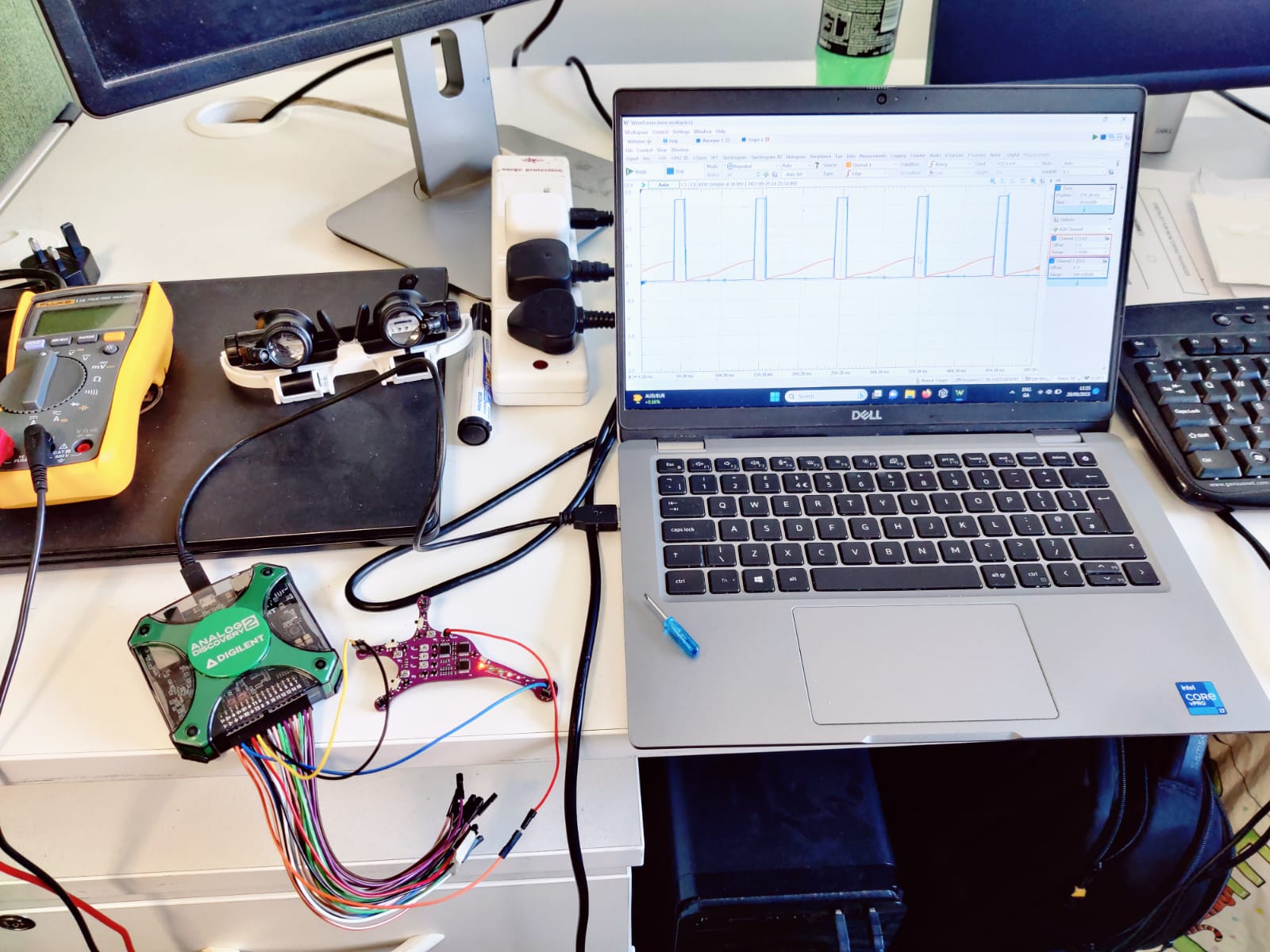}}
\hspace{1cm}
\subfloat[]{
\includegraphics[width=6cm, trim = {19cm 0 0 0}, clip]{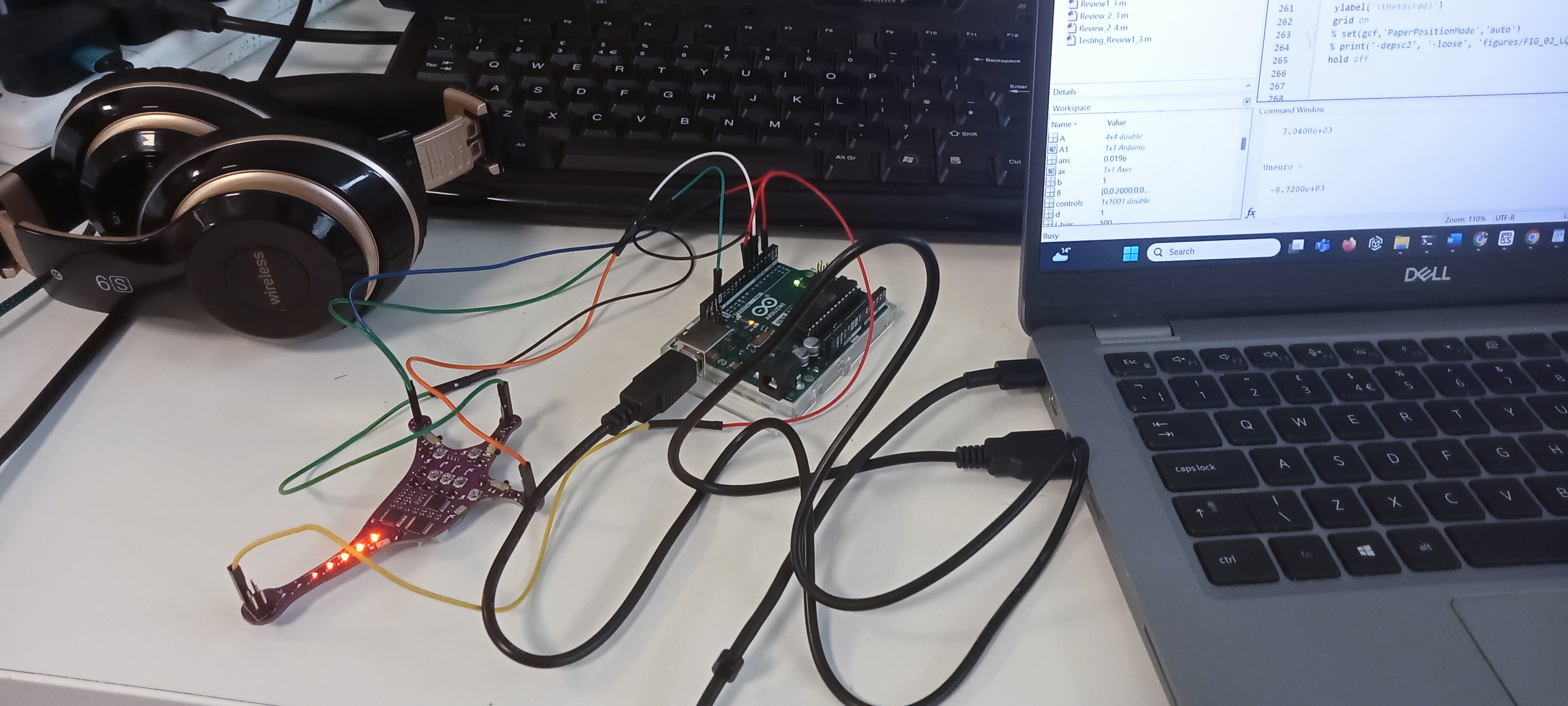}}
%\hspace{1cm}
\caption{Experimental setup for Lu.i. (a) shows the Lu.i board with the inputs and output labeled, (b) shows the Lu.i board attached to a Digilent DAQ connected to the laptop running the multiplier experiment, (c) \textcolor{black}{ shows Lu.i as a controller connected to a laptop running the control algorithm on a cartpole simulator, via an Arduino Uno.}}
\label{fig:Luiexpt1}
\end{figure}

\paragraph{\textbf{Lu.i neuron}}
Lu.i is an educational neuron board \cite{gerstner2014neuronal,Luiweb}, developed by the Kirchhoff-Institute for Physics, Heidelberg University, Germany. It is based on the LIF model of a spiking neuron and each Printed Circuit Board (PCB) has three synaptic inputs (see Fig.~\ref{fig:Luiexpt1}a). Details about the schematics and the components can be found at~\cite{Luiweb} and the relevant features for our work have been summarized in Table~\ref{tab:Luitable}. Each input has a switch to toggle between positive and negative dendritic weights.

\paragraph{\textbf{Testing Lu.i for Weighted summation operation}}
A preliminary experiment was performed to assess the efficacy of using Lu.i for multiplying two decimal numbers. One of these numbers is a variable, in this case the spike-encoded state variables, and is taken as an input through the dendrite, and the second number is a constant corresponding to the weight on the dendrite. The purpose of this verification is to evaluate the percentage error and performance before its application as a feedback multiplier for control. The Lu.i neuron was interfaced with a laptop using a Digilent Analog Discovery 2. The oscilloscope and function generator options were used from this data acquisition system ~\cite{digilent,grout2015introductory}.
The setup is shown in Figure~\ref{fig:Luiexpt1}. 
The output frequency of spikes from Lu.i should ideally be equal to the scalar product of the frequency of the pulse wave provided at the input and the weight assigned to that synaptic input. The power consumed by the Lu.i board was measured from the voltage across its battery and the current drawn by the PCB from the battery, monitored by a series ammeter. The weight on the Lu.i board dendrite ($R_w)$, equivalent to the potentiometer's resistance as mentioned in Table \ref{tab:Luitable}, was estimated at $10$ Hz input, by measuring the output frequency $f_{out}$ and assuming that the resistance $R_{w} = \dfrac{f_{out}}{10}$.   

\begin{table}
\caption{Features of interest of the Lu.i neuron board.}
\label{tab:Luitable}
\begin{center}
%\item[] 
\begin{tabular}{@{}lll}\br
Parameters&Number&Description\\
\mr
Synaptic inputs&3&Positive/negative\\
Tunable parameters&6 potentiometers&3 weights  + synaptic time constant \\ & &  + membrane time constant \\  & & + leakage voltage\\
Data type&Spikes&Spatio-temporal voltage/\\& & current spikes\\
Power source&3 volts DC&Battery\\
Active components&15&7 OpAMPs + 8 MOSFETs\\
Passive components&27& 12 resistors + 6 potentiometers\\ & &  + 2 capacitors + 7 LEDs \\
\br
\end{tabular}
\end{center}
\end{table}

%Lastly, the potentiometer resistance measures  \(\approx 1\Omega/V\), meaning that for feedback control, matching the feedback coefficients to the potentiometer voltage is sufficient to ensure the effectiveness of the feedback mechanism.

\paragraph{\textbf{Testing Lu.i as a feedback controller}}
For testing Lu.i for feedback control, an Arduino UNO connected to a laptop running the MATLAB cartpole simulator was used to send and receive a spike train with the desired frequency in real-time (see Figure \ref{fig:Luiexpt1}c). 
The weight potentiometers were tuned based on the calculated values of the LQR feedback coefficients $K$. The tuning was done within the limits of human error by providing square wave pulses to the three inputs with a certain time delay between themselves and recording the change in the membrane potential (since \(\Delta V_{mem} \propto I \times R_w\).

% \begin{figure}
% \centering
% \includegraphics[width=8cm]{Images/Lui_controller_expt.jpg}
% \caption{Experimental setup for testing Lu.i as a feedback controller}
% \label{fig:Luiexpt2}
% \end{figure}

\textcolor{black}{In simulations, a minimum of two neurons was necessary for balancing the pole on a cart. In order to perform the same task using a single neuron in hardware, it requires some pre-processing and post-processing. The implementation was done following the steps below:
\begin{enumerate}
\item The potentiometers of the Lu.i board were tuned manually to match the approximate ratio given by the MATLAB Riccati Equation Solver. This ensures successful control even though the desired values of the performance metrics may not be achieved. 
\item The cartpole was simulated in steps of $5ms$ (simulation time) laps. Within the span of each time step, the force on the cart remained constant. 
\item This control was calculated by computing the sum of spikes for 0.5 seconds (real-time). Thus the Lu.i board provided $2$ inferences per second.
\item The input to the dendrites was biased to yield only positive values. This is because for a negative value of synaptic current, the Lu.i board does not return a spike. Thus a positive bias has been provided to the input signal. This input was then multiplied with 10 kHz to yield an input spike frequency proportional to the value of the state variable. 
\item Since a single Lu.i neuron was being used, the sign of the post-synaptic current was pre-calculated using the formula $o=kx$. The input spikes were then multiplied by the sign to ensure a positive post-synaptic current, summed over all time steps. These spikes were then sent to the Arduino board via USB.
\item The Arduino UNO sent the spikes to Lu.i via $3$ pins and the spikes generated at the axon were received by the Arduino via $1$ pin from the Digital Pin Set.
 \item Arduino computed the sum of spikes for $0.5s$. This value (proportional to the spike output frequency) was sent back to Matlab via USB, which was then multiplied by the sign of $o$ previously computed, to get $o^{\prime}\propto kx$. This was further scaled/linearly decoded and multiplied with $-1$ to get the control force
$F\approx -kx$. The scaling was done via trial and error to obtain the desired value of the control force.
\end{enumerate}}

\textcolor{black}{A point to be noted here is that this weighted-summation operation can also be performed at the input connections of a non-spiking neuron. However, the focus is to verify whether the spiking version can perform this disjunctive sum operation, with an acceptable accuracy for the given control task, provided that the input is rate
encoded into spikes. The idea is that, if, by using nearly the same number of spiking neurons as
non-spiking ones, a near-optimal control is achievable, then it can be argued (either from the datasheet
or from physical measurements) that the power demand of a spiking hardware would be orders of
magnitude less that the conventional neural networks for the target application, at the cost of some
computational accuracy.}

\subsection{\textbf{Linear control using spiking neuron ensembles}}\label{Multiple neuron}
From the previous section, it is seen that two neurons, or a single neuron with some off-chip processing, can perform LQR control of a cartpole, through rate-based encoding scheme. However, for more precise control and to get the desired control performance, a population of neurons, where each neuron spikes for a certain range of control values, is desirable. This is called population encoding scheme \cite{bekolay2014nengo}. In this work, Nengo Neural Engineering Framework(NEF) based on Python, is used to simulate such populations of neurons.
First, an experiment was performed with a cartpole using $2$ to $2048$ neurons, showing the variation in control and neuromorphic performance metrics. Based on the obtained insights, an ensemble with $100$ neurons is chosen to control multi-linked pendulums on a cart, both on simulations and on the Loihi neuromorphic hardware. 

\subsubsection{\textbf{Simulations}}
\paragraph{\textbf{CPU Specifications}}:
\textcolor{black}{These simulations were run using a Python IDE, running on a 64 bit Ubuntu 22.04.4 LTS OS in a machine with an Intel® Core™ i7-8700K CPU @ 3.70GHz × 12, with a 32 GB RAM.}
\paragraph{\textbf{Balancing a pole on a cart using varying numbers of neurons in the NEF}}
In this set of experiments, the balancing of a pole on a cart is studied using $2$ spiking LIF neurons, using the NEF. While using 2 neurons, the ensemble behaves similar to rate-based encoding, with each neuron spiking for either positive or negative inputs with a frequency proportional to the input value. Then the number of neurons is varied up to $2048$, in powers of $2$, in order to monitor the variation of control performances, as well as power consumption and on-chip area utilization, besides other neuromorphic metrics. This is followed by a parametric study in which the parameters of the spiking neuron ensemble are varied and the effects on the control performance are studied.

\paragraph{\textbf{Validating control performance of a multi-linked pendulum on a cart with a $100$ neuron ensemble}}

Based on the insights gained, a validation study employing a fixed number of neurons in the ensemble is conducted for the feedback control of a cart-mounted multi-link pendulum system, with configurations ranging from one to four pendulum links.
Table~\ref{tab:expcontrol} lists the experiments carried out in simulations and hardware, the corresponding system configurations and initial conditions. Nengo NEF~\cite{dewolf2020nengo} is used to run the control algorithm, while the system to be controlled is designed in Python with the Lagrangian equations and Runge-Kutta $4^{th}$ order solver. Here, $\theta = 0$ corresponds to the unstable equilibrium position (UEP) as described in the equations.
The number of connections going into the neuron ensemble in Figure~\ref{fig:GUIBD} depends on the dimension of the state vector. For example, for a cartpole, four inputs are needed, and for a double pendulum on a cart, six are needed. The weights on the connections are the elements of the feedback coefficient $K_{1 \times 6}$. 
% For a mass variation from 0.5 kg to 5 kg, a realistic variation of pole length would be from 1.1 m to 2 m because the moment of inertia is proportional to the mass of the bob and the square of the length of the link. 
The control energy tuning parameter $R$ is chosen to be equal to $20$, and the state variable tuning parameter $Q$ is chosen such that the coefficients for the angles are $10^4$ and for the angular velocities are $10^3$.

\textcolor{black}{This exploratory analysis informs the trade-off to be made between the control performance desired
and the number of neurons needed in the ensemble population. Having an approximate knowledge of the least number of neurons that can perform a specific control task reduces the resource utilization, power consumption, on-chip space allocation, and computational burden.}

\begin{table}
\caption{List of LQR control experiments using Nengo SNNs. Different systems with various characteristics were tested. DPC stands for double pendulum on a cart, TPC for triple pendulum on a cart and 4lPC for 4 linked pendulum on a cart.}
\begin{center}
\fontsize{10pt}{10pt}\selectfont
\begin{tabular}{@{}*{9}{l}}

\br
 &  \textbf{System}         & \textbf{DOFs}     & \textbf{Perturbation}  & \textbf{Neurons} & \textbf{Synapses} & \textbf{Bob} & \textbf{Pole} & \textbf{Software}
\cr 
 &        & \textbf{(n)}         & \textbf{(rad)}     & \textbf{}  & \textbf{} & \textbf{mass} & \textbf{length} & \textbf{/Hardware}
 \cr
 &        & \textbf{}         & \textbf{}     & \textbf{}  & \textbf{} & \textbf{(kg)} & \textbf{(m)} & \textbf{}\cr
\mr
% {\color[HTML]{000000} \textbf{1}} & {\color[HTML]{000000} cartpole}                 & {\color[HTML]{000000} 2}           & {\color[HTML]{000000} 0.2}                       & {\color[HTML]{000000} 30}                          & {\color[HTML]{000000} 4}   
% & {\color[HTML]{000000} \(\{0.5 \, to \,5\}\)}   
% & {\color[HTML]{000000} \(2\)}   
% \\
{\color[HTML]{000000} \textbf{1}} & {\color[HTML]{000000} Cartpole}                 & {\color[HTML]{000000} 2}           & {\color[HTML]{000000} $0.2$}                       & {\color[HTML]{000000} $100$}                          & {\color[HTML]{000000} $4$}   
& {\color[HTML]{000000} \(1\)}   
& {\color[HTML]{000000} \(2\)}  
& {\color[HTML]{000000} Nengo + Loihi}  
\\ 
{\color[HTML]{000000} \textbf{2}} & {\color[HTML]{000000} DPC}                 & {\color[HTML]{000000} 3}           & {\color[HTML]{000000} \(\{0.2,0.18\}\)}                       & {\color[HTML]{000000} $100$}                          & {\color[HTML]{000000} $6$}   
& {\color[HTML]{000000} \(0.5\)}   
& {\color[HTML]{000000} \(1\)} 
& {\color[HTML]{000000} Nengo + Loihi} 
\\ 

{\color[HTML]{000000} \textbf{3}} & {\color[HTML]{000000} TPC}                 & {\color[HTML]{000000} 4}           & {\color[HTML]{000000} \(\{0.2,0.18,0.16\}\)}                       & {\color[HTML]{000000} $100$}                          & {\color[HTML]{000000} $8$}   
& {\color[HTML]{000000} \(1/3\)}   
& {\color[HTML]{000000} \(2/3\)}  
& {\color[HTML]{000000} Nengo}
\\ 

{\color[HTML]{000000} \textbf{5}} & {\color[HTML]{000000} 4lPC}                 & {\color[HTML]{000000} 5}           & {\color[HTML]{000000} \(\{0.2,0.18,0.16,0.14\}\)}                       & {\color[HTML]{000000} $100$}                          & {\color[HTML]{000000} $10$}   
& {\color[HTML]{000000} \(0.25\)}   
& {\color[HTML]{000000} \(0.5\)}  
& {\color[HTML]{000000} Nengo}
\\ 
\br
\end{tabular}
\end{center}

\label{tab:expcontrol}
\end{table}

\subsubsection{\textbf{Hardware}}
The experiment with a multi-linked pendulum on a cart system were implemented on Intel's Loihi neuromorphic hardware, using the Nengo-Loihi backend, and the performance metrics were recorded.
\textcolor{black}{Loihi was fabricated using Intel’s $14-nm$ FinFET process, incorporating $2.07$ billion transistors and $33$ MB of SRAM across its $128$ neuromorphic cores and three $x86$ cores, all within a die area of 60 mm². The chip operates across a supply voltage range of $0.50$ V to $1.25$ V.The chip includes 16 MB of synaptic memory. Using its most compact 1-bit synapse representation, Loihi achieves a synaptic density of $2.1$ million unique variables per $mm^2$. Loihi’s peak neuron density is $2,184$ neurons per mm².}
Table \ref{tab:expcontrol} captures the experiments performed on Loihi.

\section{Results}\label{sec:results}
The results from the experiments conducted as described in Tables \ref{tab:exp} and \ref{tab:expcontrol} are presented in the same order as the experiments described in Section~\ref{sec:experiments}.
\subsection{\textbf{Balancing of a cartpole using 2 LIF neurons}}
\subsubsection{\textbf{Simulations}}
\begin{figure}
\centering
\subfloat[]{\includegraphics[width=8cm]{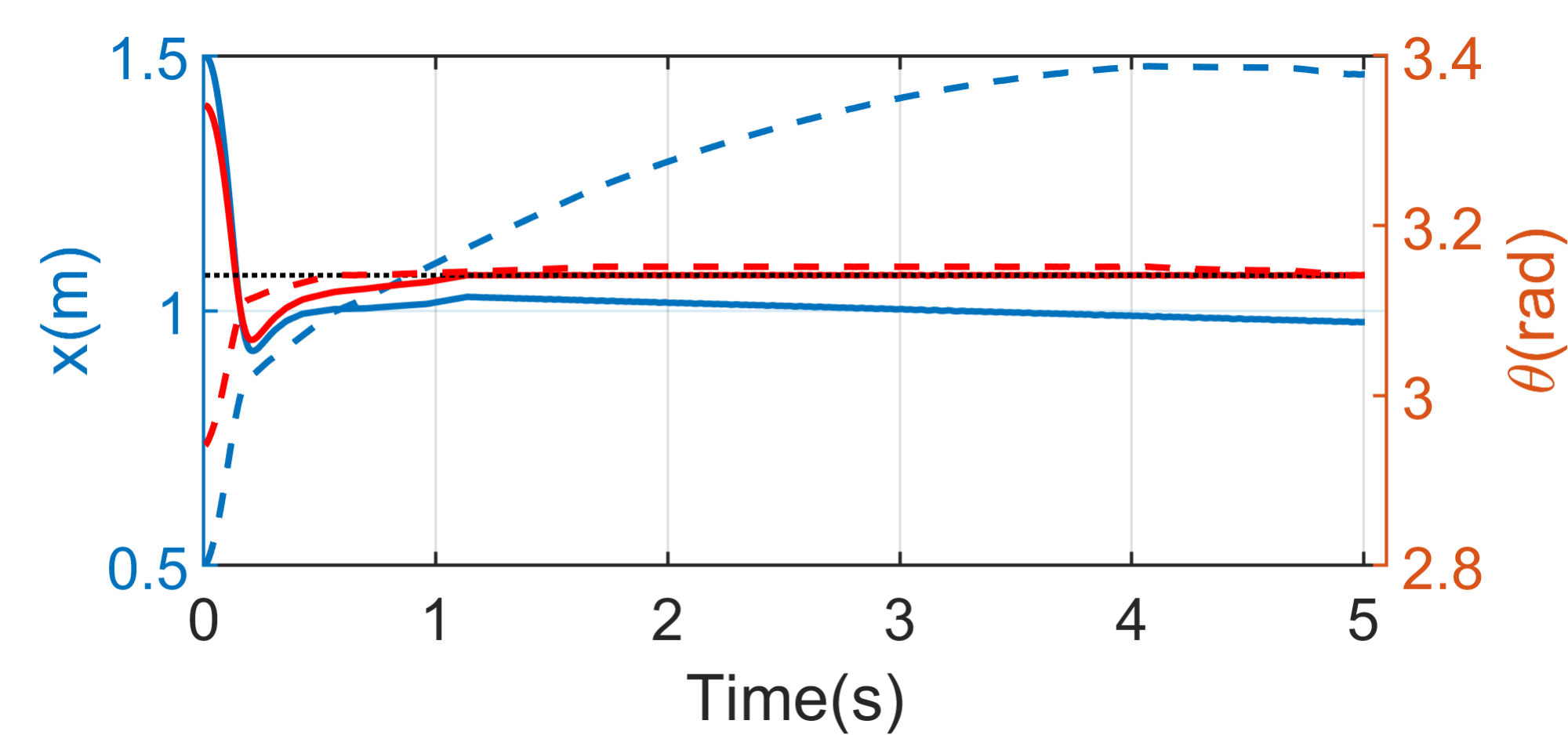}}
%\hfill
%\hspace{0.5mm}
\subfloat[]{\includegraphics[width=8cm]{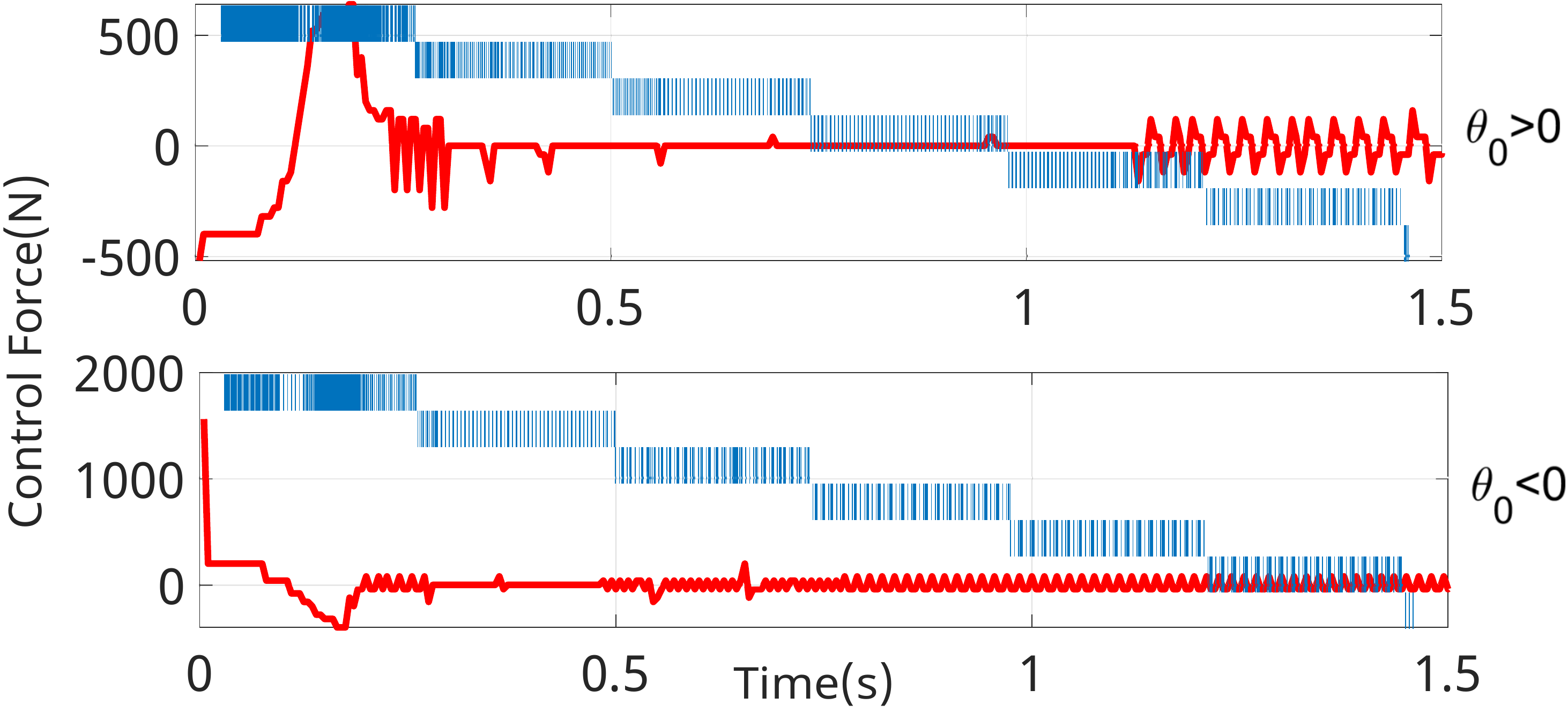}}
%\hfill
\hspace{1mm}
\subfloat[]{\includegraphics[width=8cm]{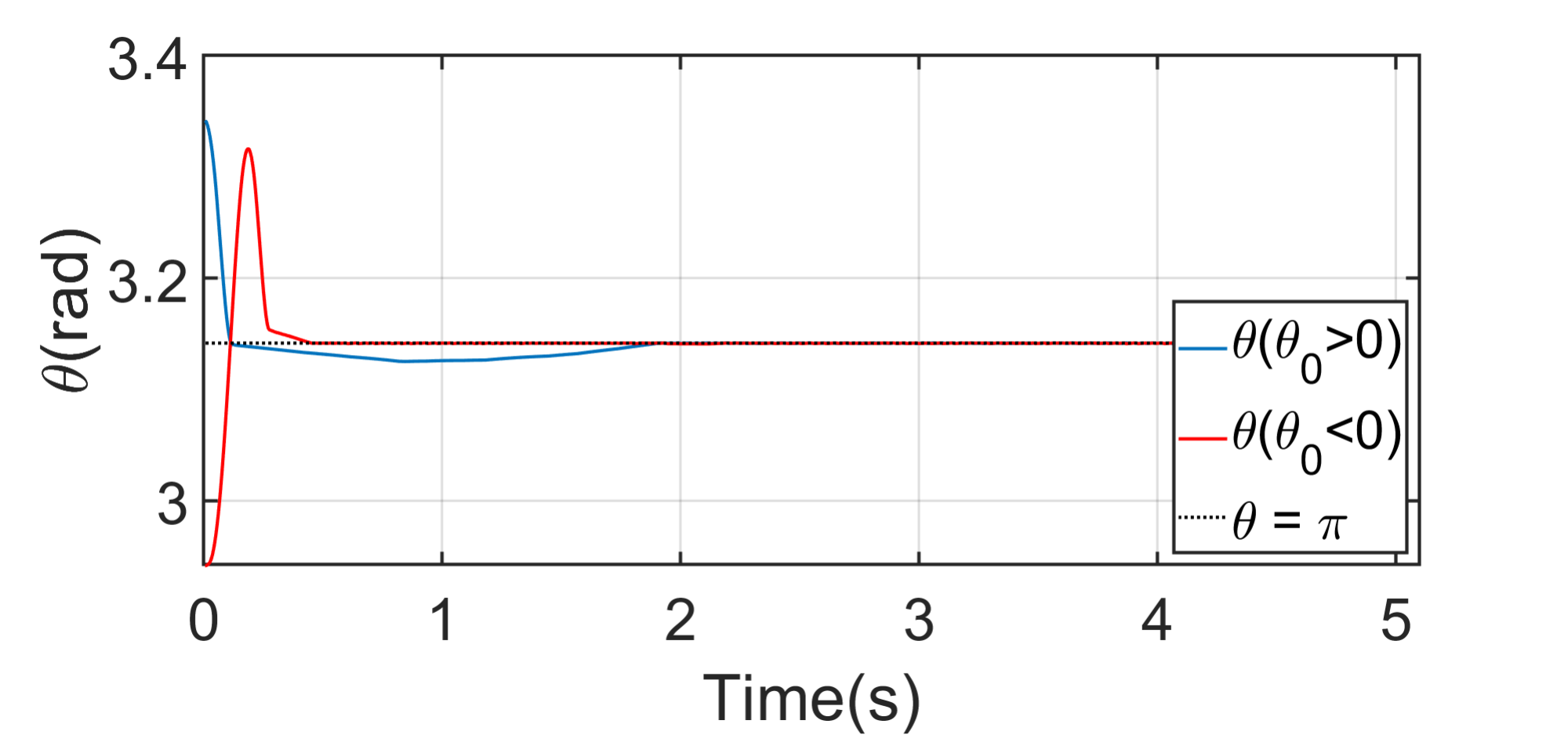}}
%\hfill
%\hspace{0.5mm}
\subfloat[]{\includegraphics[width=8cm]{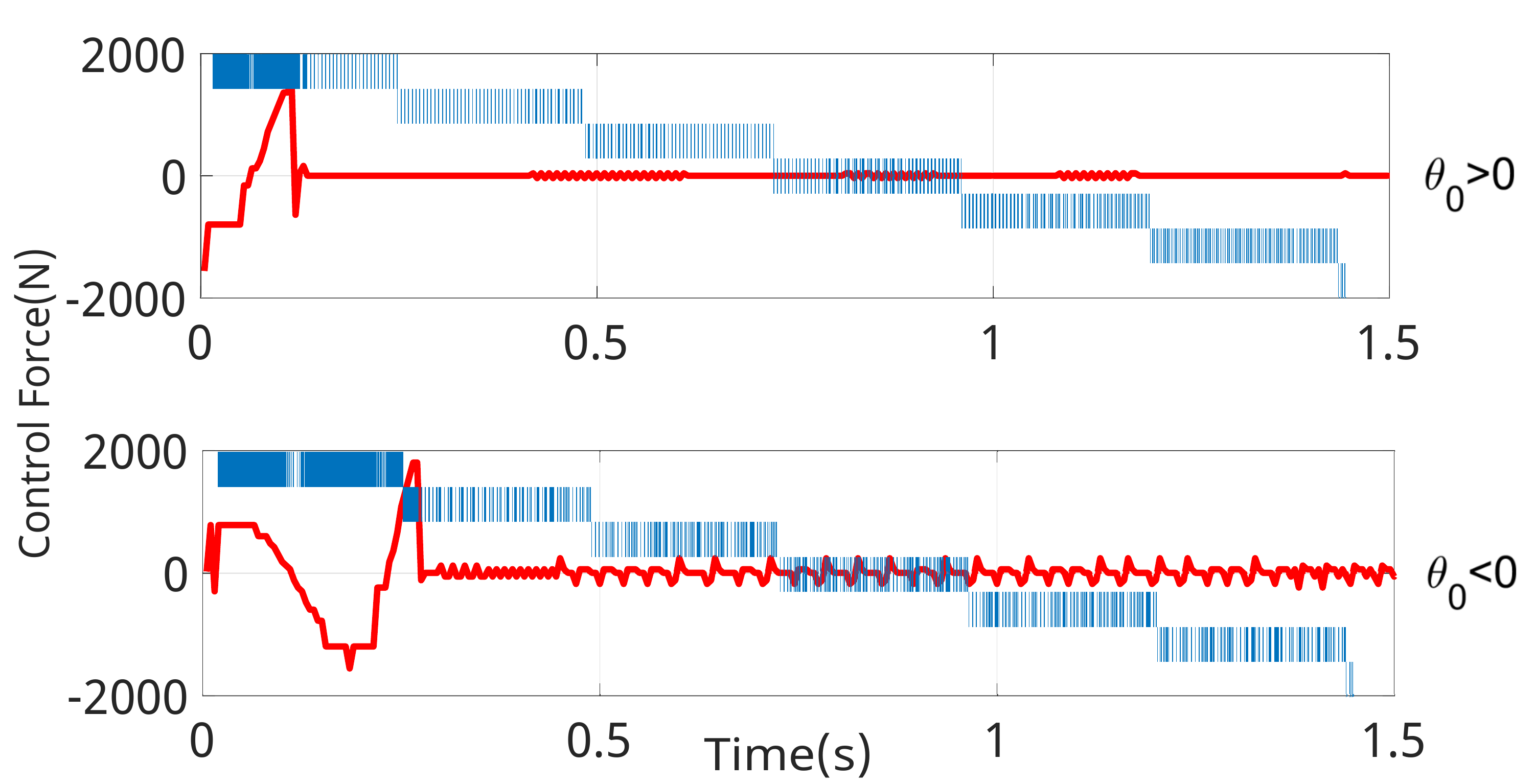}}
%\hfill
\hspace{1mm}

\caption{\textcolor{black}{Plots for balancing a cartpole using 2 MATLAB simulated neurons with 4 and 3 dendrites. (a) and (c) show the pole angle evolution for positive and negative initial angles of the pole for 4 and 3 dendrites respectively, while (b) and (d) show the control values and the spike output frequency for the respective cases.}}
\label{fig:2_LIF_control}
\end{figure}

Figure~\ref{fig:2_LIF_control} (a) shows the state variables $x$ and $\theta$ over time for positive and negative perturbations around UEP, using two simulated LIF neurons and four synaptic inputs that receive the cartpole state. Figure~\ref{fig:2_LIF_control} (c) shows the evolution of $\theta$ over time for positive and negative perturbations around UEP, using two simulated LIF neurons and three synaptic inputs that receive the cartpole states except for the cart position $x$. The blue segments in Figure~\ref{fig:2_LIF_control} (b), (d) depict the spike patterns an the red line corresponds to the control values input to the system to keep it around the UEP. The spike rasters have been divided into segments for enhanced visibility. On zooming into the plots, it can be seen that the spike density is high when the control value is high, and decreases as the control force diminishes, as expected for linear frequency-based decoding scheme.

Table~\ref{tab:comparisonmet} shows a comparative study of the metrics obtained from the two experiments. To analyse the performance, rise time $T_r$ and overshoot (PO) are chosen as the transient characteristics, and steady-state error (SSE) and settling time $T_s$ (time taken to reach a variation of less than 5\% about the UEP) as the steady-state characteristics as in \cite{nise2020control}. Lower values of these metrics indicate superior control. The main difference between the two simulations is that, for the control with $2$ neurons and $3$ dendrites, the state information is omitted, hence only the pole angle $\theta$ is controlled and not the cart position $x$. However, control time metrics like $T_r$ and $T_s$ are superior for the 3-dendrite case since only one state variable is controlled.

% \begin{table}
% \caption{LQR control performance metrics for single or couple of simulated LIF neurons used to control the cartpole compared to classical LQR control without using neurons.}
% \lineup
%     \begin{center}
%     \fontsize{9pt}{9pt}\selectfont
    
%     \begin{tabular}{@{}*{10}{l}}
%     \br
%          & \multicolumn{3}{c}{\textbf{LQR}}  & \multicolumn{3}{c}{\textbf{2 Neurons 3 Synapses}} & \multicolumn{3}{c}{\textbf{1 Neurons 3 Synapses}} \\
%         \textbf{Metrics} & \crule{3} & \crule{3} & \crule{3}\\
%          & \emph{Positive} & \emph{Negative} & \emph{Average} & \emph{Positive} & \emph{Negative} & \emph{Average} & \emph{Positive} & \emph{Negative} & \emph{Average} \cr
%         \mr 
%         {$T_r$ (s)} & \00.41 & \00.41 & \00.41 & \00.36 & \00.95 & \00.66 & \00.34 & \00.36 & \00.35 \cr
%         {PO (\%)} & 14.37 & 14.37 & 14.37 & 16.157 & 39.59 & 27.87 & 68.45 & 19.56 & 44.00 \cr 
%         {$T_s$ (s)} & \01.92 & \01.92 & \01.92 & \04.2 & \04.76 & \04.48 & \02.49 & \01.03 & \01.76 \cr
%         {SSE} & -0.0001 & \00.0001 & \00.0001 & \00.0084 & -0.0018 & \00.0051 & \00.0110 & \00.0137 & \00.0124 \cr 
%         \br
%     \end{tabular}
%     \end{center}
%     \label{tab:comparisonmet}
% \end{table}

\begin{table}
\caption{\textcolor{black}{Control performance metrics for the pole angle during the control of a cartpole using 2 neurons and 4 dendrites and 2 neurons and 3 dendrites in MATLAB simulation, and with the Lu.i single neuron board. The averages are calculated by using the absolute values for the positive and negative perturbations.}}
\lineup
    \begin{center}
    \fontsize{9pt}{9pt}\selectfont
    
    \begin{tabular}{@{}*{10}{l}}
    \br
           &  \multicolumn{3}{c}{\textbf{2 Neurons 4 Synapses}} & \multicolumn{3}{c}{\textbf{2 Neurons 3 Synapses}} & \multicolumn{3}{c}{\textbf{Lu.i}} \\
        \textbf{Metrics}  & \crule{3} & \crule{3} & \crule{3}\\
         & \emph{Positive} & \emph{Negative} & \emph{Average} & \emph{Positive} & \emph{Negative} & \emph{Average} & \emph{Positive} & \emph{Negative} & \emph{Average} \cr
        \mr 
        {$T_r$ (s)} & \00.13 & \00.68 & \00.141 & \00.11 & \00.11 & \00.11 & \01.76 & \02.11 & \01.94 \cr
        {PO}  & \00.4 & \00.05 & \00.22 & \00.1 & \00.8 & \00.45 & \00.35 & \00.25 & \00.30 \cr 
        {$T_s$ (s)} & \01.25 & \04.8 & \03.03 & \00.465 & \01.89 & \01.77 & 14.52 & 14.56 & 14.54 \cr
        {SSE}  & $\approx$ 0 & $\approx$ 0 & $\approx$ 0 & -0.0025 & -0.0025 & \00.0025 & \00.05 & -0.05 & \00.05 \cr 
        \br
    \end{tabular}
    \end{center}
    \label{tab:comparisonmet}
\end{table}

\textcolor{black}{Please note that the spike density increases with an increase in the control
value and gets rarefied as the control value reduces. It is not expected that the control performance will change based on the number of dendrites, because
the reduction in dendrites reduces the system state input to the neuron, and hence, the number of
system states controlled. In this work, for the 3 dendrite neuron, the information about the cart
position $x$ is eliminated, so position control is not possible in this case. Since, for the 3 dendrite situation, only one
variable needs to be controlled, the performance metrics are observed to be slightly better than the 4
dendrite case.}
% \begin{figure}
% \centering
% \subcaptionbox{}{\includegraphics[width=7cm]{Images/LQR-eps-converted-to.pdf}\label{fig:LinImage2a}}
% \hfill
% \subcaptionbox{}{\includegraphics[width=7cm]{Images/Neurons2_Dendrite3-eps-converted-to.pdf}
% \label{fig:LinImage2b}}
% \hfill
% \subcaptionbox{}{\includegraphics[width=7cm]{Images/Neurons_1_dendrites_3-eps-converted-to.pdf}
% \label{fig:LinImage2c}}
% \hfill
% \subcaptionbox{}{\includegraphics[width=7cm]{Images/Lu.i-eps-converted-to.pdf}
% \label{fig:LinImage2d}}
% \caption{Time evolution plots for $\theta$ for balancing a cartpole system (a) in MATLAB without using neurons (b) with 2 simulated spiking neurons having 3 synaptic inputs, (c) with 1 simulated spiking neurons having 3 synaptic inputs and (d) with Lu.i single neuron board having three synaptic inputs. In each case, $\theta_0$ refers to the initial perturbed angle which is either positive or negative.}
% \label{fig:LinImage2}
% \end{figure}

\subsubsection{\textbf{Hardware}}

\paragraph{\textbf{Testing Lu.i as multiplier}}
In order to determine its suitability for being used as a feedback controller, Lu.i was tested for weighted sum of inputs. It was observed that the board consumes a total of around $35 mW$ of power for a $10 Hz$ input frequency.

% \begin{table*}
% \begin{center}
% \fontsize{9pt}{9pt}\selectfont
% \begin{tabular}{|c|c|c|c|c|p{2.5cm}|}
% \hline
% {\color[HTML]{000000} \textbf{Input(Hz)}}   & 
% {\color[HTML]{000000} \textbf{Weight(V)}}       & {\color[HTML]{000000} \textbf{Output(Hz)}}  &
% {\color[HTML]{000000} \textbf{Expected output (Hz)}}  
% & {\color[HTML]{000000} \textbf{Error(\(\%\))}} 
% & {\color[HTML]{000000} \textbf{Power Consumption(mW)}}  
% \\ \hline
% {\color[HTML]{000000} 10}             & 
% {\color[HTML]{000000} 1.42}          &
% {\color[HTML]{000000} 14.73}         &
% {\color[HTML]{000000} 14.24}          &
% {\color[HTML]{000000} 3.46}          &
% {\color[HTML]{000000} 34.31} 
% \\ \hline

% {\color[HTML]{000000} 10}             & 
% {\color[HTML]{000000} 0.71}          &
% {\color[HTML]{000000} 7.39}         &
% {\color[HTML]{000000} 7.12}          &
% {\color[HTML]{000000} 3.81}          &
% {\color[HTML]{000000} 35.13} 
% \\ \hline

% \end{tabular}
% \end{center}
% \caption{Results for the experiment of using Lu.i as a multiplier. Two observations are recorded by varying the synaptic weight value for proof of concept and recording the power consumption}
% \label{tab:Luiresult1}
% \end{table*}

\begin{figure}
\centering
\subfloat[]{\includegraphics[width=7.5cm,trim = {0 0.2cm 0 0cm},clip]{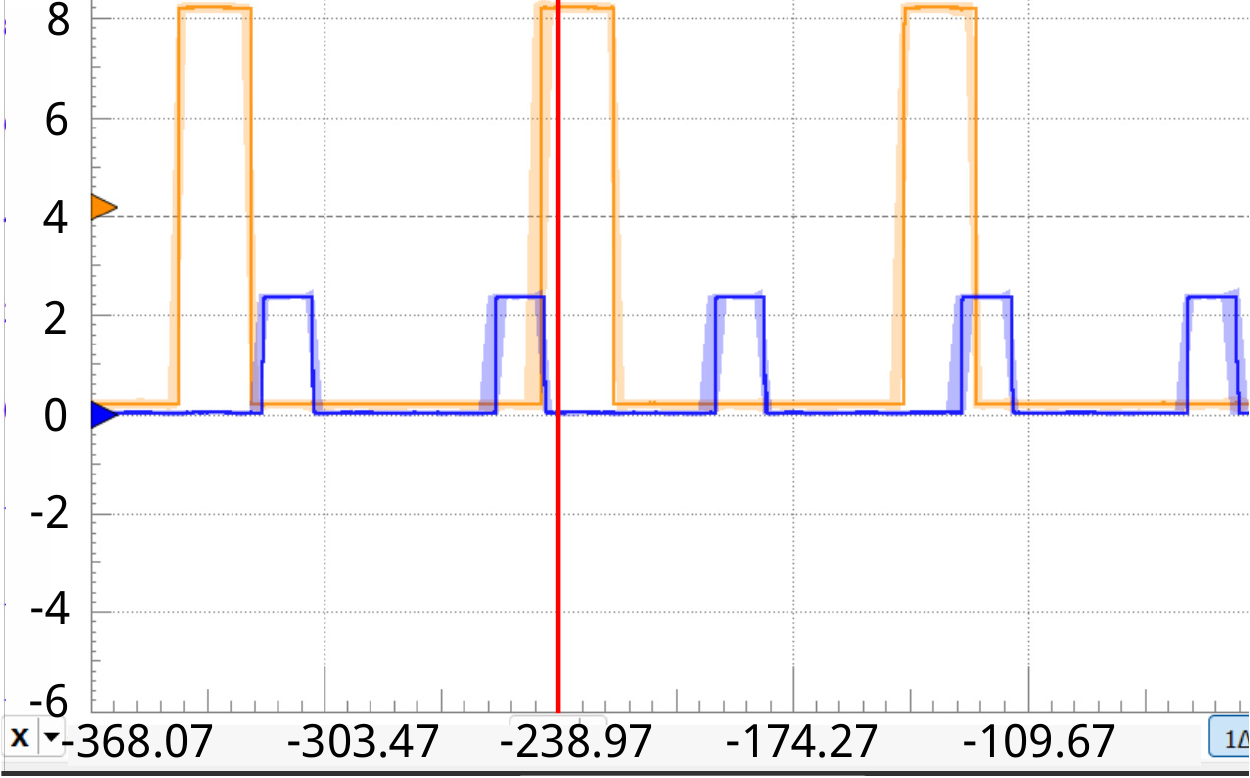}\label{fig:Luiexpt1a}}\hspace{6mm} \subfloat[]{\includegraphics[width=7cm,trim = {0 0cm 0 0.45cm},clip]{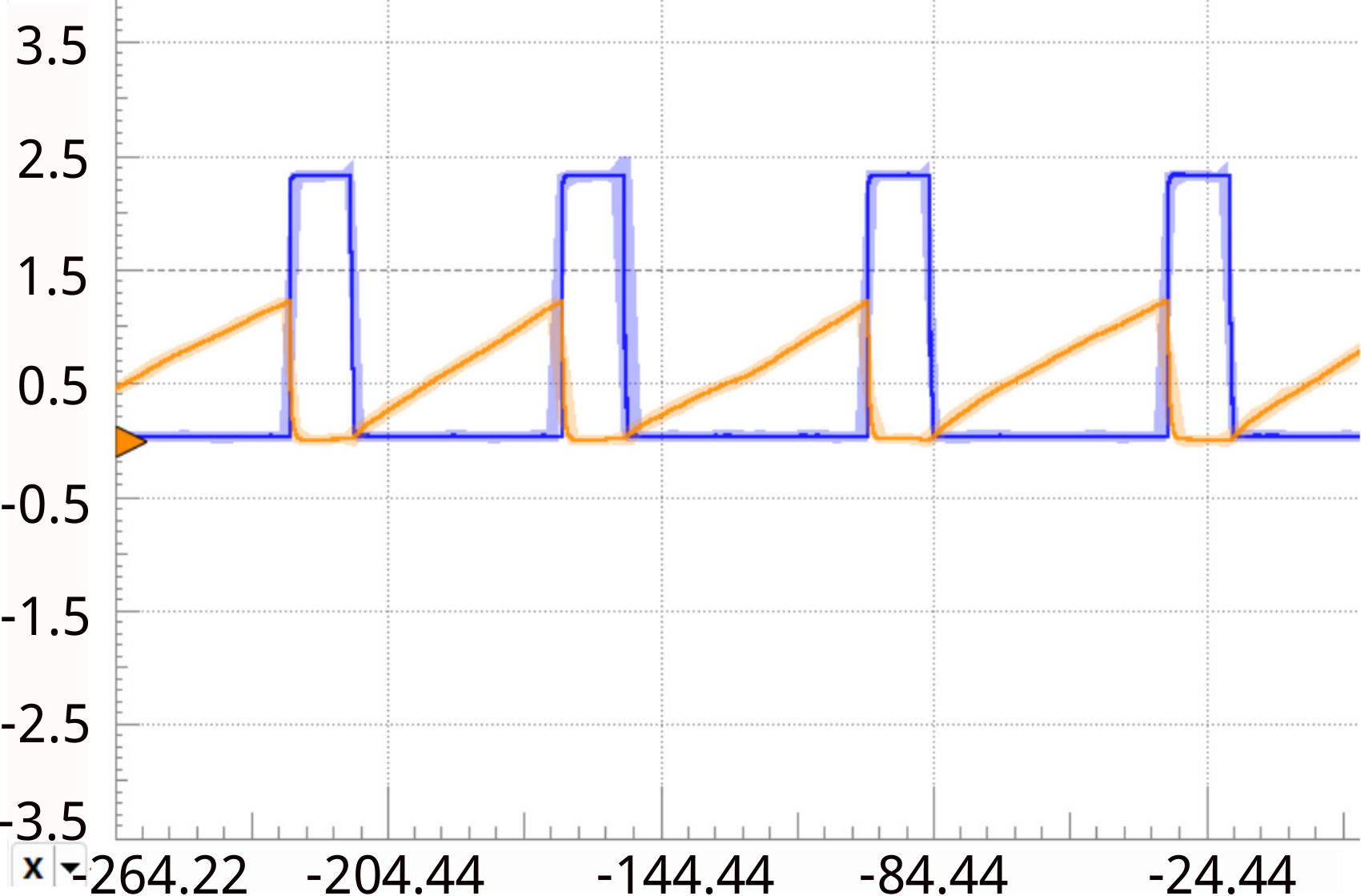}\label{fig:Luiexpt1b}}
\caption{\textcolor{black}{Snapshots of oscilloscope window for multiplication using Lu.i. (a) Input spikes in orange (Channel 1) and the output spikes in blue (Channel 2).
X-axis: major div = 64.6 ms/div; Y-axis: major div = 2V/div.
(b) Membrane potential and output spikes in channels 1 and 2 respectively. 
X-axis: major div = 60.0 ms/div; Y-axis: major div = 1V/div.}}
\label{fig:Digiwave}
\end{figure}

During the testing, waveforms were recorded with the Digilent scope, and one such visualization of the signals for a $10$ Hz input pulse wave and multiplier weight of $1.42$ is shown in Figure~\ref{fig:Digiwave}. Figure ~\ref{fig:Digiwave}(a) shows the input voltage pulses in orange, and the output voltage pulses in blue. Figure ~\ref{fig:Digiwave}(b) shows the membrane potential of Lu.i in orange, and the output voltage pulses in blue. It can be observed in Figure ~\ref{fig:Digiwave}(a) that the frequency of the output signal is about $1.47$ times the input, which is in close agreement with the expected value of $1.42$.

\begin{figure}
\centering
\includegraphics[width=7cm]{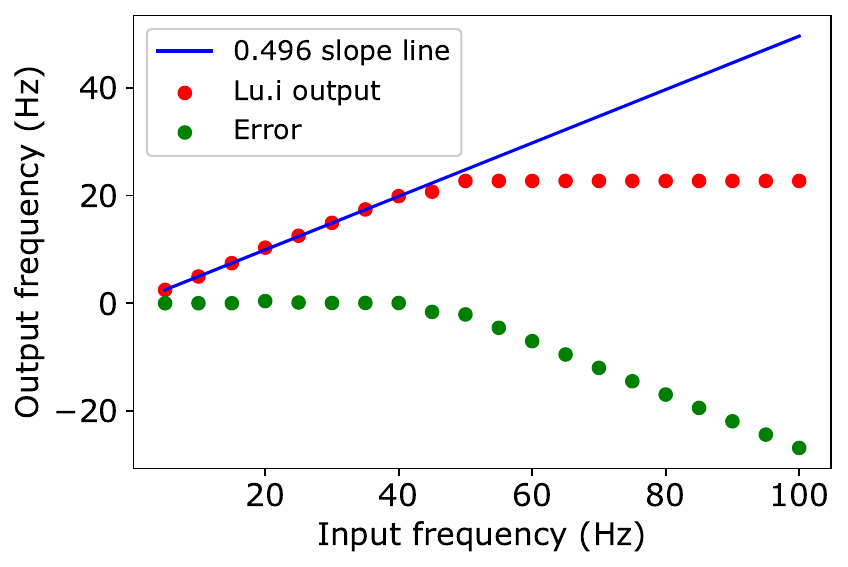}
\caption{Linear behaviour of the Lu.i output with the input frequency.}
\label{fig:Luilinearity}
\end{figure}

Figure~\ref{fig:Luilinearity} shows Lu.i output has a linear trend when varying the input frequency considering a synaptic weight multiplier of $0.496$. A pulse wave was fed at the input with a duty cycle proportional to the frequency. Therefore, the total duration of \emph{high} state of the pulse wave per second is proportional to the input frequency, and consequently the power input to the circuit.
Eight samples of input frequency were tested from $5$ to $40$ Hz. If the duration of each pulse (high state) is kept constant, and the frequency is increased, then the duty cycle of the wave increases moving from an approximate spike train to a DC signal. Since an LIF neuron acts as an integrator, the rate of change of membrane potential, and hence the output spiking frequency depends on the RMS value of the input voltage. Hence, once the input duty cycle becomes sufficiently large, the output frequency saturates. Figure ~\ref{fig:Luilinearity} shows such an example for a pulse duration of $20$ ms.

% \begin{table}
% \begin{center}
% \begin{tabular}{|c|c|c|c|c|c|}
% \hline
% {\color[HTML]{000000} \textbf{Input(Hz)}}   & 
% {\color[HTML]{000000} \textbf{Weight(V)}}       & {\color[HTML]{000000} \textbf{Output(Hz)}}  &
% {\color[HTML]{000000} \textbf{Expected output (Hz)}}  
% & {\color[HTML]{000000} \textbf{Error(\(\%\))}} 
% & {\color[HTML]{000000} \textbf{Power Consumption(mW)}}  
% \\ \hline
% {\color[HTML]{000000} 10}             & 
% {\color[HTML]{000000} 1.42}          &
% {\color[HTML]{000000} 14.73}         &
% {\color[HTML]{000000} 14.24}          &
% {\color[HTML]{000000} 3.46}          &
% {\color[HTML]{000000} 34.31} 
% \\ \hline

% {\color[HTML]{000000} 10}             & 
% {\color[HTML]{000000} 0.71}          &
% {\color[HTML]{000000} 7.39}         &
% {\color[HTML]{000000} 7.12}          &
% {\color[HTML]{000000} 3.81}          &
% {\color[HTML]{000000} 35.13} 
% \\ \hline

% \end{tabular}
% \end{center}
% \caption{Results for the experiment of using Lu.i as a multiplier. Two observations are recorded by varying the synaptic weight value for proof of concept and recording the power consumption}
% \label{tab:Luiresult1}
% \end{table}

\begin{table}
\caption{\textcolor{black}{Results for the experiment of using Lu.i as a multiplier. Two observations are recorded by varying the synaptic weight value for proof of concept and recording the power consumption.}}
\label{tab:Lui_power_table}
\lineup
\begin{center}
%\item[] 
\begin{tabular}{@{}llllll}\br
\textbf{Input(Hz)}&\textbf{Weight}&\textbf{Output(Hz)}&
\textbf{Expected (Hz)}&\textbf{Error($\%$)}&\textbf{Power($mW$)}\\
\mr
10&1.42&14.73&14.248&3.46&34.31\\
10&0.71&7.39&7.120&3.81&35.13\\ 
\br
\end{tabular}
\end{center}
\end{table}

\noindent The obtained results yield two insightful observations. Firstly, the multiplication error is \(\le 4\%\), a value acceptable for feedback control applications. This is because the feedback multipliers are chosen based on the desired response of the system, both in terms of error and control energy spent. Slight variations in operands and hence the output hold no significant concerns in this context as the primary objective is only to stabilize the pole in its UEP without strict constraints on the response characteristics such as settling time, overshoot, or control energy. Previous works on CMOS-based analog multiplier circuits \cite{van2000design,popa2013improved} show an acceptable linearity error range of \(<1\%\) to \(2\%\). This combined with the synaptic non-linearity, membrane non-linearity of the neuron, and target application makes \(\le 4\%\) an acceptable error for this work.
Secondly, temporal encoding of the data allows for the same hardware to be used to process a diverse range of inputs if the spike duty cycle can be appropriately chosen to push the saturation effect to a higher frequency range. This contrasts
with digital circuits, where the number of components needed increases with the number of operand bits, leading to a strict constraint on the range of data permitted. Thus neuromorphic hardware can be much more scalable compared to non-neuromorphic digital chips. A representative data of the power consumption of the Lu.i board is given in Table \ref{tab:Lui_power_table}. Since the number of synaptic operations depends on the input frequency, the power consumption should ideally be the same for the same input frequency as seen from the table.
%The power is measured from the product of the battery voltage and the current recorded by a series ammeter.

\paragraph{\textbf{Testing Lu.i as feedback controller}}
\begin{figure}
\centering
\subfloat[]{\includegraphics[width=8cm]{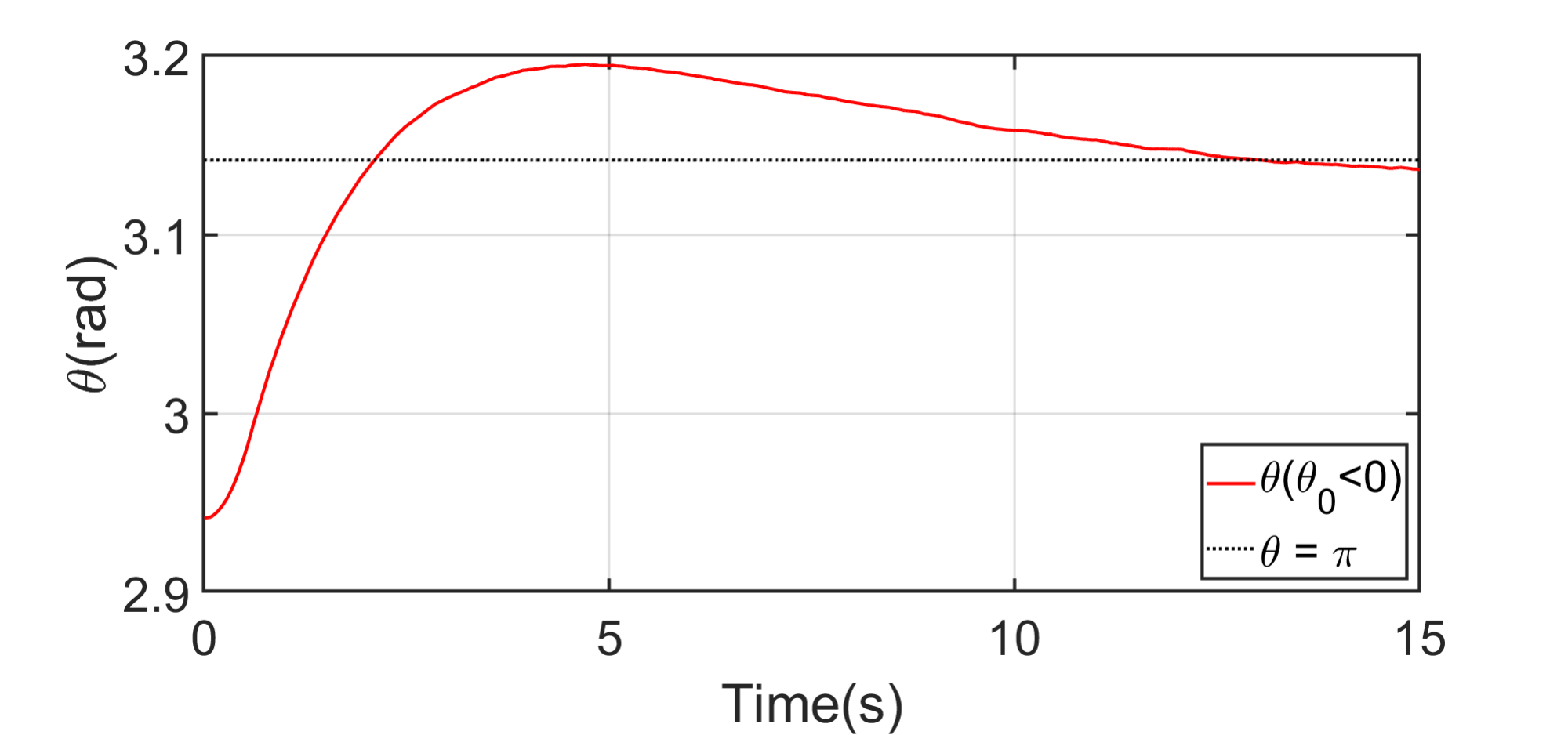}}
%\hfill
\subfloat[]{\includegraphics[width=8cm]{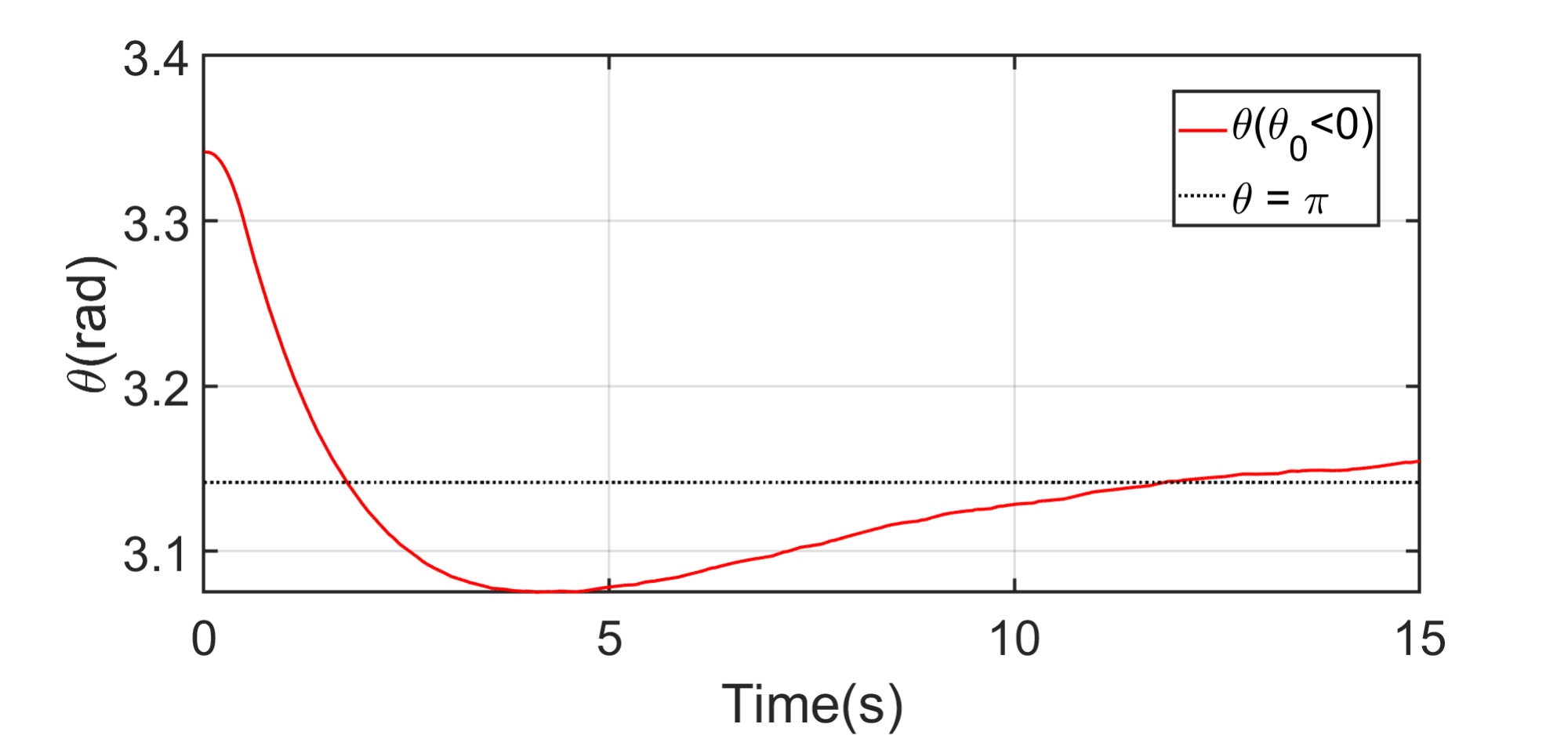}}
\hspace{1mm}
\subfloat[]{\includegraphics[width=8cm]{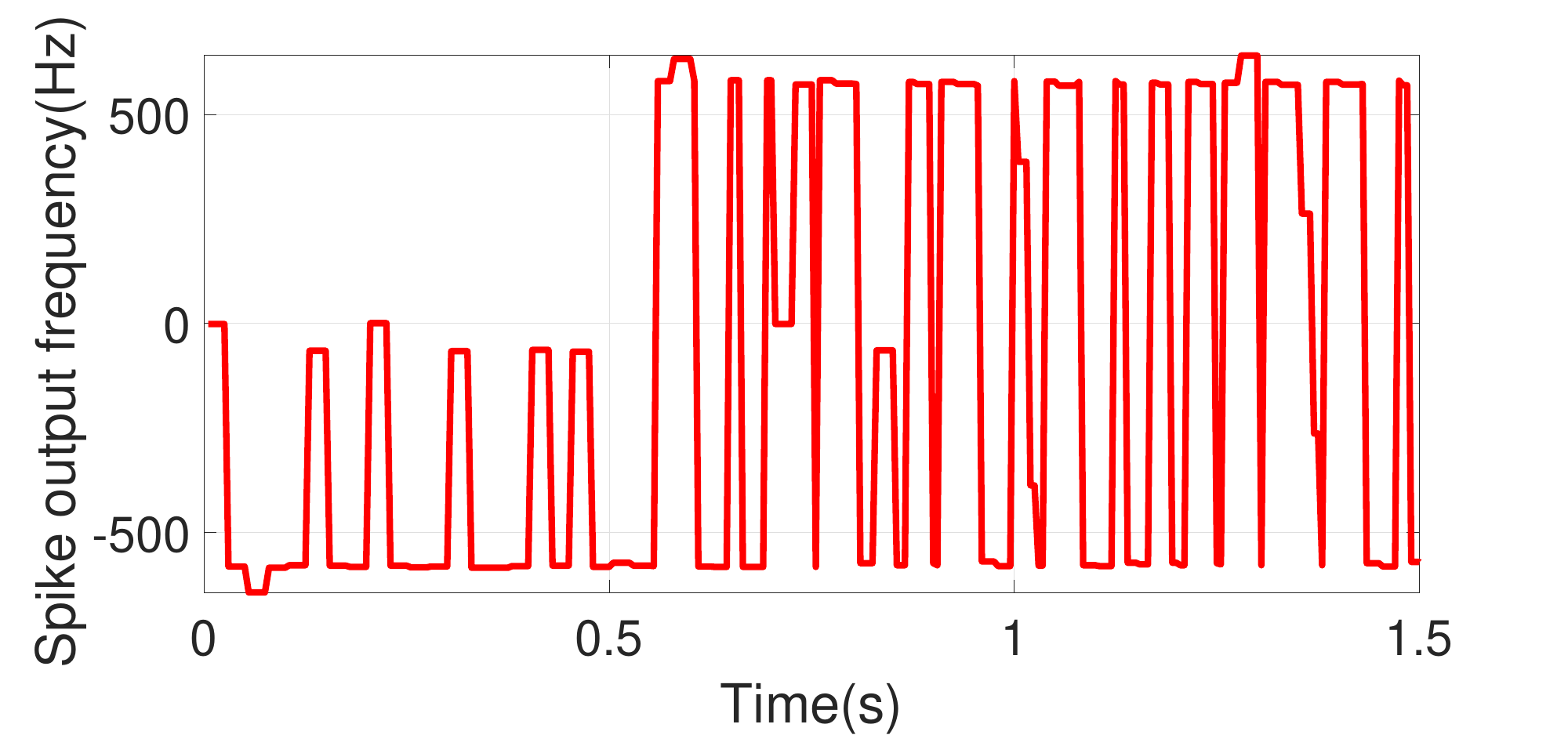}}
\subfloat[]{\includegraphics[width=8cm]{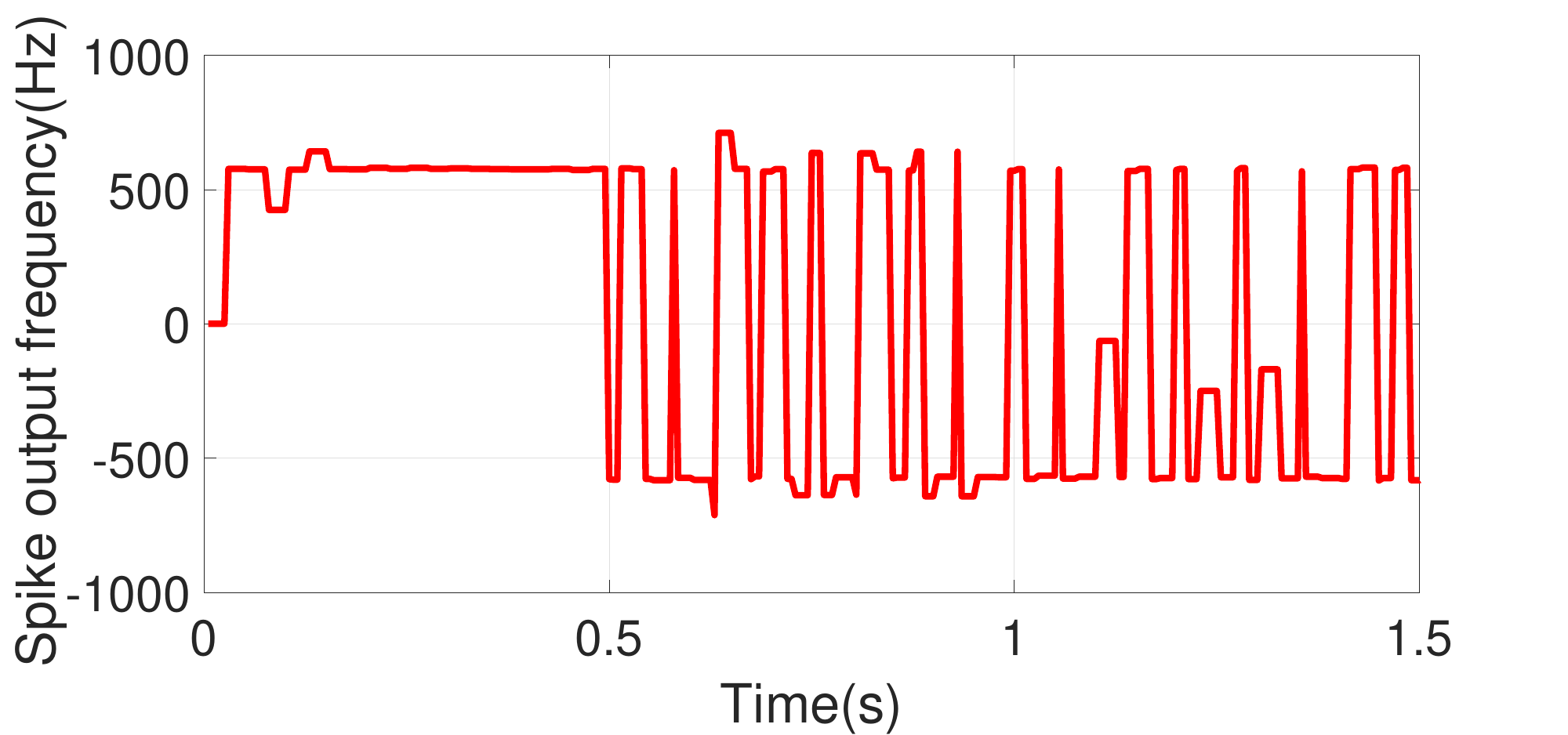}}
%\label{fig:LinImage2d}}
\hspace{1mm}
\subfloat[]{\includegraphics[width=8cm]{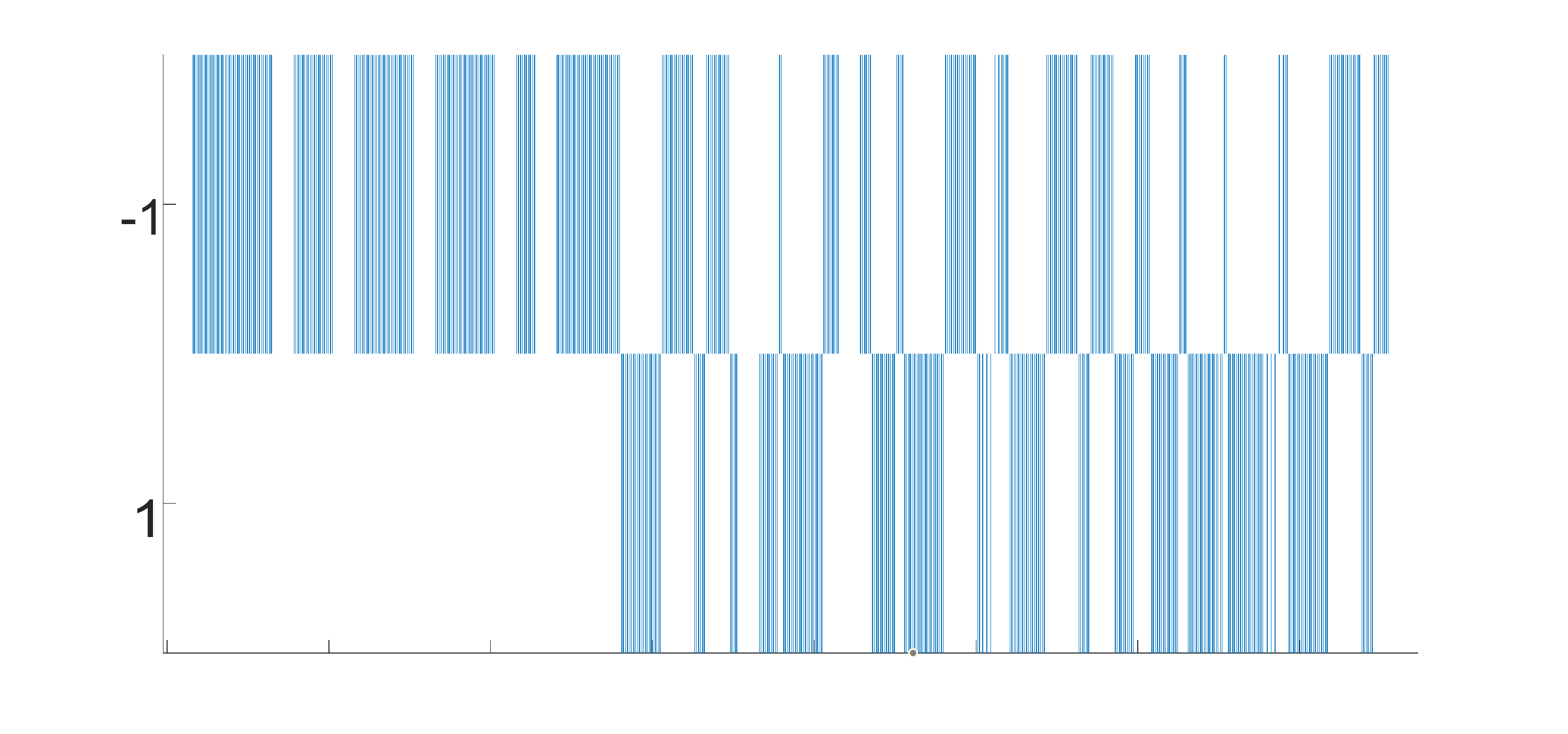}}
\subfloat[]{\includegraphics[width=8cm]{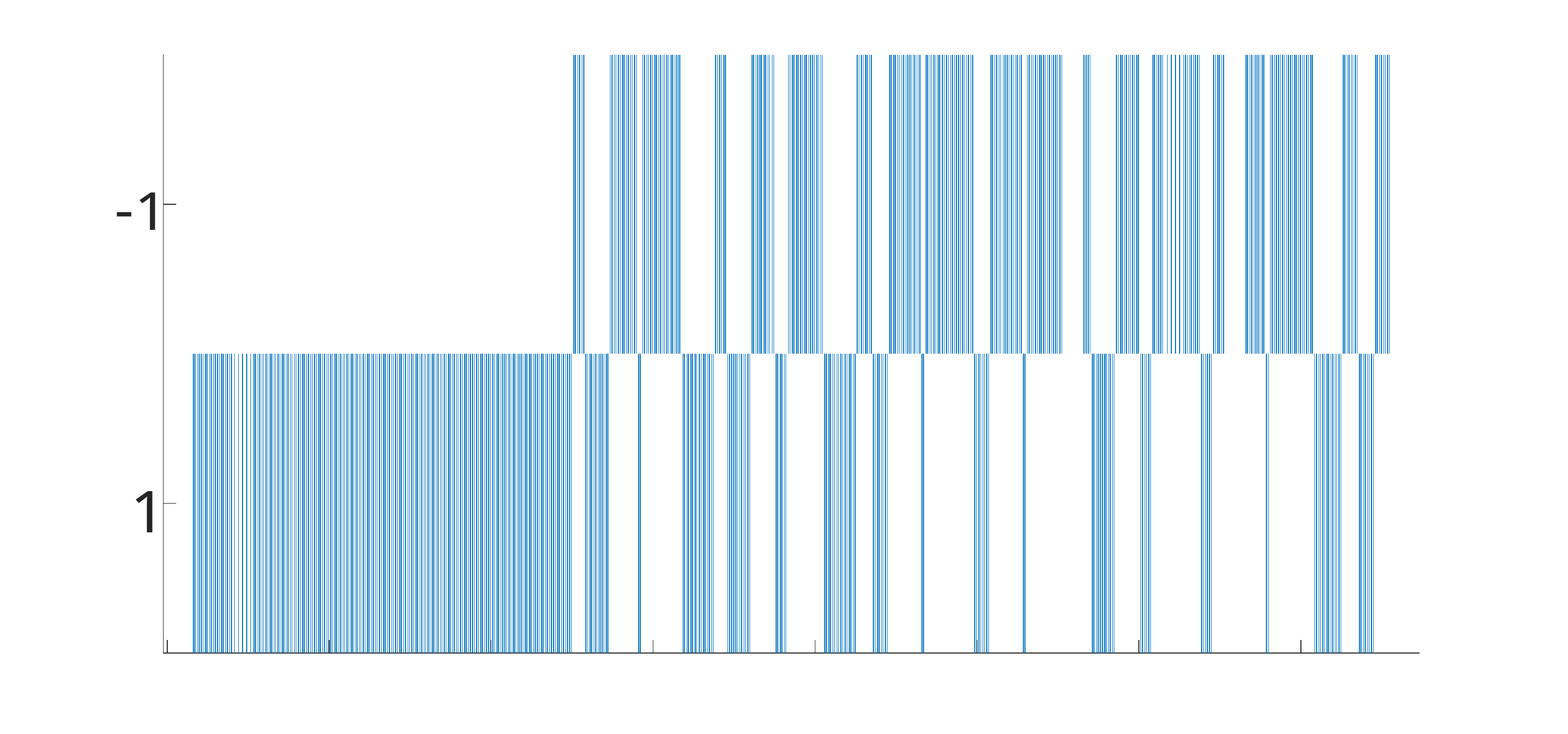}}
%\label{fig:LinImage2d}}
\hspace{1mm}
\caption{\textcolor{black}{Plots for balancing a cartpole using the Lu.i analog neuron board. The left column shows the plots for the negative initial angle. The right column shows the plots for a positive initial angle. From top to bottom, the rows show the angle evolution over time, the spike output frequency, and the spike raster plot respectively. The control value is a negative scaled version of the spike output frequency.}}
\label{fig:Lui_control}
\end{figure}

Figure~\ref{fig:Lui_control} shows the evolution of pole angle $\theta$, the spike output frequency (which is proportional to the control force $F$), and the spike rasters with time, for both positive and negative starting angles. It is clear that the control is successful, i.e., the pole is successfully steered towards the UEP, starting from the initial position.  

% For about 1.6 seconds, the pole can be stabilized for both positive and negative perturbations. However, after 1.6 seconds (which is not shown in the figure), the Lu.i board emitted spikes at a constant frequency irrespective of the input to the neuron which caused the control to fail. 
% The main reasons behind this behaviour are:
% \begin{enumerate}
%     \item Our encoding scheme cannot represent frequencies greater than a threshold based on our chosen high pulse period (shown in Figure ~\ref{fig:Luilinearity}). This causes the Lu.i board to emit a constant output frequency beyond a certain value of the inputs.
%     \item The weights set on the Lu.i board did not accurately match the ratio given by MATLAB, due to human error. Hence, the control values and the resulting response characteristics did not match exactly with the simulation results. 
%     \item Due to imperfect feedback constants resulting from imprecise weights on the board, the linear velocity \(\dot{x}\) and the angular velocity \(\dot{\theta}\) of the pole exceeds the boundary condition of the state vector for which the linearity assumption holds (from Equation \ref{cartpoleEOM}). Hence the pole falls over after 1.6 seconds, and the neuron gives an unexpected output.
   
% \end{enumerate}
Table \ref{tab:comparisonmet} shows the control performance metrics with the Lu.i board.
The experiment shows that the convergence occurs in reasonable time demonstrating that a single spiking neuron or two spiking neurons can be used for LQR control. The metrics reflect that for the implementation on a spiking hardware, the control performance is poorer compared to simulations. The time metrics are much larger for Lu.i which is expected due to the following reasons:
\begin{itemize}
    \item Spiking hardware provides poorer accuracy and high noise levels (which is not considered in simulations).
    \item The hardware implementation uses a single neuron to approximate excitatory and inhibitory behaviours, limiting its performance.
    \item Real-time rate encoding for the hardware is less precise than in MATLAB.
    \item The communication overhead between 
    Lu.i, Arduino, and the laptop is substantial compared to running everything in a single MATLAB simulation.
\end{itemize} 
After the $15s$ simulation time, $\theta$ is seen to have small oscillations about $\theta = \pi$, within the limits of the steady-state error.
Although 2 neurons with a rate-based encoding can control systems with linear state variable feedback type control, unlike a non-spiking neuron, to achieve precise end-to-end control with spiking neurons, the number of such neurons needed would be dependent on characteristics such as range and frequency components of the signal to be represented. Hence a population of neurons, each sensitive to a certain range of control values, is needed for precise control, as discussed in the next set of experiments.

\subsection{\textbf{Rate to Population Encoding for Neurons to Balance a Cartpole}}
From the previous section, it is seen that 2 neurons can adequately balance a pole on a cart system. Although rate-based encoding using a single neuron can achieve linear state variable feedback control (steer the pole on the cart to the UEP starting from the initial perturbed state)  and is robust to noise coming from the fluctuations of the parameters of the board/chip or from encoding imperfections (i.e., when the spike encoding of the state or the values of current generated by the neurons, vary within a small interval between successive trials, the deviations in the metrics of control don't exceed a certain threshold), it is not very precise (we cannot achieve minor changes in KPIs).
To address this, population-based encoding for control is preferred. Here, each neuron encodes a small percentage of the signal, increasing the overall precision. This is because each neuron is sensitive to a small fragment of the signal, and hence collectively, an ensemble of neurons can better represent the entire signal. Here, the Neural Engineering Framework is used for simulating populations of neurons. Although population encoding requires more on-board resources (number of neurons and synapses), it is useful when the precision is more important than the resource/energy constraints.

\subsubsection{\textbf{Simulation}}
\begin{figure}
\centering
\includegraphics[width=6cm]{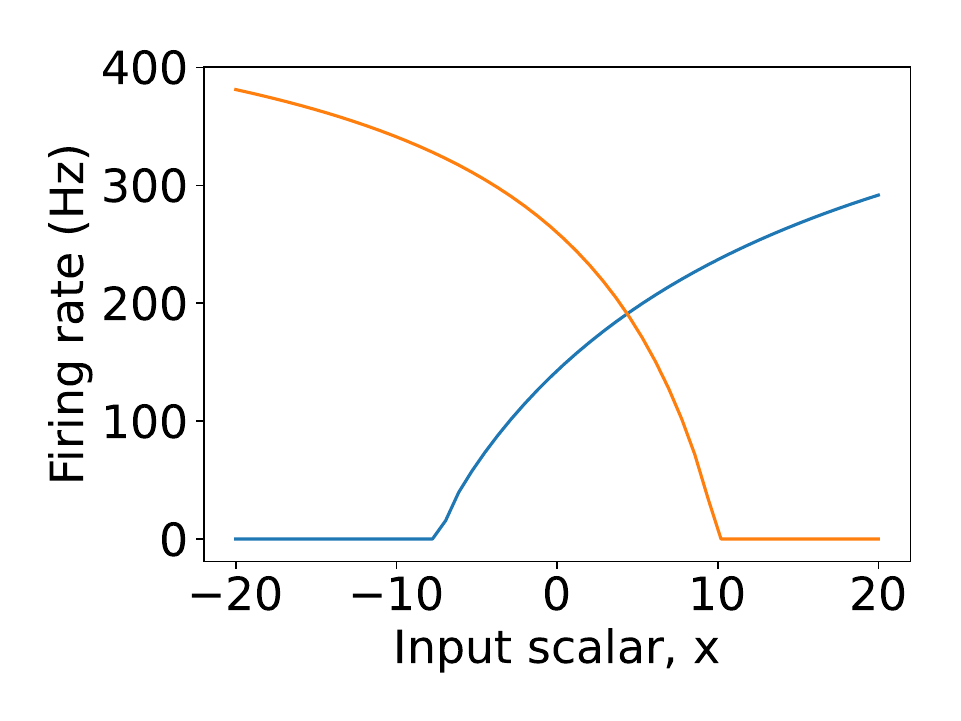}
\caption{\textcolor{black}{Tuning curves for 2 neuron control in Nengo for positive starting angle of the pole.}}
\label{fig:Tuning2neurons}
\end{figure}

\begin{figure}
\centering
\subfloat[]{\includegraphics[width=6cm]{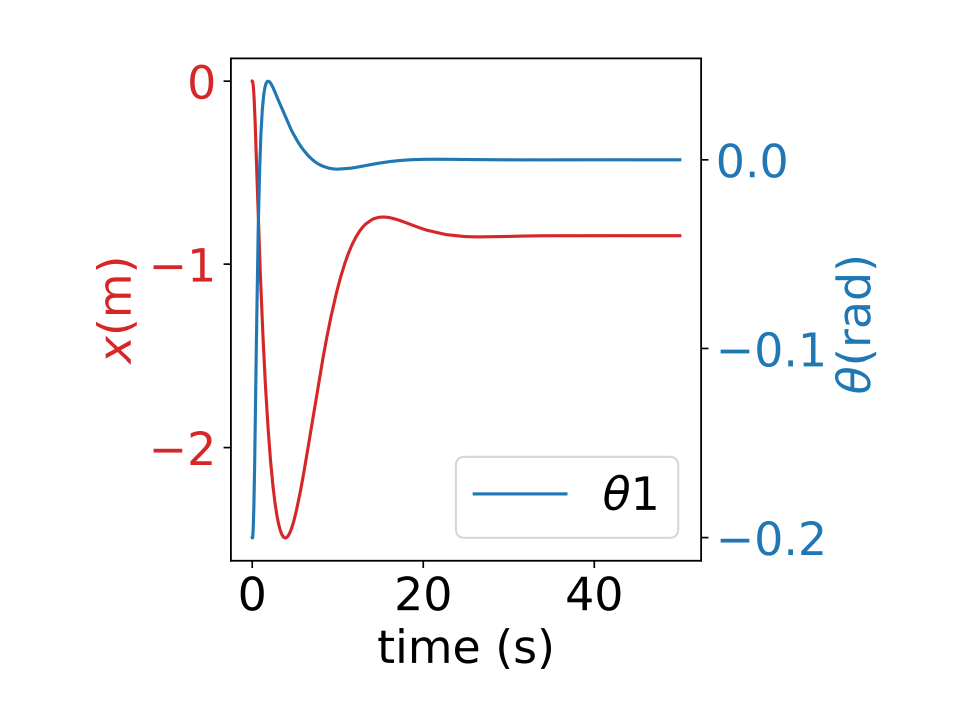}}
%\hfill
\subfloat[]{\includegraphics[width=6cm]{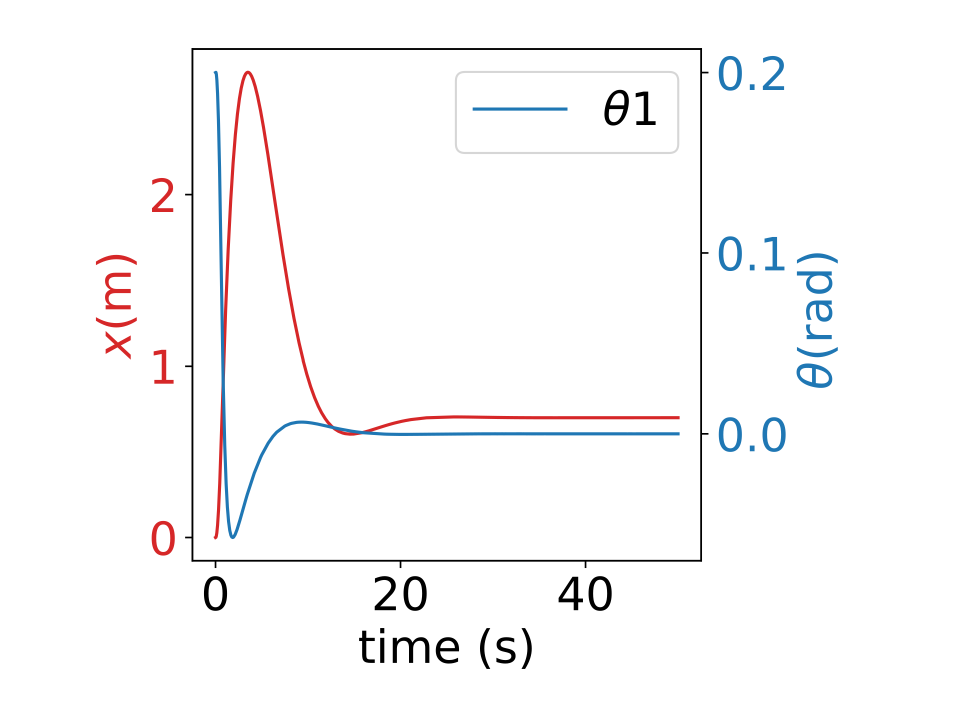}}
\hspace{1mm}
\subfloat[]{\includegraphics[width=6cm]{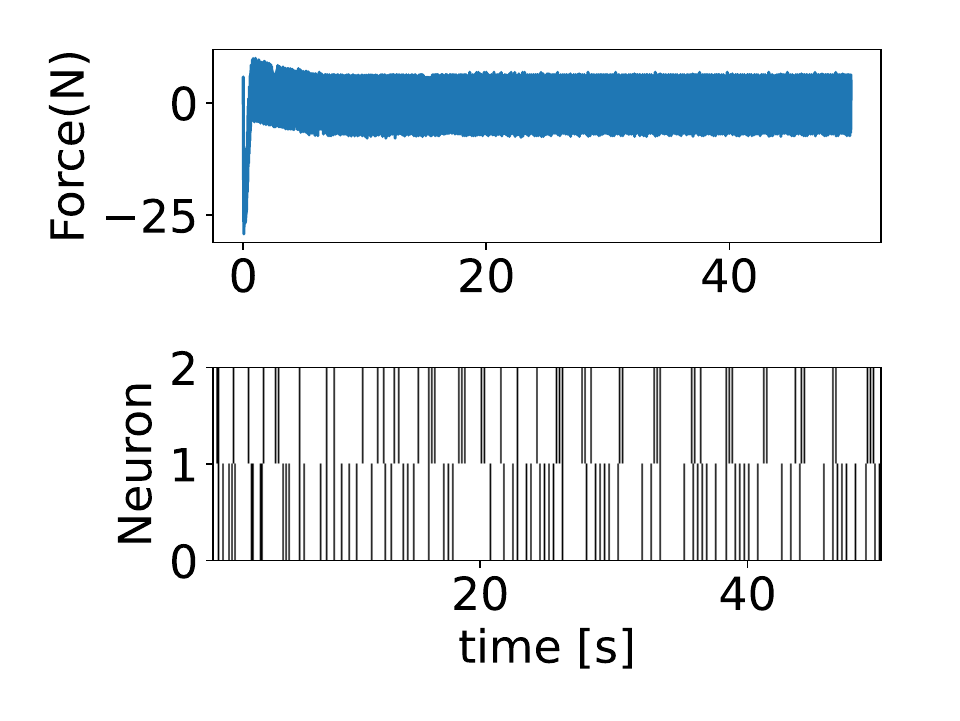}}
\subfloat[]{\includegraphics[width=6cm]{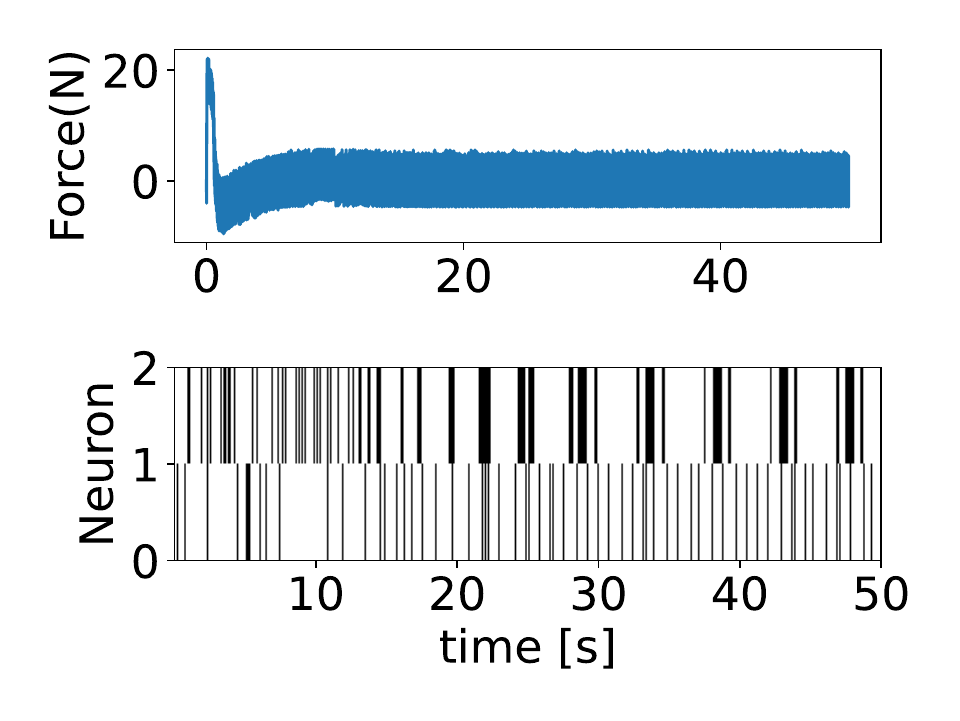}}
%\label{fig:Nengo1neuron}}
\hspace{1mm}

\caption{\textcolor{black}{Plots for balancing a cartpole using $2$ neuron in the NEF.(a) and (b) show pole angle and cart position evolution over time for positive and negative initial values respectively. (c) and (d) show the corresponding control value and spike pattern.}}
\label{fig:Nengo_2_neurons_control}
\end{figure}

\begin{figure}[ht]
\centering
\subfloat[]{\includegraphics[width=9cm]{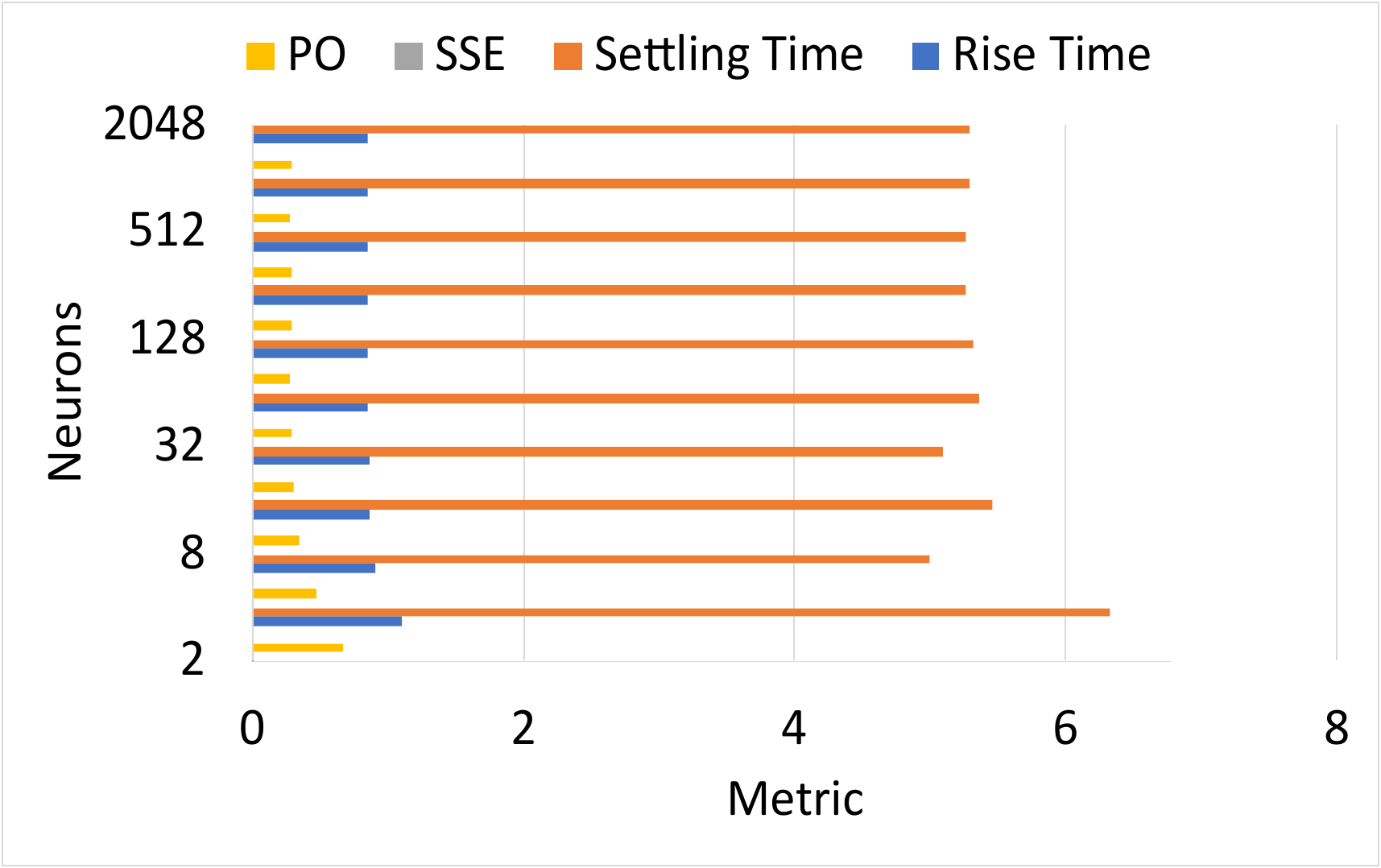}}
\subfloat[]{\includegraphics[width=8.5cm,trim = {0 0 0 1cm},clip]{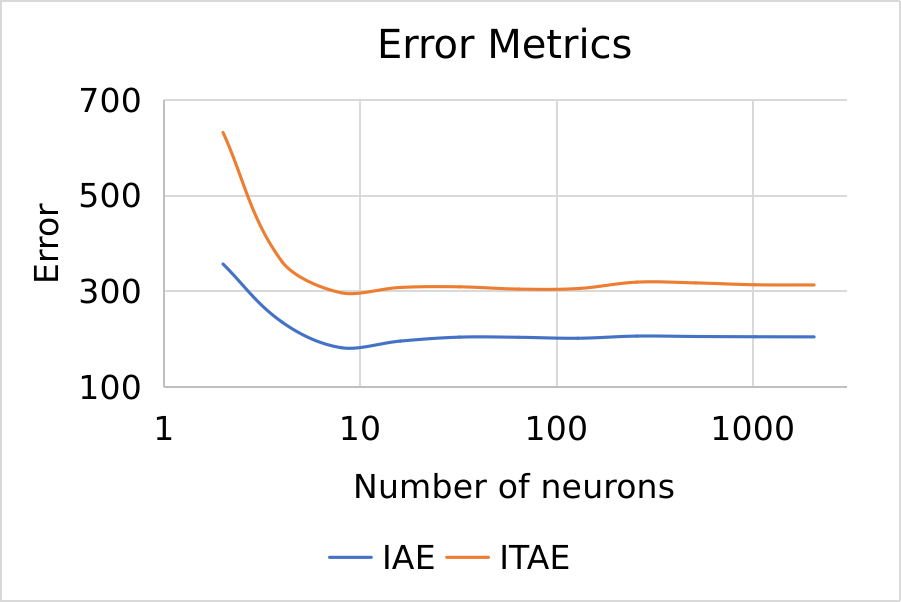}}
\caption{\textcolor{black}{Change of control metric values with the number of neurons for a cartpole.}}
\label{fig:Nengo_neuron_metrics}
\end{figure}

\begin{figure}[ht]
\centering
\includegraphics[width=10cm,trim = {0 0 0 1.5cm},clip]{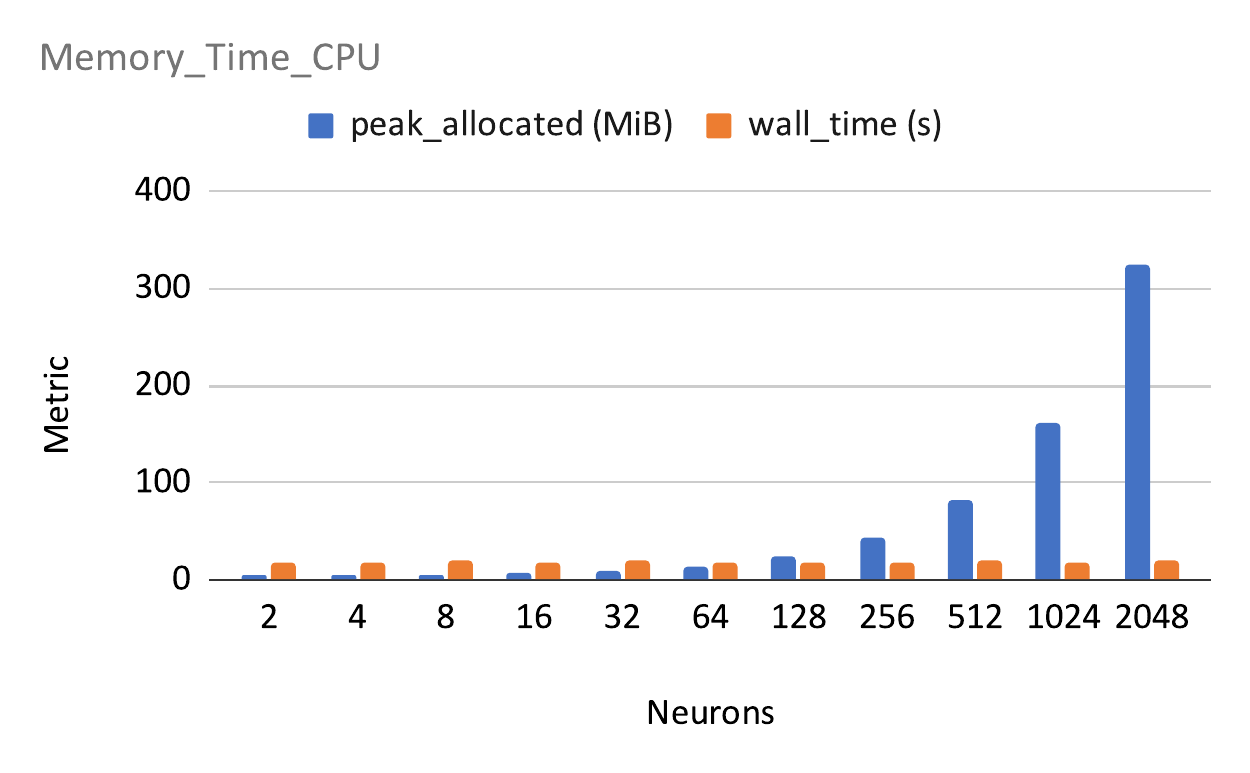}
\caption{\textcolor{black}{Change of neuromorphic metrics with the number of neurons for a cartpole when run on CPU.}}
\label{fig:Nengo_CPU_metrics}
\end{figure}

\begin{figure}[ht]
\centering
\subfloat[]{\includegraphics[width=10cm, trim = {0 0 0 1.5cm}, clip]{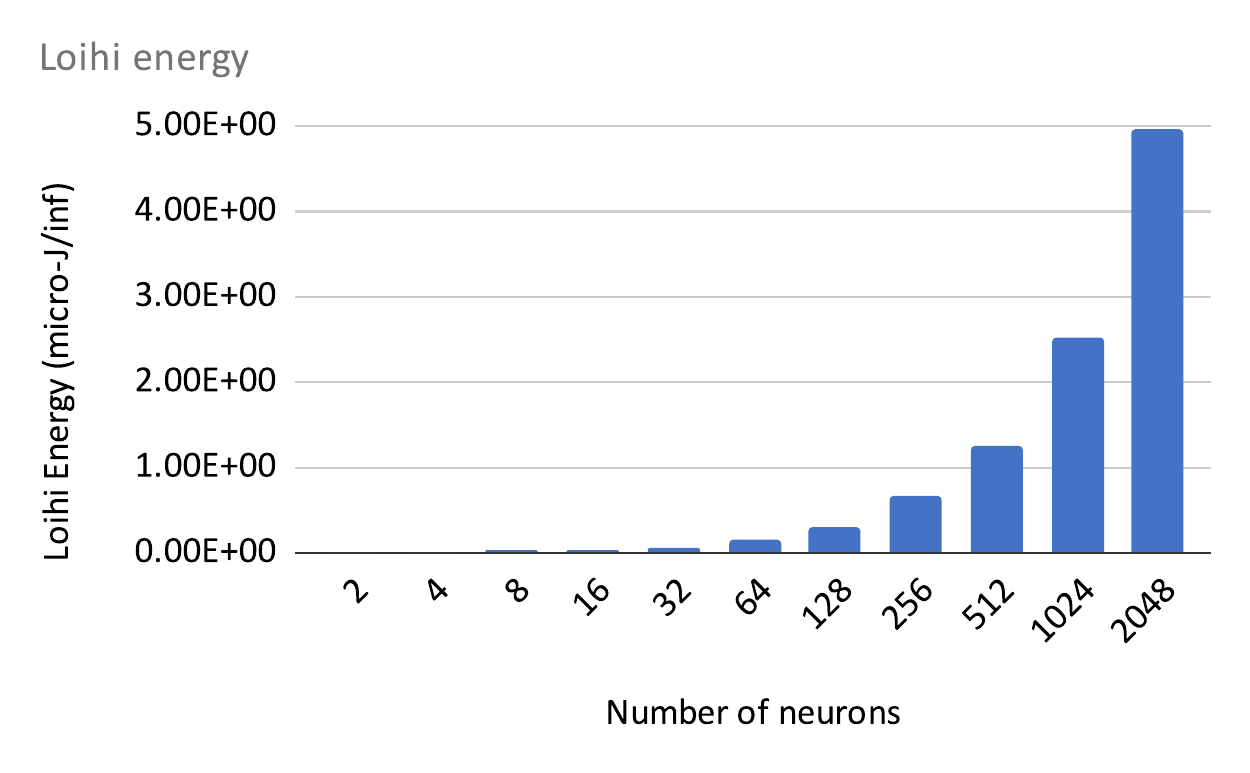}}

\subfloat[]{\includegraphics[width=10cm, trim = {0 0 0 1.5cm}, clip]{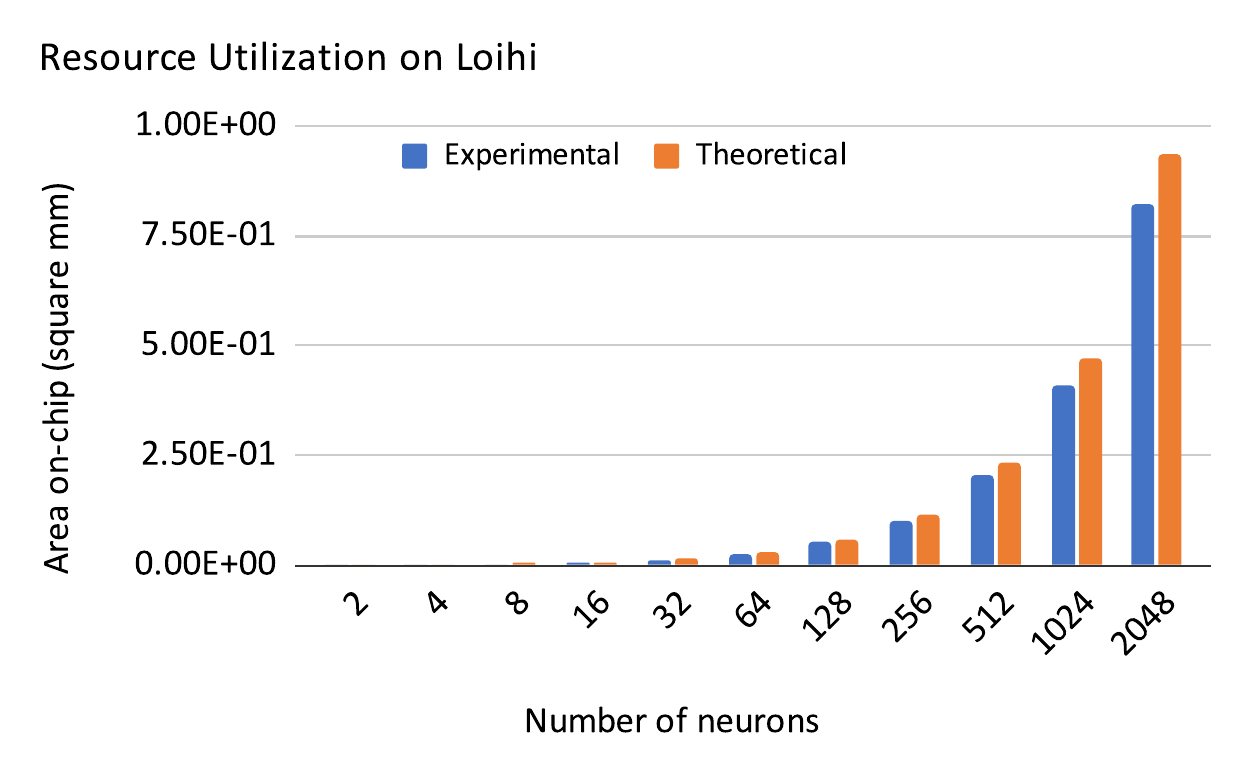}}
\caption{\textcolor{black}{Change of (a) energy consumption and (b) chip area with the number of neurons for a cartpole.}}
\label{fig:Loihi_neuron_energy}
\end{figure}

\begin{figure}[ht]
\centering
\subfloat[]{\includegraphics[width=6cm]{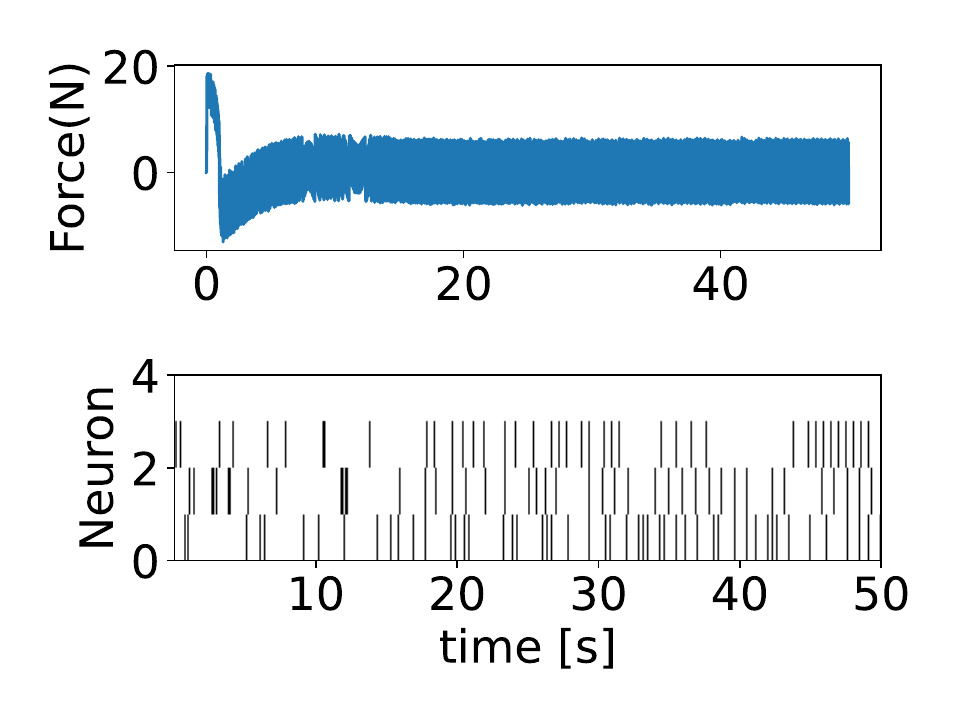}}
\subfloat[]{\includegraphics[width=6cm]{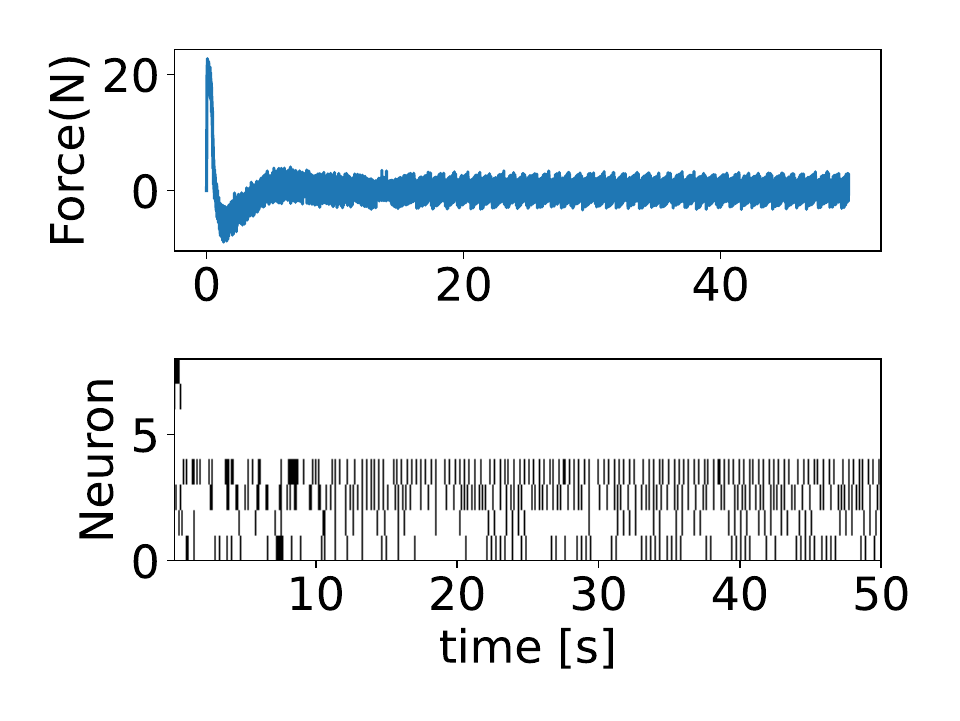}}
\hspace{1mm}
\subfloat[]{\includegraphics[width=6cm]{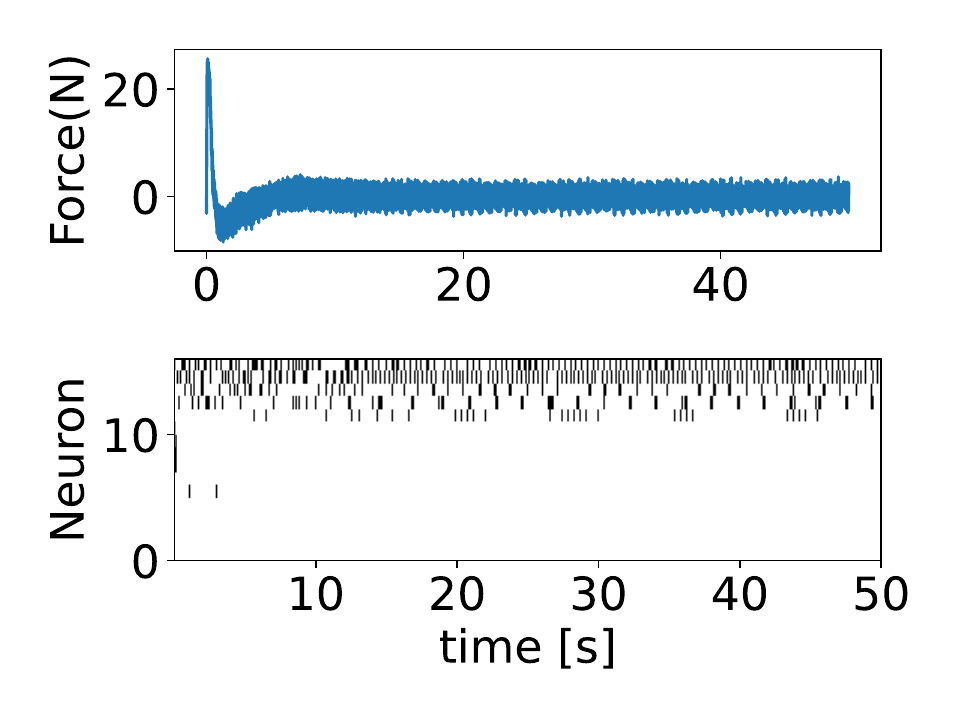}}
\subfloat[]{\includegraphics[width=6cm]{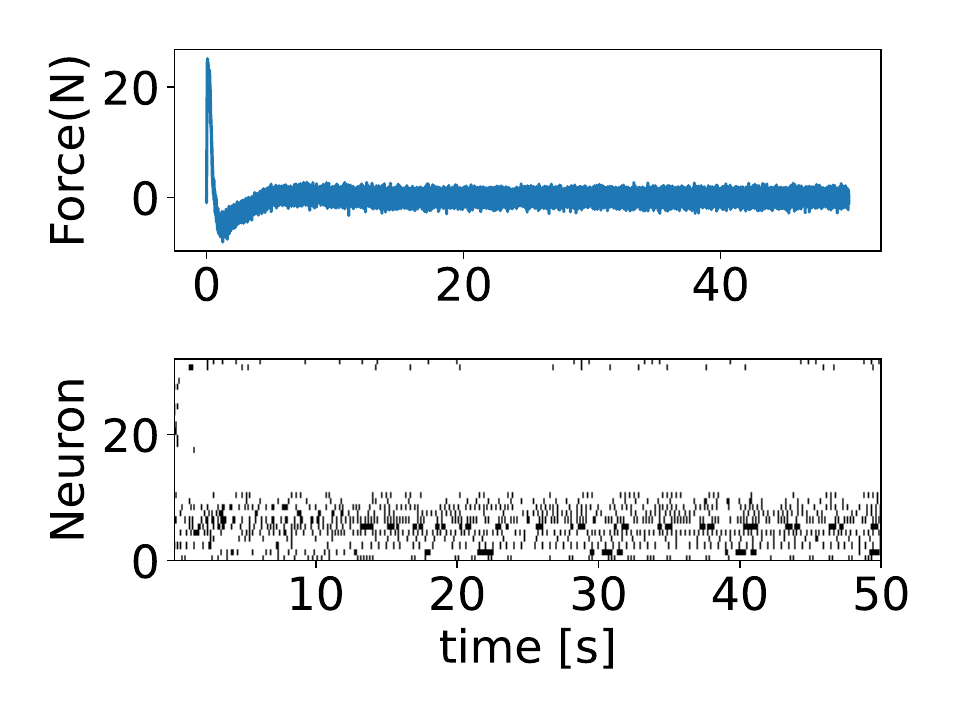}}
% \subfloat[]{\includegraphics[width=6cm]{Images/Positive_perturbation_control_spikes_64_neurons_svg-tex.pdf}}
% \subfloat[]{\includegraphics[width=6cm]{Images/Positive_perturbation_control_spikes_128_neurons_svg-tex.pdf}}
%\hfill
\hspace{1mm}

\caption{\textcolor{black}{Control signal profile and spike raster plots for balancing a cartpole using a Nengo ensemble with  (a) 4, (b) 8, (c) 16, and (d) 32 neurons respectively.}}
\label{fig:Multi_nueron_control}
\end{figure}

\paragraph{\textbf{Balancing a cartpole with a varying number of neurons in the NEF}}
In this section, the results for controlling a cartpole using an ensemble of neurons in NEF, and the change in control and neuromorphic performance metrics with the number of neurons, are presented. Figure \ref{fig:Tuning2neurons} shows the tuning curve for a 2-neuron ensemble in Nengo, used to control a cartpole. From this figure, it is seen that one neuron fires mostly for positive inputs and the other for negative inputs. Clearly, this behavior resembles rate-based encoding, similar to 2-neuron control in MATLAB discussed before. Figure \ref{fig:Nengo_2_neurons_control} shows the evolution of $x$ and $\theta$ for balancing a pole on a cart with $2$ neurons for both positive and negative initial angles, and the corresponding filtered control signals and spike raster plots. Now, the number of neurons is increased in powers of $2$ from $2$ to $2048$, and the change in control performance is tabularized in Table \ref{tab:NengoMetricTable}. The same has been graphically illustrated in Figure \ref{fig:Nengo_neuron_metrics}. 
\textcolor{black}{A few neuromorphic metrics for the simulation on CPU are highlighted in Table \ref{tab:sim_perf}, and the energy and resource-utilization for the implementation of such control on the Loihi neuromorphic chip are highlighted in Table \ref{tab:core_area_metrics}}. It is seen that the control performance improves up to around $128$ neurons and then saturates. 
\textcolor{black}{The IAE (Integral of Absolute Error) and ITAE (Integral of Time-weighted Absolute Error) metrics show a minimum at around $10$ neurons and then saturate. Clearly, the neuron requirement is around $100$ or even less for the smooth control signal profile needed for balancing a cartpole. From Figure \ref{fig:Nengo_CPU_metrics}
it is seen that the peak allocated memory increases with an increase in number of neurons, althought the wall-time remains more or less constant.
Also, as expected, the energy and on-board resources (neuron area on chip) keep increasing with the number of neurons as seen in Figure \ref{fig:Loihi_neuron_energy}. The theoretical energy and chip area are calculated from the unit values provided in the paper on Intel's Loihi neuromorphic board \cite{davies2018loihi}. An important point to note here is that, for the simulations, there are $1000$ inferences per second, so for a duration of $10s$, the number of inferences is $10^4$. Thus, for a $128$ neuron ensemble, the Loihi energy consumption is $3$ orders of magnitude less than CPU which decreases upto $5$ orders of magnitude for a $2$ neuron ensemble. The theoretical chip area is compared with the actual chip area requirement by experimentally determining the percentage of neuromorphic cores utilized for the job run. It is seen that though the experimental chip area requirement is slightly less than the theoretical one, both increase comparably with an increase in the number of neurons}. Therefore, to limit resource utilization and to design ASICs for a particular application on the edge, a suitable number of neurons needs to be chosen for a particular control problem. Figure \ref{fig:Multi_nueron_control} shows the control force and spike raster plots for control with $4$, $8$, $16$, and $32$  neurons. Two important observations emerge from these plots:

\begin{enumerate}
\item The noise levels in the control signals decrease with an increase in the number of neurons, as expected. This increases the precision in control.
\item As the number of neurons increases, certain neurons spike, while others remain dormant as seen from the spike raster plots. This is the reason behind the saturation of control performance after a certain point, since the number of neurons actively spiking gets fixed.
\item \textcolor{black}{Although the control performance saturates after around $100$ neurons, the peak memory allocation on CPU and the percentage core allocation on Loihi increase drastically. This creates a need to limit the number of neurons needed for the application at hand without compromising with the control performance.}
\end{enumerate}

\begin{table}[h]
\caption{\textcolor{black}{Control performance of a cartpole for varying numbers of neurons.}}
\label{tab:NengoMetricTable}
\lineup
\begin{center}
%\item[] 
\begin{tabular}{@{}llllllll}\br
\textbf{Neurons}&\textbf{P.O.}&$\mathbf{T_r}$&
$\mathbf{T_s}$&\textbf{SSE}&\textbf{IAE}&\textbf{ITAE}&\textbf{ISC}\\
\textbf{}&\textbf{($\%$)}&$(s)$&
$(s)$&$(\%)$&$(rad.s)$ &$rad.s^2$&$(N^2.s)$ \\
\textbf{}& & & & $(\times10^{-5})$&$(\times 10^{-3})$&$(\times 10^{-3})$&$(\times 10^{2})$\\
\mr
2&  67.5&   1.483&  6.778&  226& 357.289&	632.429 &	5.74\\
4&  46.8&   1.104&  6.320&  315& 235.768&	362.054 &	3.05\\ 
8&  35.1&   0.906&  4.994&  265& 182.391&	297.242 &	24.7\\
16& 30.6&   0.863&  5.454&  27.8& 196.206&	308.334 &	2.40\\
32& 29.7&   0.863&  5.093&  63.3& 204.459&	309.658 &	2.11\\ 
64& 29.7&   0.856&  5.365&  34.0& 204.155&	304.939 &	2.06\\ 
128&  28.4&   0.857&    5.323&  3.77& 202.119&	306.029 &	2.01\\
256&    28.5&   0.857&  5.267&  22.3& 206.671&	319.478 &	2.01\\
512&    28.2&   0.854&  5.262&  28.0& 205.808&	317.848 &	1.99\\
1024&   28.4&   0.856&  5.293&  11.8& 205.257&	313.961 &	1.99\\
2048&   28.4&   0.857&  5.283&  0.267& 205.009&	313.597 &	1.99\\
\br
\end{tabular}
\end{center}
\end{table}

\begin{table}[h]
\centering
\caption{\textcolor{black}{Neuromorphic performance metrics across different neuron counts on CPU.}}
\begin{tabular}{@{}llllllll}
\toprule
 \textbf{Neurons} & \textbf{Sim} & \textbf{Wall} & \textbf{Real} & \textbf{Max} & \textbf{Peak} & \textbf{Spikes/} & \textbf{Estimated} \\
\textbf{} & \textbf{time} & \textbf{time} & \textbf{time} & \textbf{memory} & \textbf{allocated} & \textbf{neuron} & \textbf{CPU} \\
\textbf{} & \textbf{(s)} & \textbf{(s)} & \textbf{factor} & \textbf{RSS (MiB)} & \textbf{(MiB)} & \textbf{($\times 10^5$)} & \textbf{energy (J)} \\
\midrule
2    & 10.000 & 18.417 & 0.543 & 340.398 & 4.665 & 5.975 & 2.739 \\
4    & 10.000 & 18.944 & 0.528 & 882.414 & 4.996 & 9.590 & 1.597 \\
8    & 10.000 & 19.488 & 0.513 & 890.055 & 5.585 & 11.886 & 3.352 \\
16   & 10.000 & 18.486 & 0.541 & 897.070 & 6.805 & 7.422 & 3.585 \\
32   & 10.000 & 20.459 & 0.488 & 901.821 & 9.264 & 6.815 & 5.345 \\
64   & 10.000 & 18.700 & 0.535 & 905.172 & 14.161 & 10.406 & 4.395 \\
128  & 10.000 & 18.708 & 0.534 & 919.140 & 23.906 & 8.376 & 6.058 \\
256  & 10.000 & 19.113 & 0.523 & 961.339 & 43.460 & 8.831 & 5.984 \\
512  & 10.000 & 19.776 & 0.507 & 987.973 & 82.568 & 8.206 & 1.329 \\
1024 & 10.000 & 19.052 & 0.525 & 1098.559 & 160.778 & 8.537 & 2.109 \\
2048 & 10.000 & 21.079 & 0.474 & 1317.242 & 323.659 & 8.795 & 3.334 \\
\bottomrule
\end{tabular}
\label{tab:sim_perf}
\end{table}

\begin{table}[h]
\centering
\caption{\textcolor{black}{Loihi neuromorphic ore utilization and area metrics for increasing neuron counts.}}
\begin{tabular}{@{}llllll}
\toprule
\textbf{Neurons} & \textbf{Core} & \textbf{Number} & \textbf{Experimental} & \textbf{Theoretical} & \textbf{Estimated}\\
 & \textbf{Utilization} & \textbf{of Cores} & \textbf{Area on chip} & \textbf{Area on chip} &\textbf{Energy} \\
  & \textbf{($\%$)} & \textbf{} & \textbf{($mm^2$)} & \textbf{($mm^2$)} & \textbf{($\mu J/inf$)}\\
\midrule
2    & 0.20  & 1 & $8.20 \times 10^{-4}$  & $9.00 \times 10^{-4}$ & $9.14\times 10^{-3}$\\
4    & 0.40  & 1 & $1.64 \times 10^{-3}$  & $1.80 \times 10^{-3}$ &$1.08 \times 10^{-2}$\\
8    & 0.80  & 1 & $3.28 \times 10^{-3}$  & $3.70 \times 10^{-3}$ &$1.96 \times 10^{-2}$\\
16   & 1.60  & 1 & $6.56 \times 10^{-3}$  & $7.30 \times 10^{-3}$ &$2.88 \times 10^{-2}$\\
32   & 3.10  & 1 & $1.27 \times 10^{-2}$  & $1.46 \times 10^{-2}$ &$6.60\times 10^{-2}$\\
64   & 6.20  & 1 & $2.54 \times 10^{-2}$  & $2.93 \times 10^{-2}$ &$1.56 \times 10^{-1}$\\
128  & 12.50  & 1 & $5.13 \times 10^{-2}$  & $5.86 \times 10^{-2}$ &$2.95 \times 10^{-1}$\\
256  & 25.0 & 1 & $1.03 \times 10^{-1}$  & $1.17 \times 10^{-1}$ &$6.60\times 10^{-1}$\\
512  & 50.0 & 1 & $2.05 \times 10^{-1}$  & $2.34 \times 10^{-1}$ &$1.26$\\
1024 & 100.0 & 1 & $4.10 \times 10^{-1}$  & $4.69 \times 10^{-1}$ &$2.52$\\
2048 & 100.0 & 2 & $8.20 \times 10^{-1}$  & $9.38 \times 10^{-1}$ &$4.96$\\
\bottomrule
\end{tabular}
\label{tab:core_area_metrics}
\end{table}

\paragraph{\textbf{Effects of Variation of NEF Parameters on the Control Performance}}
\textcolor{black}{Besides the number of neurons in an ensemble, other parameters like the encoders, intercepts of the tuning curves, and maximum firing rates can be varied to check the effects on the control performance. In this capacity, this section talks about manually altering the tuning curves of a neural ensemble and recording the changes in the control performances thereof.}

\textcolor{black}{The intercepts are chosen in three different ways. First they are equally spaced (linear spacing) within a varying interval (Table \ref{tab:performance_metrics_linspace}), then a normal distribution is chosen with mean at $0$, and varying standard deviation (Table \ref{tab:performance_metrics_normal}), and finally, the intercepts are sampled from an uniform distribution within a varying interval (Table \ref{tab:performance_metrics_uniform}), for a $100$ neuron ensemble.}

    \begin{table}[htbp]
    \caption{\textcolor{black}{Performance Metrics for linearly spaced intercepts with a varying range}}
    \centering
    \begin{tabular}{cccccccc}
    \toprule
    \textbf{Range} & \textbf{Ts} & \textbf{Tr} & \textbf{PO} & \textbf{SSE} & \textbf{IAE} & \textbf{ITAE} & \textbf{ISC} \\
\textbf{} & \textbf{($s$)} & \textbf{($s$)} & \textbf{($\%$)} & \textbf{($\%$)} & \textbf{($rad.s$)} & \textbf{($rad.s^2$)} & \textbf{($N^2.s$)} \\
     \textbf{} & \textbf{} & \textbf{} & \textbf{} & \textbf{$(\times 10^{-3})$} & \textbf{($\times 10^{-3}$)} & \textbf{($\times 10^{-3}$)} & \textbf{($\times 10^{2}$)} \\
\midrule
    -1.0 to 1.0  & 5.181 & 0.814 & -25.491 & 2.200 & 207.132 & 321.833 & 2.044 \\
    -0.75 to 0.75 & 5.130  & 0.815 & -25.516 & -1.670 & 206.592 & 318.035 & 2.029 \\
    -0.5 to 0.5  & 5.149 & 0.816 & -25.348 & -0.634 & 204.111 & 313.622 & 2.034 \\
    -0.25 to 0.25 & 4.725 & 0.819 & -28.664 & -0.250 & 201.531 & 287.758 & 2.111 \\
    -0.1 to 0.1  & 6.065 & 0.847 & -17.316 & -92.0  & 197.172 & 365.274 & 3.616 \\
    \bottomrule
    \end{tabular}
    \label{tab:performance_metrics_linspace}
    \end{table}

    \begin{table}[htbp]
    \caption{\textcolor{black}{Performance Metrics for intercepts sampled from a normal distribution centred at $0$ with varying standard deviations.}}
    \centering
    \begin{tabular}{cccccccc}
    \toprule
   \textbf{STD} & \textbf{Ts} & \textbf{Tr} & \textbf{PO} & \textbf{SSE} & \textbf{IAE} & \textbf{ITAE} & \textbf{ISC} \\
\textbf{} & \textbf{($s$)} & \textbf{($s$)} & \textbf{($\%$)} & \textbf{($\%$)} & \textbf{($rad.s$)} & \textbf{($rad.s^2$)} & \textbf{($N^2.s$)} \\
     \textbf{} & \textbf{} & \textbf{} & \textbf{} & \textbf{$(\times 10^{-3})$} & \textbf{($\times 10^{-3}$)} & \textbf{($\times 10^{-3}$)} & \textbf{($\times 10^{2}$)} \\
\midrule
    0.0 & 6.081 & 0.848 & -17.215 & -85.7 & 197.437 & 367.653 & 3.671 \\
    0.1 & 5.318 & 0.806 & -27.390 & -3.280 & 213.380 & 341.667 & 2.176 \\
    0.2 & 5.276 & 0.815 & -24.272 & 1.190  & 205.378 & 326.067 & 2.031 \\
    0.3 & 5.002 & 0.818 & -26.016 & -0.453 & 204.000 & 312.081 & 2.019 \\
    0.4 & 5.141 & 0.816 & -25.142 & -0.334 & 204.259 & 313.208 & 2.0166 \\
    \bottomrule
    \end{tabular}
    \label{tab:performance_metrics_normal}
    \end{table}

\begin{figure}[ht!]
\centering
\subfloat[]{\includegraphics[width = 0.4\textwidth]{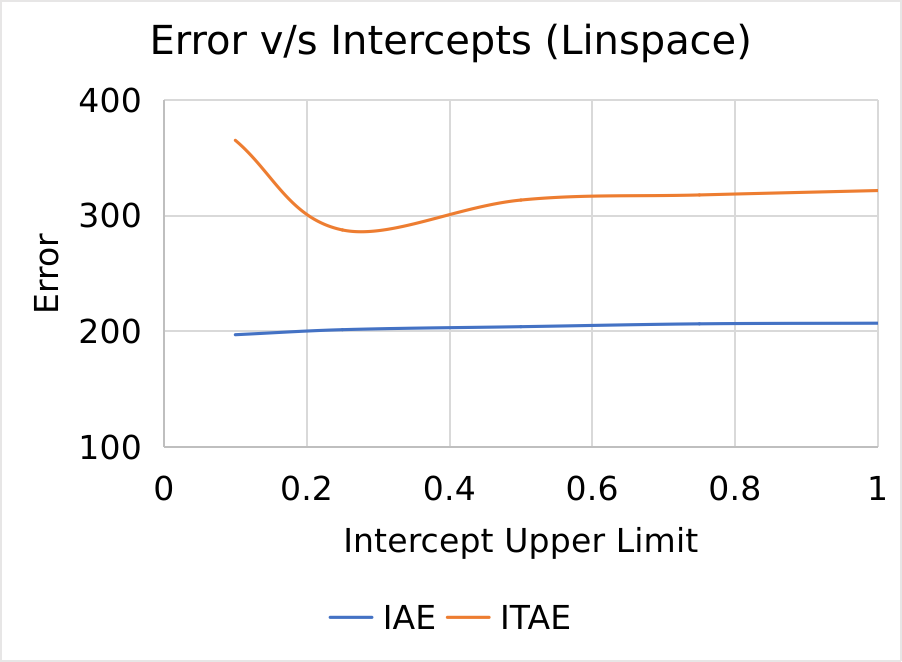}}
\subfloat[]{\includegraphics[width =0.4\textwidth]{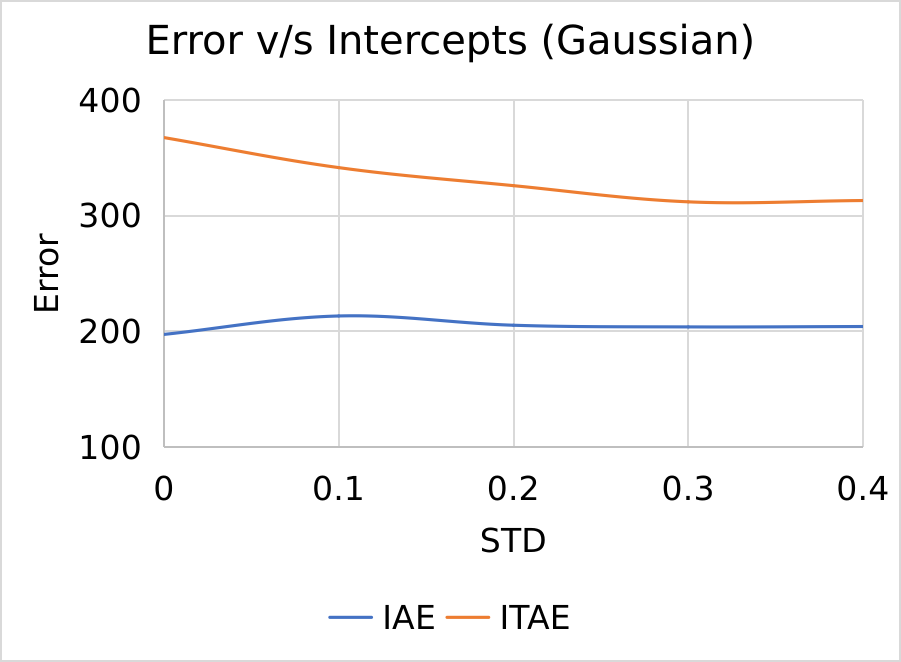}}
\hspace{0.5mm}
\subfloat[]{\includegraphics[width =0.4\textwidth]{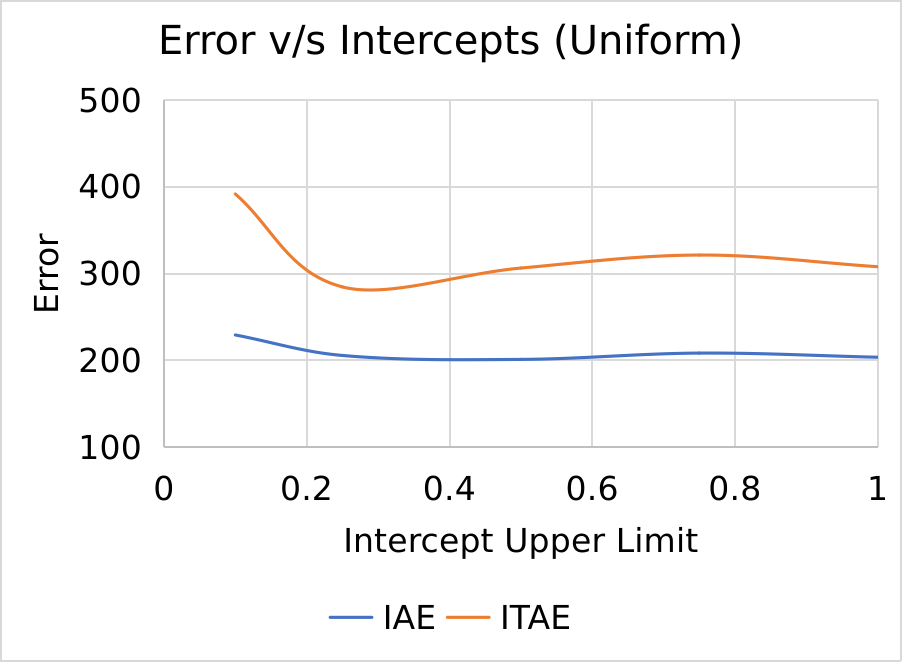}}
\caption{\textcolor{black}{Variation of IAE and ITAE with (a) Range of linearly sampled intercepts about $0$, (b) Standard deviation of intercepts sampled from a normal distribution centred at $0$, (c) Range of intercepts sampled from a uniform distribution. The y-axis shows errors scaled by $10^3$.}}\label{fig:Intercept}
\end{figure}

    \begin{table}[htbp]
    \caption{\textcolor{black}{Performance Metrics for intercepts sampled from a uniform distribution with a varying range.}}
    \centering
    \begin{tabular}{cccccccc}
    \toprule
    \textbf{Range} & \textbf{Ts} & \textbf{Tr} & \textbf{PO} & \textbf{SSE} & \textbf{IAE} & \textbf{ITAE} & \textbf{ISC} \\
\textbf{} & \textbf{($s$)} & \textbf{($s$)} & \textbf{($\%$)} & \textbf{($\%$)} & \textbf{($rad.s$)} & \textbf{($rad.s^2$)} & \textbf{($N^2.s$)} \\
     \textbf{} & \textbf{} & \textbf{} & \textbf{} & \textbf{$\times 10^{-3}$} & \textbf{($\times 10^{-3}$)} & \textbf{($\times 10^{-3}$)} & \textbf{($\times 10^2$)} \\
\midrule
    -1.0 to 1.0  & 5.077	& 0.817	& -25.803	&-1.560	& 203.601	&307.952	&2.053\\
    -0.75 to 0.75 & 5.176	&0.817	&-26.4210	&-0.299	&208.375	&321.426	&2.020
 \\
    -0.5 to 0.5  &5.115	&0.822	&-24.724	&0.443	&201.103	&306.424	&2.029

\\
    -0.25 to 0.25 & 4.443&	0.819&	-30.902&	-1.400&	205.643&	284.705&	2.151
 \\
    -0.1 to 0.1  & 5.420&	0.816&	-28.106&	254.0&	229.290&	391.814&	2.244
\\
    \bottomrule
    \end{tabular}
    \label{tab:performance_metrics_uniform}
    \end{table}
\textcolor{black}{From the tables, it is seen that most of the metrics show a high value for intercepts placed very closely about the origin, and saturates to a lower value as the intercepts are spread out in the interval between $-1$ and $1$. This is because, for extremely low values of intercepts, every neuron fires for the entire range of input data, which reduces the precision of control. Spacing them out causes some neurons to fire for a smaller range of inputs with much higher precision. However, beyond a certain spacing, the minor increase in precision does not reflect in the control values, thus saturating the performance.  Also, for the intercepts spaced linearly and from a uniform distribution, there is a minimum around $0.3$. This is because with larger intercepts, more neurons fire only for the higher inputs, and the precision for lower values of control decreases. This is not observed in the normal distribution as most intercepts are concentrated around the mean ($0$). These results are visually apparent from Figures \ref{fig:Intercept}}. 

% Moreover, the control energy is large for very small values of intercept, where almost all neurons fire for the entire range of inputs.

\textcolor{black}{It is also seen that the control effort is very high for low values of intercept where all neurons fire for the entire range of inputs, as compared to the case where the intercepts are more distributed over a wide range}.

\begin{table}[htbp]
\caption{\textcolor{black}{Performance Metrics for Different Ranges of maximum firing rates}}
\centering
\begin{tabular}{cccccccc}
\toprule
\textbf{Limits} & \textbf{Ts} & \textbf{Tr} & \textbf{PO} & \textbf{SSE} & \textbf{IAE} & \textbf{ITAE} & \textbf{ISC} \\
\textbf{} & \textbf{($s$)} & \textbf{($s$)} & \textbf{($\%$)} & \textbf{($\%$)} & \textbf{($rad.s$)} & \textbf{($rad.s^2$)} & \textbf{($N^2.s$)} \\
     \textbf{} & \textbf{} & \textbf{} & \textbf{} & \textbf{$\times 10^{-3}$} & \textbf{($\times 10^{-3}$)} & \textbf{($\times 10^{-3}$)} & \textbf{($\times 10^2$)} \\
\midrule
200 to 400 & 5.093 & 0.817 & -25.696 & 1.180 & 205.886 & 314.376 &  2.029 \\
250 to 400 & 5.050 & 0.822 & -25.833 & -0.961 & 203.182 & 305.407 & 2.021 \\
300 to 400 & 5.159 & 0.814 & -25.121 & -0.216 & 205.321 & 317.080 & 2.018 \\
350 to 400 & 5.080 & 0.816 & -25.744 & 1.430 & 204.742 & 310.726  & 2.019 \\
400 to 400 & 5.146 & 0.816 & -25.484 & 0.075 & 205.110 & 314.930  & 2.014 \\
200 to 300 & 5.136 & 0.814 & -25.392 & 1.140 & 206.634 & 319.050  & 2.061 \\
250 to 300 & 5.160 & 0.817 & -25.882 & -4.850 & 206.187 & 315.121 & 2.032 \\
300 to 300 & 5.146 & 0.815 & -25.927 & 2.300 & 207.626 & 320.225  & 2.048 \\
200 to 200 & 5.234 & 0.825 & -23.774 & 0.854 & 202.276 & 315.596  & 2.066 \\
\bottomrule
\end{tabular}
\label{tab:performance_metrics_firing}
\end{table}

\textcolor{black}{However, there is no pronounced effect of firing rate on the error or effort metrics. The time metrics are seen to increase with a decrease in firing rate, because the higher the firing rate, the higher is the precision of control, and faster is the convergence.}

\paragraph{\textbf{Ensemble of LIF neurons for the feedback control of a higher DOF mechanical system}}

\begin{figure}[ht]
    \centering
    \subfloat{\includegraphics[width=6cm]{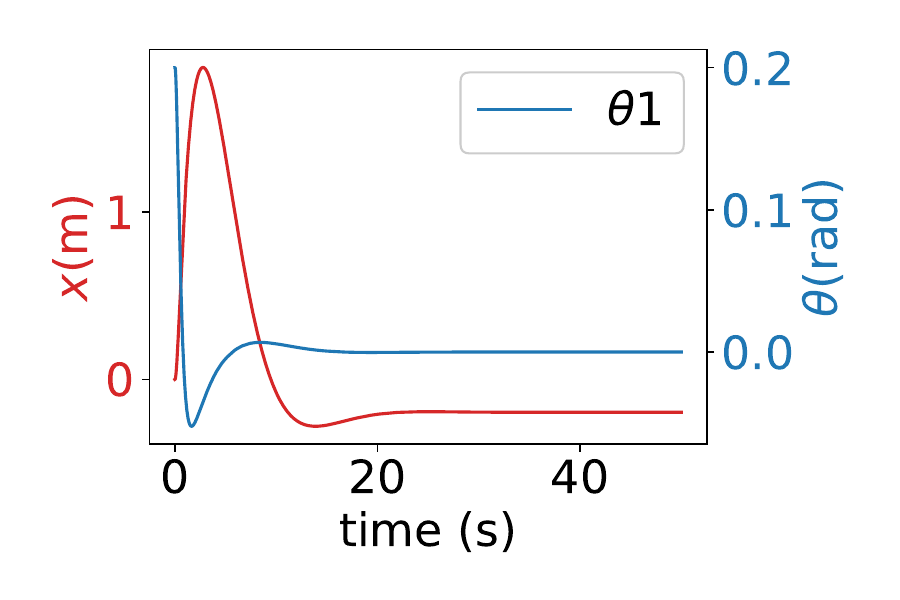}}
    \subfloat{\includegraphics[width=6cm]{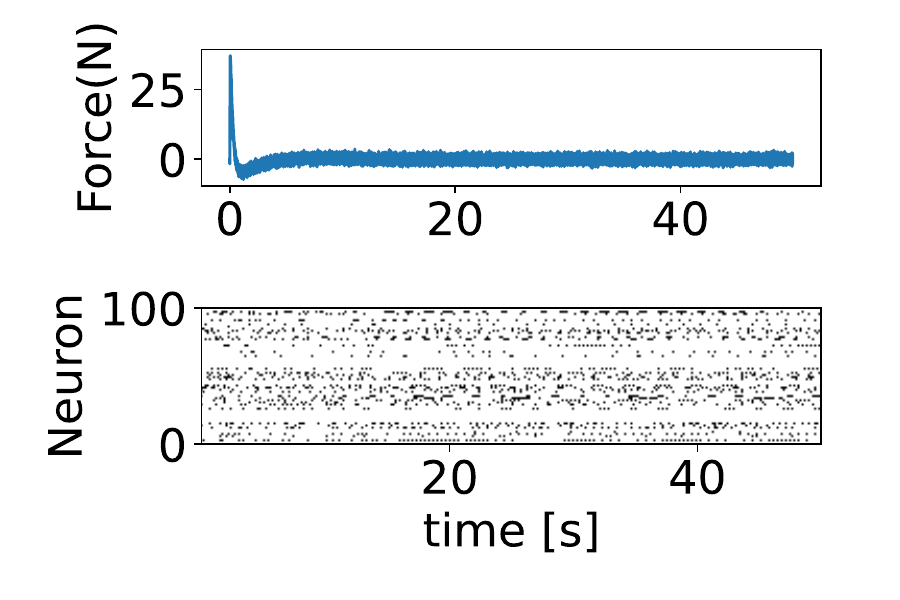}}
    \hfill
    \subfloat{\includegraphics[width=6cm]{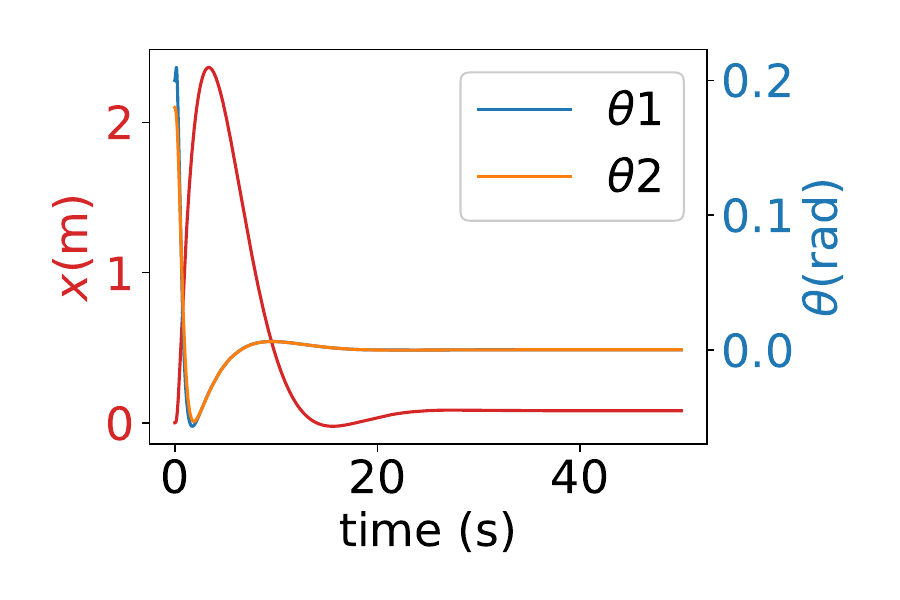}}
    \subfloat{\includegraphics[width=6cm]{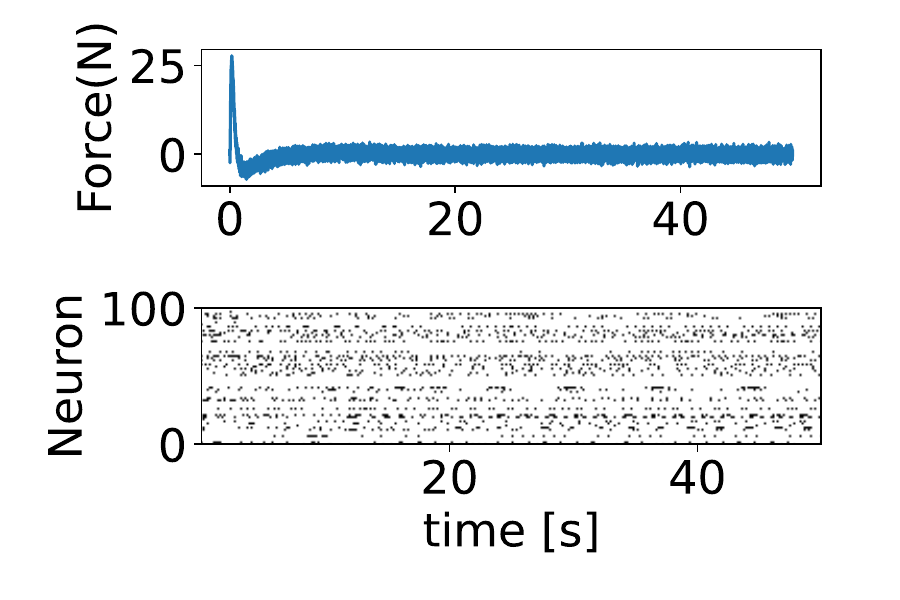}}
    \hfill
    \subfloat{\includegraphics[width=6cm]{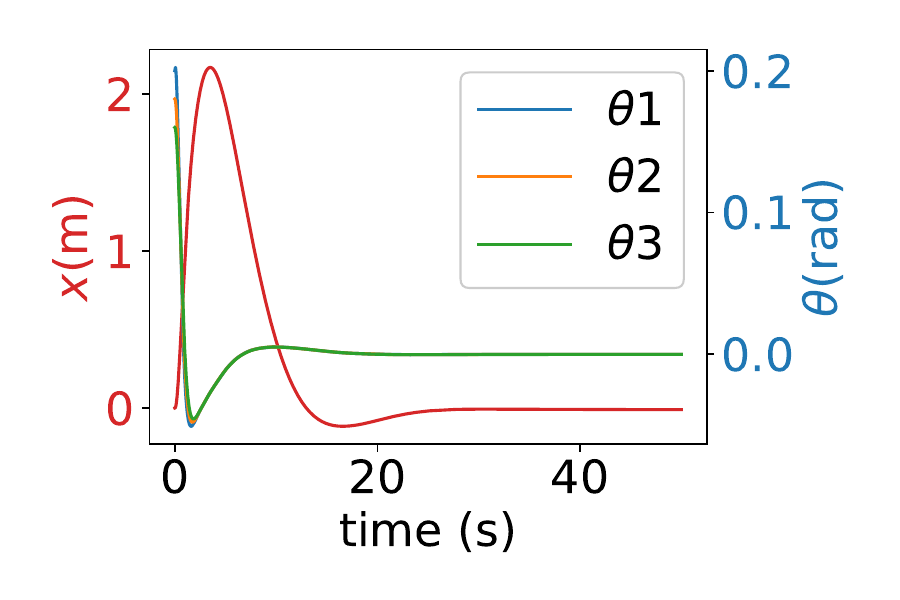}}
    \subfloat{\includegraphics[width=6cm]{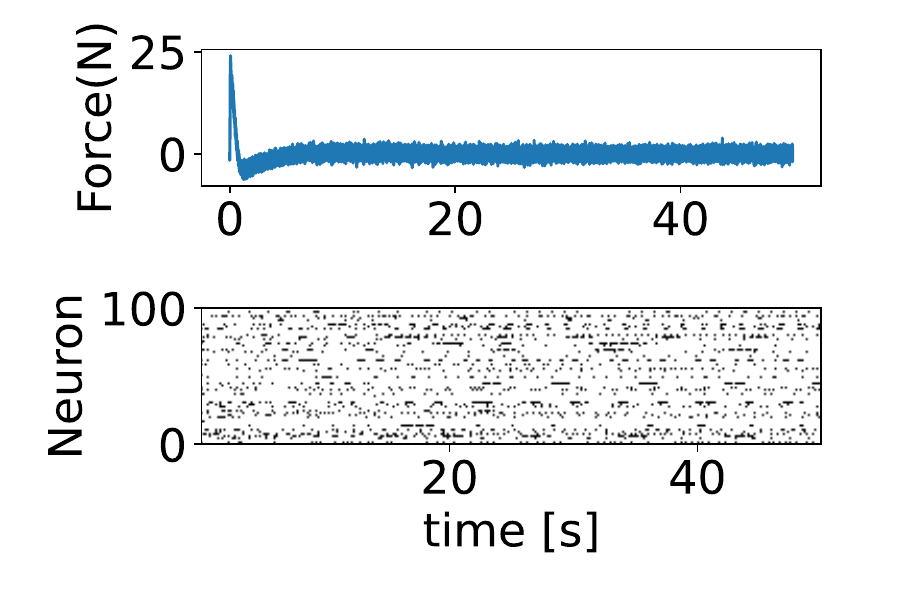}}
    \hfill
    \subfloat{\includegraphics[width=6cm]{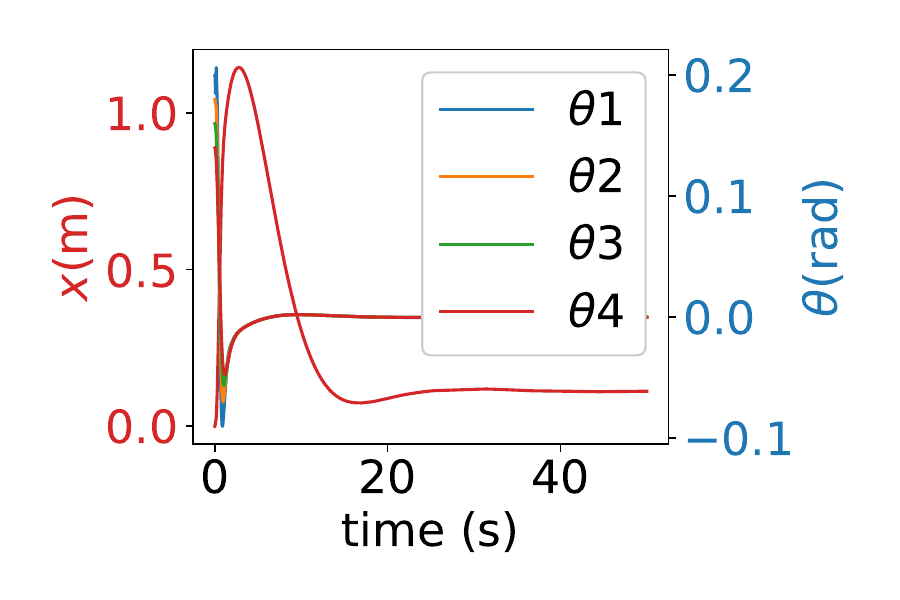}}
    \subfloat{\includegraphics[width=6cm]{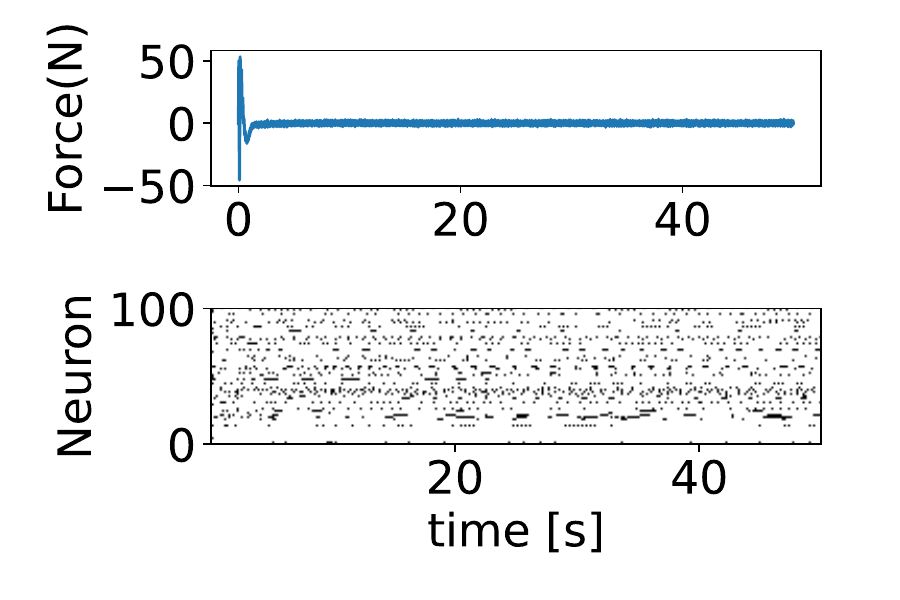}}
    \caption{\textcolor{black}{Plots for multi-linked pendulum on a cart system. The left column shows the evolution of $\theta$(s) and $x$, and the right column shows the control force profile and spike raster plot. From top to bottom, the figures show the curves for a cartpole, a double pendulum on a cart, a triple pendulum on a cart, and a four-linked pendulum on a cart.}}
    \label{fig:Multilinked_pend}
\end{figure}

\begin{figure}[ht]
    \centering
    \subfloat{\includegraphics[width=5cm]{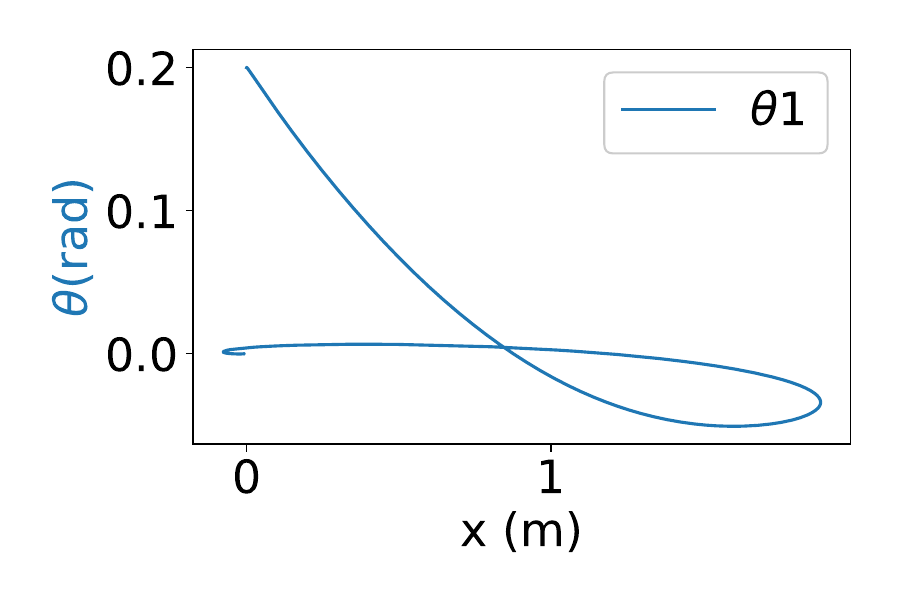}}
    \subfloat{\includegraphics[width=5cm]{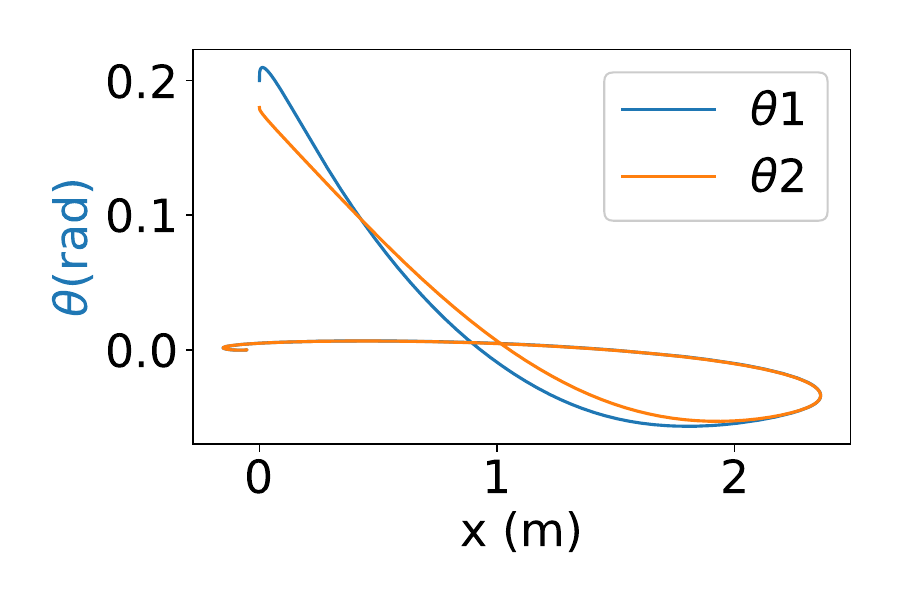}}
    \subfloat{}
    \hfill
    \subfloat{\includegraphics[width=5cm]{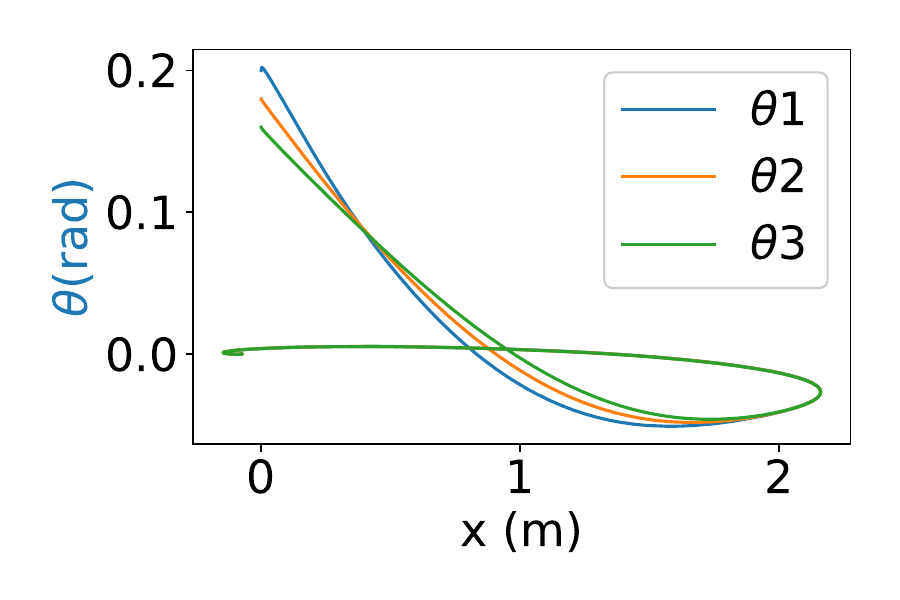}}
    \subfloat{\includegraphics[width=5cm]{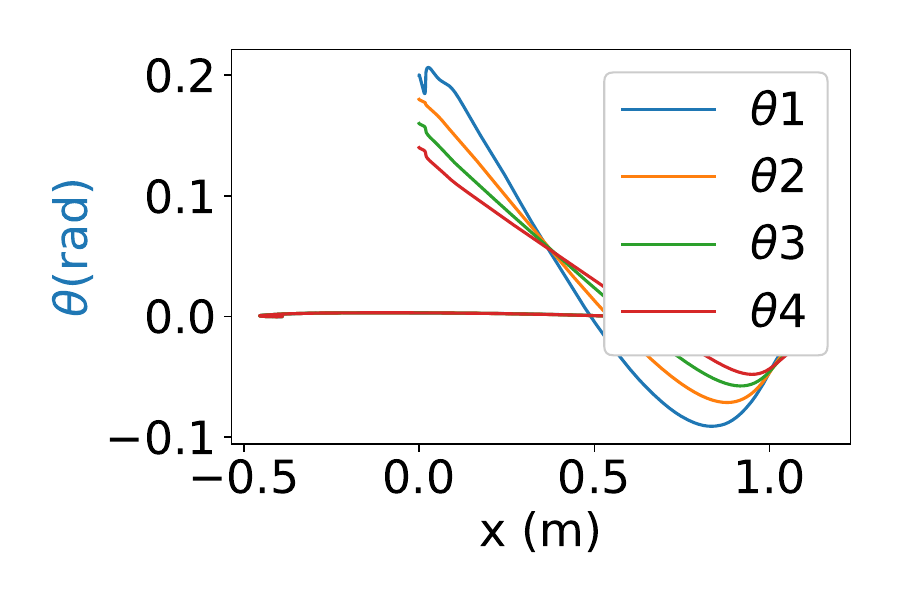}}
    
    \caption{\textcolor{black}{Phase portraits of pole angle versus cart position for multi-linked pendulum on a cart system. (a), (b), (c) and (d) illustrate the phase portraits for a cartpole, a double-linked, triple-linked and 4-linked pendulum on a cart, respectively.}}
    \label{fig:Phase_Multilinked_pend}
\end{figure}

%\subsubsection{\textbf{Balancing an n-linked pole on a cart using NEF in simulation}}
The plots for experiments in Table~\ref{tab:expcontrol} are shown in  Figure~\ref{fig:Multilinked_pend} and Figure~\ref{fig:Phase_Multilinked_pend}. The angles of the links with respect to the UEP, in the multi-linked pendulum scenario are chosen to be different to avoid loss of generality. Although, theoretically, an under-actuated multi-linked pendulum has full linear controllability, the following conditions were used to choose the angles of the links with respect to the UEP:
\begin{enumerate}
\item $\theta_1 = 0.2^c \approx 11.46^\circ$ 
\item $ \theta_n =  \theta_1 - 0.02^\circ \times (n-1)$
\newline
where, $\theta_1$ is the angle of the link directly attached to the cart and $\theta_n$ is the $n^{th}$ link from the cart.
\end{enumerate}
In the Lagrangian equations, for the general case (see Equation~\ref{GenLinEOM}) there are $(\theta_i-\theta_j)$ terms present along with the dot products of angles. In the linearity assumption, these terms are neglected, so the angles are to be chosen carefully so that the time derivatives and angle differences are not too large. 
Table ~\ref{tab:ensemble_metric} shows the control performance metrics for this set of experiments. 
% For the multi-linked pendulum scenario, it is seen that the overshoot is higher but the rise time is lower for the links closer to the cart and vice versa. This is expected as the links closer to the cart, and hence closer to the point of actuation, need to have drastic control movements so that all the links are balanced. For the steady-state characteristics, all the link angles for the multi-linked pendulum on a cart system have the same value as expected. The difference in such angles is only visible for the transients.

The overshoot for a particular link is seen to be proportional to the starting angle of the link, as expected. Moreover, the time metrics and the steady state error of all the links are roughly the same. A slight increase in time metrics is observed as the number of links is increased, which is obvious, since with more un-actuated DOFs, the control becomes more difficult.

\begin{table}
\caption{\textcolor{black}{LQR control performance metrics for Nengo neural ensemble-based control of the cartpole system across varying number of links.}}
    \begin{center}
    \fontsize{9pt}{9pt}\selectfont
    \lineup
    \begin{tabular}{@{}*{11}{l}}
    \br
         & \textbf{Cartpole} & \centre{2}{\textbf{DPC}}  & \centre{3}{\textbf{3lPC}} & \centre{4}{\textbf{4lPC}} \\ \cr
        \ns
        \textbf{Metrics} & \crule{1} & \crule{2} & \crule{3} & \crule{4} \cr
        \textbf{} & \centre{1}{\emph{$\theta$}} & \centre{1}{\emph{$\theta_1$}} & \centre{1}{\emph{$\theta_2$}} & \centre{1}{\emph{$\theta_1$}} & \centre{1}{\emph{$\theta_2$}} & \centre{1}{\emph{$\theta_3$}} & \centre{1}{\emph{$\theta_1$}} & \centre{1}{\emph{$\theta_2$}} & \centre{1}{\emph{$\theta_3$}} & \centre{1}{\emph{$\theta_4$}} \cr
        \mr
        {Rise time (s)} & \00.813  & \multicolumn{2} {c}{\00.976} & \multicolumn{3}{c}{\01.057} & \multicolumn{4}{c}{\00.765} \cr
        {PO (\%)} & \00.255  & \00.28 & \00.26 & \00.245 & \00.230 & \00.220 & \00.415 & \00.315 & \00.26 & \00.22 \cr
        {Settling time (s)} & 16.747  & \multicolumn{2}{c}{18.130} & \multicolumn{3}{c}{22.44} & \multicolumn{4}{c}{22.01} \cr
        {SSE} & \00.0005  & \multicolumn{2}{c}{\0\00.0006} & \multicolumn{3}{c}{\0\0\00.0005} & \multicolumn{4}{c}{\0\0\00.0006} \cr
        \br
    \end{tabular}
    \end{center}
    
    \label{tab:ensemble_metric}
\end{table}

\subsubsection{Hardware}
\begin{figure}[ht]
    \centering
    \subfloat{\includegraphics[width=5cm]{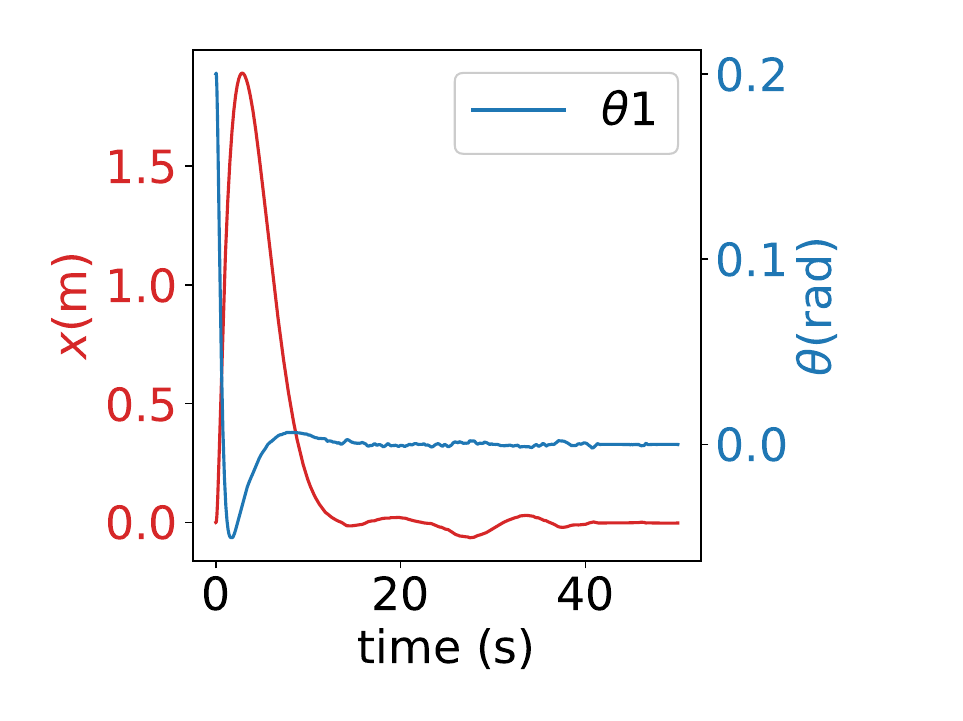}}
    \subfloat{\includegraphics[width=5cm]{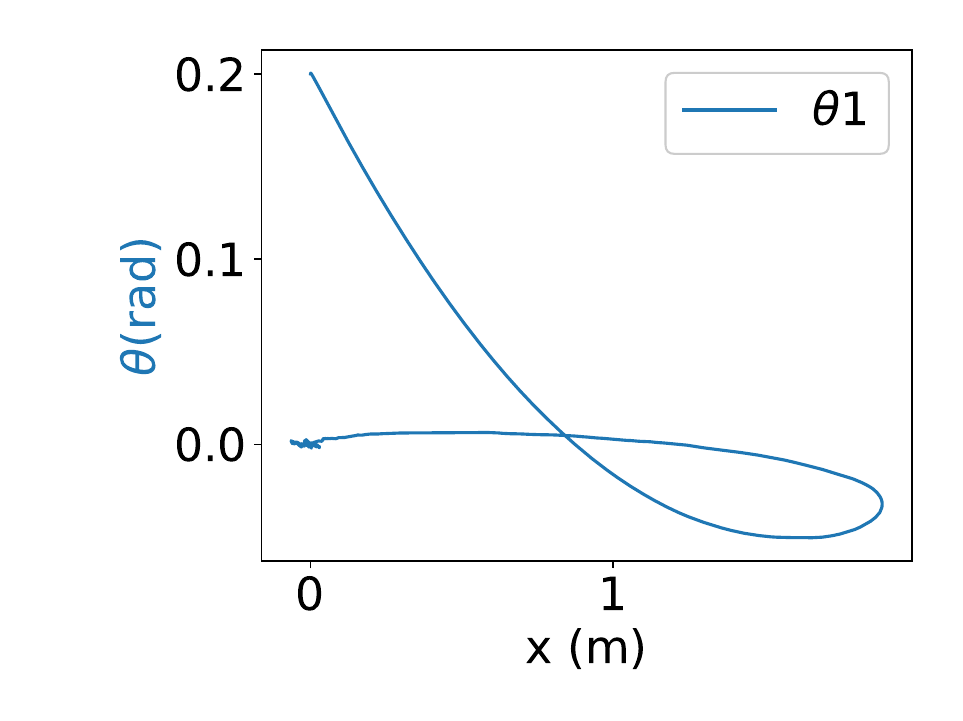}}
    \subfloat{\includegraphics[width=5cm]{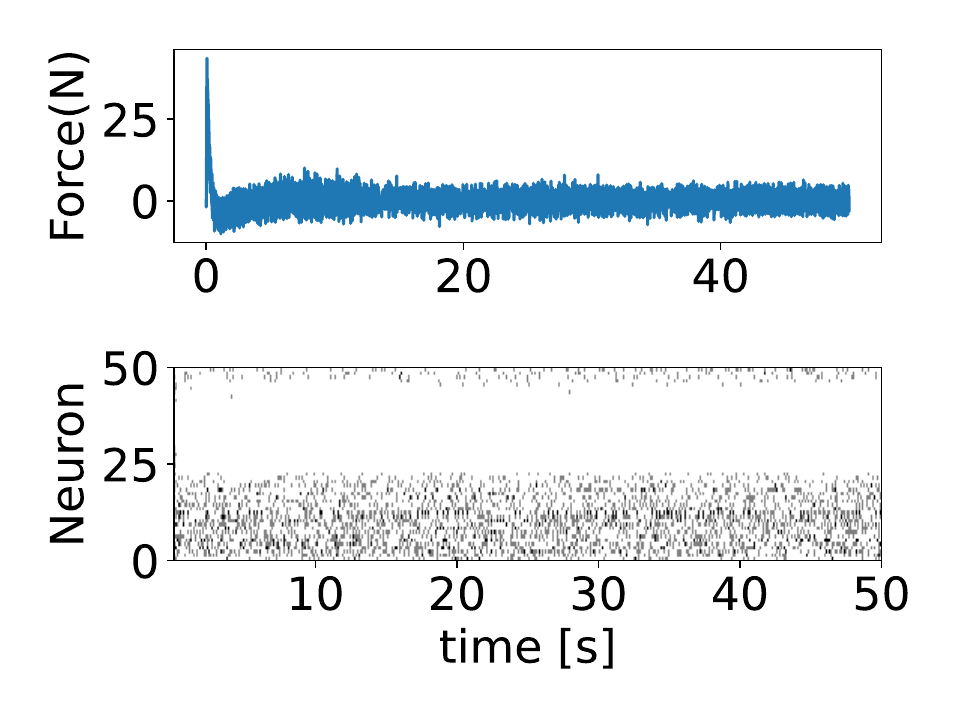}}
   
    \subfloat{\includegraphics[width=5cm]{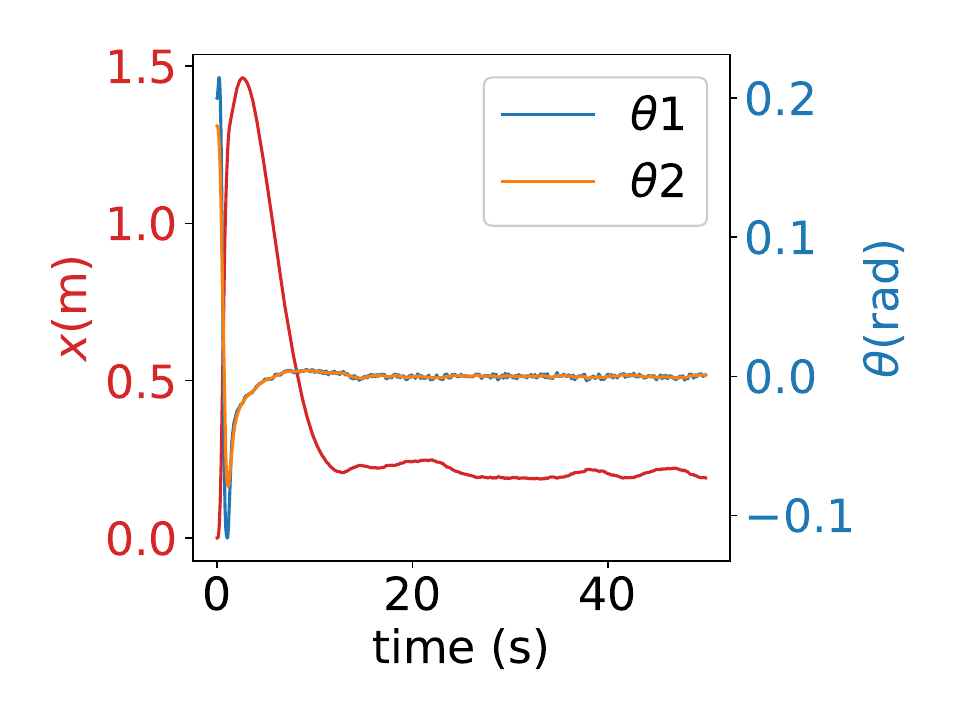}}
    \subfloat{\includegraphics[width=5cm]{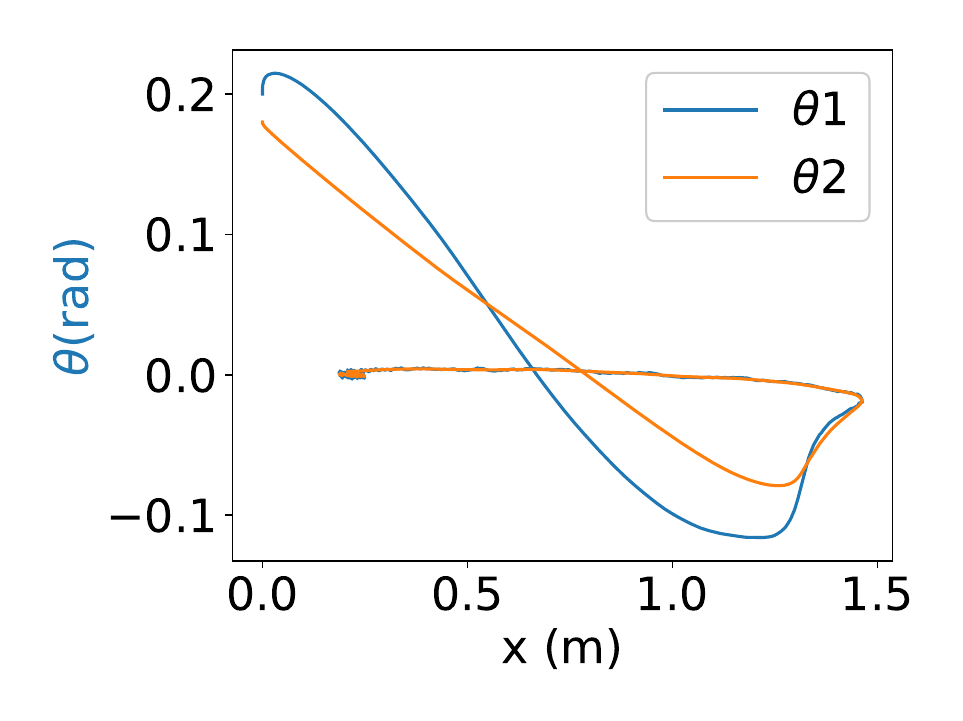}}
    \subfloat{\includegraphics[width=5cm]{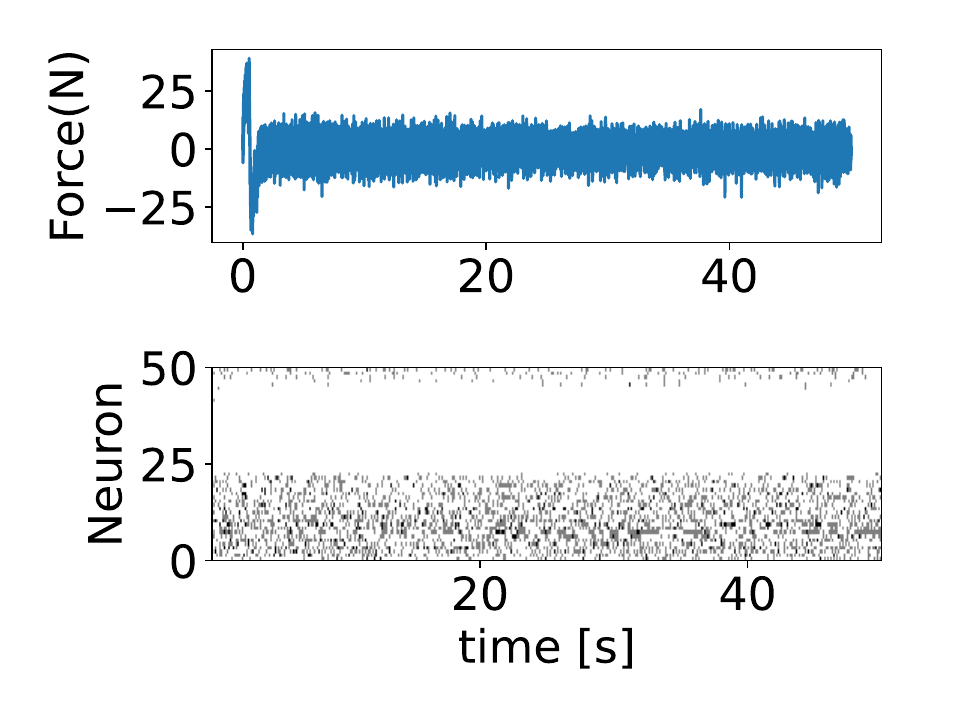}}
    
    \caption{\textcolor{black}{Plots for control multi-linked pendulum on a cart system. The left column shows the evolution of $\theta$(s) and $x$, the middle column shows the phase portrait of $\theta$(s) versus $x$, and the right column shows the control force profile and spike raster plot. From top to bottom, the figures show the the curves for a cartpole and a double pendulum on a cart.}}
    \label{fig:Loihi_Multilinked_pend}
\end{figure}

This subsection outlines the results of implementing control of a multi-linked pendulum on a cart system using spiking neurons implemented on Intel's Loihi neuromorphic hardware. Figure \ref{fig:Loihi_Multilinked_pend} shows the state evolution, phase portrait, and spike raster plots for a cartpole and a double pendulum on a cart system. Table \ref{tab:Loihi_metric} shows the transient and steady-state metrics for the experiments with Loihi. The following observations emerge from the results shown:

\begin{enumerate}
\item Overall, the performance metrics for on-chip implementation is slightly poorer than the simulation. This is because noise levels on the chip are much more than the simulation.
\item The links with a higher starting angle have more overshoot as expected.
\item Control fails for triple, and 4-linked pendulum on a cart, i.e., the poles fall over and are not balanced on the cart for these systems. Clearly, the bounds of accuracy that Loihi provides do not cater to the precision of control demanded by these under-actuated higher-DOF cartpole systems.
\end{enumerate}

\begin{table}
\caption{\textcolor{black}{LQR control performance metrics for Nengo neural ensemble-based control of the cartpole system across varying number of links on Loihi.}}
    \begin{center}
    \fontsize{9pt}{9pt}\selectfont
    \lineup
    \begin{tabular}{@{}*{4}{l}}
    \br
         & \textbf{Cartpole} & \centre{2}{\textbf{DPC}} \\ \cr
        \ns
        \textbf{Metrics} & \crule{1} & \crule{2} \\ \cr
        \textbf{} & \centre{1}{\emph{$\theta$}} & \centre{1}{\emph{$\theta_1$}} & \centre{1}{\emph{$\theta_2$}} \\ \cr
        \mr
        {Rise time (s)} & \00.858  & \multicolumn{2} {c}{\00.810}  \cr
        {PO (\%)} & \00.255  & \00.57 & \00.368 \cr
        {Settling time (s)} & 18.552  & \multicolumn{2}{c}{20.455} \cr
        {SSE} & \00.0005  & \multicolumn{2}{c}{\0\00.0003} \cr
        \br
    \end{tabular}
    \end{center}
    
    \label{tab:Loihi_metric}
\end{table}

\subsubsection{Comparison with other Models of Control}
\textcolor{black}{This section demonstrates a comparison of the spiking LQR performance metrics with spiking PID, both using $100$ spiking neurons. The main drawback of LQR is that the $K_p$, $K_i$, and $K_d$ coefficients should be tuned to match the control metrics desired for the system, based on certain heuristics. However for optimal control methods like LQR, one can specify state and control costs ($Q$ and $R$ matrices), and solve the algebraic Riccati equation to get the desired value of the feedback coefficients. To establish a fair comparison, the $K_p$ and $K_d$ values are chosen to match the corresponding LQR coefficients for the pole angle and pole angular velocity, and then the $K_i$ value is progressively increased from $0$ to $1$ to see the effect of adding the integral controller to the PD controller. Here, $K_p$ is fixed at $190.61$ and $K_d$ is chosen to be $78.26$. The case of $K_i=0$ corresponds to the case of LQR pole angle control. Table \ref{tab:ki_metrics} summarises the 7 performance metrics of control across varying $K_i$.}

    \begin{table}[htbp]
    \caption{\textcolor{black}{Performance Metrics for Different Values of $K_i$ in PID control of the pole angle of a cartpole.}}
    \centering
    \begin{tabular}{cccccccc}
    \toprule
    \textbf{$K_i$} & \textbf{PO} & \textbf{Tr} & \textbf{Ts} & \textbf{SSE} & \textbf{IAE} & \textbf{ITAE} & \textbf{ISC} \\
    \textbf{} & \textbf{($\%$)} & \textbf{($s$)} & \textbf{($s$)} & \textbf{($\%$)} & \textbf{($rad.s$)} & \textbf{($rad.s^2$)} & \textbf{($N^2.m^2.s$)} \\
     \textbf{} & \textbf{} & \textbf{} & \textbf{} & \textbf{$\times 10^{-3}$} & \textbf{($\times 10^{-3}$)} & \textbf{($\times 10^{-3}$)} & \textbf{($\times 10^2$)} \\
    \midrule
    $0.0$  & $0.0$          & $0.0$   & $10.0$  & $0.710$ & $121.813$ & $56.698$& $1.505$ \\
    $0.1$ & $0.0$          & $0.0$   & $10.0$  & $0.041$  & $118.855$ & $52.695$ & $1.509$ \\
    $0.2$ & $0.0$          & $0.0$  & $10.0$  & $0.191$ & $119.033$ & $52.945$ & $1.514$ \\
   $ 0.3$ & $-0.531$ & $2.174$ & $2.177$ & $-0.749$ & $116.432$ & $49.198$ & $1.524$ \\
    $0.4$ & $-0.559$ & $2.133$ & $2.174$ & $-0.462$ & $116.038$ & 48.866 & 1.527 \\
    $0.5$ & $-1.328 $ & $1.840$ & $2.112$ & $-0.982$ & $113.502$ & $45.933$ & $1.533$ \\
    $0.6$ & $-0.908$  & $2.019$ & $2.147$ & $-1.115$ & $114.432$ & $47.233$ & $1.539$ \\
    $0.7$ & $-1.351$ & $1.808$ & $2.102$ & $-1.568$ &$ 112.723$ & $45.357$ & $1.538$ \\
    $0.8$ &$ -2.399$ & $1.617$ & $2.050$ & $-1.982$ & $111.746$ & $44.489 $& $1.545$ \\
   $ 0.9$ & $-1.974$  & $1.636$ & $2.044$ & $-1.002$ & $110.730$ & $43.550$ & $1.558$ \\
   $ 1.0$ & $-3.743$  & $1.472$ & $1.990$ & $-3.885$ & $110.387$ & $43.681$ & $1.558$ \\
    \bottomrule
    \end{tabular}
    \label{tab:ki_metrics}
    \end{table}

\begin{figure}[ht!]
    \centering
    \subfloat[]{\includegraphics[width = 0.4\textwidth]{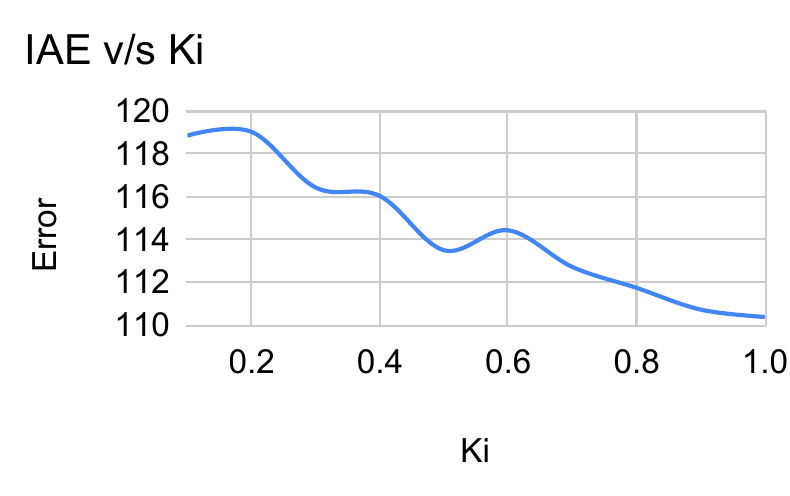}}
    \subfloat[]{\includegraphics[width =0.4\textwidth]{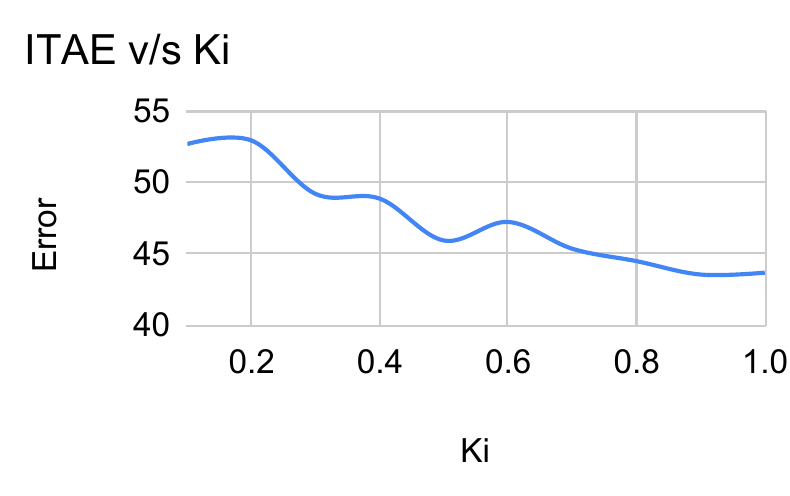}}
    \hspace{0.5mm}
    
    \caption{\textcolor{black}{Variation of (a) IAE and (b) ITAE with $k_i$ for the spiking PID control of the pole angle of a cartpole.}}\label{fig:PID}
    \end{figure}

  \textcolor{black}{It is observed that the error metrics decrease and the response becomes faster (seen as a decrease in the rise and peak time), with an increase in the value of $K_i$ (See Figure \ref{fig:PID}. However, to achieve faster control, the response overshoots. Hence, the overshoot increases with an increase in $K_i$. Overall, it is seen, that if the $K_p$ and $K_d$ values can be tuned to match the performance of an LQR controller, adding the integral control improves performance. However, the challenge of fine-tuning a PID controller as per the desired response makes LQR a better choice for linear optimal control.}

    \textcolor{black}{A detailed comparison with different models of control are highlighted in Table \ref{tab:var_models}. The errors are scaled by $10^3$.} 

\begin{table}[htbp]
\scriptsize
\centering
\caption{\textcolor{black}{Comparison of control methods for spiking and non-spiking models. In all these examples, both the position of the cart and the angle of the pole are being controlled.}}
\begin{adjustbox}{max width=4\textwidth}
\begin{tabular}{llrrrrrrr}
\toprule
\textbf{} & \textbf{Controller} & \textbf{PO} & \textbf{Tr} & \textbf{Ts} & \textbf{SSE} & \textbf{IAE} & \textbf{ITAE} & \textbf{ISC} \\
&\textbf{} & \textbf{($\%$)} & \textbf{($s$)} & \textbf{($s$)} & \textbf{($\%$)} & \textbf{($rad.s$)} & \textbf{($rad.s^2$)} & \textbf{($N^2.m^2.s$)} \\
     &\textbf{} & \textbf{} & \textbf{} & \textbf{} & \textbf{$\times 10^{-3}$} & \textbf{($\times 10^{-3}$)} & \textbf{($\times 10^{-3}$)} & \textbf{($\times 10^2$)} \\
\midrule
\multirow{4}{*}{Spiking} 
& LQR & $-26.57$ & $0.814$ & $5.008$ & $1.39E-4$ & $205.663$ & $305.787$ & $2.09$ \\
& Adaptive LQR & $-31.89$ & $0.599$ & $5.081$ & $-2.00E{-2}$ & $166.678$ & $218.951$ & $0.431$ \\
& PID & $-24.61$ & $0.681$ & $5.268$ & $-1.22E{-2}$ & $205.976$ & $320.914$ & $2.07$\\
& Neural imitator & $-52.60$ & $0.832$ & $5.67$ & $-37.5$ & $349.365$ & $528.552$ & $7.65$ \\
\midrule
\multirow{4}{*}{Non-spiking} 
& LQR & $-9.61E{-3}$ & $0.100$ & $5.122$ & $2.90E{-6}$ & $203.216$ & $313.462$ & $1.87$ \\
& PID & $-7.12E{-2}$ & $0.100$ & $5.276$ & $-9.05E{-3}$ & $210.399$ & $334.977$ & $1.95$ \\
& MPC & $-10.2$ & $0.100$ & $1.511$ & $11.1$ & $66.236$ & $16.474$ & $3.23$ \\
& SMC & $-5.41E{-4}$ & $0.002$ & $0.800$ & $4.59E{-4}$ & $9.05E{-4}$ & $3.61E{-4}$ & $7.20$ \\
\bottomrule
\end{tabular}
\end{adjustbox}
\label{tab:var_models}
\end{table}

\textcolor{black}{$4$ spiking and $4$ non-spiking control models are compared and contrasted. Between spiking LQR and spiking adaptive LQR control, it is seen that adding the adaptive block decreases the time of convergence, which is evident from the Tr metric. The integral error metrics IAE and ITAE are also seen to decrease on adding the adaptive control block. However, due to faster convergence, the spiking adaptive controller causes the response to overshoot more than the spiking LQR controller working alone. A comparison between spiking PID and spiking LQR shows that spiking PID shows lesser overshoot and a quicker response compared to LQR control, which is expected since adding an integral block to the controller keeping the proportional and derivative coefficients same as the LQR controller, causes the PID controller to be more sensitive to the integral errors, and settle down faster. However, the ITAE metric is around $5\%$ higher for PID, although the IAE metric is comparable. This is possibly because with an increase in time, the decrease in the steady-state error becomes less for PID controller, as compared to an LQR controller. 
%For the spiking PID controller, the control effort is higher than spiking LQR, since an added control signal is introduced. 
The non-spiking LQR and PID controllers show a similar trend, but the transient and steady-state metrics are much better for the non-spiking case, as compared to the spiking scenario. This is expected, since a neuromorphic controller is less precise in its outputs, compared to a non-spiking analog controller. The SMC (Sliding Mode Controller) applies "bang-bang" type control, so the controller is faster than LQR or PID, as is seen from the Tr and Ts metrics. Also, the peak overshoot is $1$ order of magnitude lower than LQR, and the steady-state-error is comparable with LQR. Even the IAE and ITAE are $6$ orders of magnitude lower compared to the LQR and PID controllers. This is because the control force for SMC is high (as seen from the control effort metric), allowing for a quick decrease in the error values. MPC optimizes the trajectory of the controller till a certain time in the future, at every time step. Hence, due to progressive optimizations, the overshoot and steady-state errors are much higher compared to LQR controller, since at every time step, the optimizer detects that value of the pole angle and tries to bring it close to the target angle.
The spiking neural imitator, based on the PES learning rule, trains a neural network using supervised learning, with the output from a standard spiking LQR controller as the target control value. As expected in learning-based control, the error and time metrics are worse compared to the spiking classical controllers. The control effort for neural imitator is seen to be about $3$ times higher than the spiking LQR controller. Despite poorer metrics, the strength of a learning-based controller lies in its ability to adapt to changes in system parameters and environmental conditions, and to control a system solely from its input-output data, without requiring accurate knowledge of the system dynamics. }

\begin{table}[ht]
\scriptsize
\centering
\caption{\textcolor{black}{Neuromorphic metrics for different models of control on CPU.}}
\begin{tabular}{lllllllll}
\toprule
 & \textbf{Controller} & \textbf{Sim} & \textbf{Wall} & \textbf{Real} & \textbf{Max} & \textbf{Peak} & \textbf{Spikes/} & \textbf{Estimated} \\
& \textbf{} & \textbf{time} & \textbf{time} & \textbf{time} & \textbf{memory} & \textbf{allocated} & \textbf{neuron} & \textbf{CPU} \\
& \textbf{} & \textbf{(s)} & \textbf{(s)} & \textbf{factor} & \textbf{RSS (MiB)} & \textbf{(MiB)} & \textbf{($\times 10^5$)} & \textbf{energy (J)} \\
\midrule
\multirow{4}{*}{Spiking}
 & LQR             & 10 & 19.597  & 0.510 & 1561.621 & 323.871 & 8.674 & 7.829 \\
 & PID             & 10 & 22.999  & 0.435 & 320.676 & 19.839 & 8.705     & 5.376 \\
 & Adaptive   & 10 & 22.227  & 0.449 & 896.492 & 29.807  & 8.684     & 7.179 \\
 & Neural Imitator & 10 & 25.138  & 0.398 & 386.996 & 19.764 & 1.743    & 9.534 \\
\midrule
\multirow{4}{*}{Non-spiking}
 & PID             & 10 & 19.278  & 0.519 & 897.164 & 3.417 & - & 3.858 \\
 & LQR              & 10 & 18.217  & 0.549 & 317.253 & 2.264 & - & 6.927 \\
 & SMC              & 10 & 14.937  & 0.669 & 922.816 & 2.089  & - & 5.796 \\
 & MPC              & 10 & 1200.048  & 0.009 & 317.429 & 7.907 & - & 495.824 \\
\bottomrule
\end{tabular}
\label{tab:neurocomparisonmet}
\end{table}

\textcolor{black}{A list of neuromorphic metrics for these models of control are listed in Table \ref{tab:neurocomparisonmet}. No improvements in computational performance metrics are seen when a neuromorphic algorithm is being run on a CPU, which is a non-neuromorphic hardware. However, based on evidences in literature\cite{davies2018loihi}, a substantial decrease in energy consumption and latency is expected when this is run on neuromorphic hardware such as Loihi or SpiNNaker.}

%%%%%%%%%%%%%%%%%%%%%%%%%%%%%%%%%%%%%%%%%%%%%%%%%%%%%%%%%%%%%%%%%%%%%%%%

\section{Conclusions}\label{sec:conclusion}
A linear controller for the stabilization of a cartpole has been demonstrated using simulated LIF neurons and actual hardware. From the nature of control, it is clear that  2 neurons (for end-to-end control) can be the lowest number of neurons necessary for such control, but in order to achieve precise control with good performance, a collection(ensemble) of neurons would be needed. Nengo is chosen as the framework for deploying an ensemble of neurons. It is seen that with 2 neurons in the ensemble, the tuning curves show a rate-encoding behavior, but on increasing the number of neurons in the ensemble, the control accuracy increases, since each neuron encodes a part of the input signal. Taking an appropriate population for the ensemble, simulations were done using Nengo, for pendulum on a cart systems with varying number of links. This work also demonstrates, through the chosen classic cartpole control example, how two different encoding schemes have different effects on the control performance. To this effect, a rate and a population-based encoding scheme have been presented and the effects on the KPIs of control and neuromorphic are shown. \textcolor{black}{The following observations emerge from this study:
\newline
\begin{enumerate}
    \item Although the control performance saturates after about $100$ neurons, the CPU memory allocation and Loihi's percentage core utilization keep increasing with the number of neurons. Hence, to operate within the bounds of power and on-board resources, limiting the number of neurons is essential. 
    \item Rate-based encoding is a good option if there are strict constraints on the number of spiking neurons and hence, the power and on-board area utilization. While rate-encoding is robust and can achieve control with very few neurons ($2$ in our case), it has larger inference times and lesser precision.
    \item Population encoded neurons provide a higher precision for control but at the cost of a higher number of neurons and synapses, which imply higher resource on-board and higher power consumption.
    \item It is seen that for a $128$ neuron ensemble, Loihi consumes $3$ orders of magnitude less power than CPU, which decreases all the way to $5$ orders of magnitude for a $2$ neuron ensemble.
    \item The fact that the same collection of neurons, ($100$ in this study) could perform balancing of a pendulum across so many experiments indicates that a range of DOF variation for the inverted pendulum systems can be handled with a constant number of neurons to reduce overhead for spiking neural processors.
    \item  Although possible in Nengo simulation, the precision and noise levels in Intel's Loihi neuromorphic board do not allow for linear control of more than 2-linked pendulum on a cart, for the chosen set of control, system and neuron parameters. 
    \item The performance metrics for different models of control indicate that LQR is a better choice for linear control over PID, when the exact coefficients for the states and their derivatives are to be determined based on the desired control performance. However, once these coefficients are known, adding an integral block improves the time and error metrics.
    \item The spiking version of these controllers shows poorer performance compared to the non-spiking ones, which is expected due to higher noise levels and lower precision of spiking networks. However, the spiking controllers are expected to consume orders of magnitude less power and compute at a much lower latency, when run on neuromorphic hardware.
    \item Although training a spiking neural network for such control increases the time to converge and the error values, learning-based control is highly favoured. This is because of its adaptability to perturbations and system parameter variations, and the ability to learn the control behaviour from data, without the need for an exact model of the system.
\end{enumerate}
}
This work can be extended to other control methodologies such as constrained LQR, where one can specify constraints on the state variables which would be important, for example, for physical cartpole systems where the length of the rails inherently limits the movement of the cart. Secondly, a linear control of these systems with the desired steady-state and transient response can be performed to check for the number of neurons needed versus the performance of control.

\section*{Acknowledgments}\label{sec7}
This publication has emanated from research conducted with the financial support of Research Ireland under Grant number 18/CRT/6049. 
%\textcolor{black}{For Open Access, the author has applied a CC BY public copyright license to any Author Accepted Manuscript version arising from this. VP: I think we can store this in Romeo-Green versions like Researchgate or on institutional sites can be acceptable by SFI the funding body. I am unsure if the CRT has additional funding for Open Access for IoP journals but check}. 
This work was inspired by activities started at the CapoCaccia Workshop toward Neuromorphic Intelligence 2023 \cite{CCNW2023}, Sardinia, Italy. The majority of the experiments were done in this lab set up in the workshop where several researchers gathered to discuss neuromorphic problems.
We would like to especially thank Sebastian Billaudelle and Yannik Stradmann from  Heidelberg University, Germany for their help and support with the Lu.i board its use in our experiments. Thanks to Felix Schmitt for help with Lu.i interfacing and related linear control of cartpole, Samia Mohinta for helping us brainstorm ideas, and Tobi Delbruck from INI labs, University of Zurich for bringing a cartpole system to the workshop for implementing various algorithms. Shreyan Banerjee would like to acknowledge Rohit Chawla, for help with Digilent software and hardware. The authors would also like to acknowledge Chiara Bartolozzi from the Italian Institute of Technology for her constant support in this work. Vikram Pakrashi acknowledges Science Foundation Ireland NexSYs 21/SPP/3756, TRaIN 22/NCF/FD/10995, Harmoni 22/FFP-P/11457 and Sustainable Energy Authority of Ireland REMOTEWIND RDD/613, TwinFarm RDD/604, FlowDyn RDD/966, and Interreg SiSDATA EAPA-0040/2022.

\newpage
\section*{\textcolor{black}{Supplementary materials}}\label{sec7}

\subsection{Control Performance Metrics}
This section describes the metrics of control studied in this work.
\begin{itemize}

  \item \textbf{Rise Time} (\(t_r\)):\\
  The time required for the system response to rise from 10\% to 90\% of its final steady-state value.
  \[
  t_r = t_{90\%} - t_{10\%}
  \]

  \item \textbf{Peak Overshoot} (\(M_p\)):\\
  The maximum overshoot relative to the steady-state value, expressed as a percentage.
  \[
  M_p = \left( \frac{y_{\text{peak}} - y_\infty}{y_\infty} \right) \times 100\%
  \]

  \item \textbf{Settling Time} (\(t_s\)):\\
  The time required for the output to remain within a specified tolerance band (typically \(\pm2\%\) or \(\pm5\%\)) around the steady-state value.
  \[
  t_s = \text{time when } |y(t) - y_\infty| < \epsilon \text{ for all } t \ge t_s
  \]

  \item \textbf{Steady-State Error} (\(e_{ss}\)):\\
  The difference between the desired output and the actual output as time approaches infinity.
  \[
  e_{ss} = \lim_{t \to \infty} |r(t) - y(t)|
  \]

  \item \textbf{Integral of Absolute Error (IAE)}:\\
  Measures the total absolute error over time.
  \[
  \text{IAE} = \int_0^\infty |e(t)| \, dt
  \]

  \item \textbf{Integral of Time-weighted Absolute Error (ITAE)}:\\
  Penalizes errors that persist over time by weighting them with time.
  \[
  \text{ITAE} = \int_0^\infty t \cdot |e(t)| \, dt
  \]

  \item \textbf{Integral of Squared Control (ISU)}:\\
    Measures the total squared control effort applied over time. Useful in penalizing aggressive control inputs.
    \[
    \text{ISC} = \int_0^\infty u^2(t) \, dt
    \]

\end{itemize}

\subsection{Simulation Performance Metrics}
This section includes the metrics of simulation performance studied in this work.
\begin{itemize}

  \item \textbf{Sim Time}:\\
  The total simulated duration of the model or system, typically measured in seconds.\\
  Example: Simulating 10 seconds of robot behavior.

  \item \textbf{Wall Time}:\\
  The actual elapsed (real-world) time required to run the simulation on a computer.\\
  Example: A 10s simulation takes 20s of wall time.

  \item \textbf{Real-Time Factor (RTF)}:\\
  The ratio of simulated time to wall-clock time. Indicates simulation speed relative to real time.
  \[
  \text{RTF} = \frac{\text{Sim Time}}{\text{Wall Time}}
  \]
  - \( \text{RTF} > 1 \): faster than real-time  
  - \( \text{RTF} < 1 \): slower than real-time

  \item \textbf{Max Memory RSS (Resident Set Size)}:\\
  The maximum amount of physical RAM occupied by the simulation process during its execution, including code, data, and shared libraries.\\
  Typically measured in mebibytes (MiB) or gibibytes (GiB).

  \item \textbf{Peak Allocated Memory}:\\
  The highest amount of memory dynamically allocated by the simulation, excluding shared libraries or OS overhead.\\
  This is often reported by memory profilers.

  \item \textbf{Estimated CPU Energy}:\\
  An estimate of the total CPU energy consumed during simulation, often measured in joules (J) or watt-hours (Wh).\\ It is measured using the following formula:\\
  \(
  Energy\,(J) = Wall\,Time\,(s) \times CPU\,Utilization \times TDP\,(W)
 \)
  , where TDP stands for the Thermal Design Power of the CPU.
  Useful for evaluating power efficiency, especially on embedded or neuromorphic systems.

\end{itemize}

\subsection{Neuromorphic Performance Metrics}
This section includes the metrics of neuromorphic performance studied in this work.
\begin{itemize}
  \item \textbf{Spikes per neuron}:\\
   This metric shows the average number of spikes emitted by each neuron in the network/ensemble during the simulation.
   \item \textbf{Core Utilization}:\\
   This metric demonstrates the percentage of each neuromorphic core on average, utilized during building the spiking neural network model on a neuromorphic chip.
   \item \textbf{Experimental Area on Chip}:\\
   This is the experimentally determined chip area allocated to the spiking neural network during simulation. It is measured using the following formula:\\
   \(Expt\,Area \, (mm^2) = Area\,per\,Core\,(mm^2) \times \%\, per\,Core \times Number\,of\,Cores\)
   \item \textbf{Theoretical Area on Chip}: \\
   This is the area on chip allocated to the spiking neural network, theoretically calculated using the datasheet for the chip using the formula: \\
   \(Theo\,Area \, (mm^2) = \frac{Number\,of\,neurons\,in\,the\,network}{Neuron\,density\,(mm^{-2})}\)
   \item \textbf{Estimated Energy}:\\
   This is an estimate of the energy consumed by the board (excluding the idle power consumption required to keep the board ON), calculated using the following formula:\\
   \(Energy\,(J/inf)=Energy/synaptic\,op\times synaptic\,ops/inf\,(J/inf)\, + \, Energy/neuron\,update\times neuron\,updates/inf\,(J/inf)\)
   
\end{itemize}

\subsection{Control Videos}
Videos of the control in action with the spiking neurons can be seen at the following github page:
\href{https://github.com/shreyanban/NCE_Reviewed_Videos}{https://github.com/shreyanban/NCE Reviewed Videos}.

% \textcolor{orange}{Please add an Appendix section discussing the metrics, and citing them}

%%%%%%%%%%%%%%%%%%%%%%%%%%%%%%%%%%%%%%%%%%%%%%%%%%%%%%%%%%%%%%%%%%%%%%%%%%%%%%%%%%%%%%%%%%%%%%

\section*{References}
\bibliographystyle{plain}
\bibliography{ref.bib}

\end{document}